\documentclass[reqno,10pt,centertags]{amsart}
\usepackage{amssymb}
\usepackage{upref}
\usepackage{hyperref}
\newcommand{\bbN}{{\mathbb{N}}}
\newcommand{\bbR}{{\mathbb{R}}}
\newcommand{\bbP}{{\mathbb{P}}}
\newcommand{\bbZ}{{\mathbb{Z}}}
\newcommand{\bbC}{{\mathbb{C}}}

\newcommand{\calB}{{\mathcal B}}
\newcommand{\calC}{{\mathcal C}}
\newcommand{\calD}{{\mathcal D}}
\newcommand{\calE}{{\mathcal E}}
\newcommand{\calF}{{\mathcal F}}
\newcommand{\calG}{{\mathcal G}}

\newcommand{\calK}{{\mathcal K}}
\newcommand{\calL}{{\mathcal L}}
\newcommand{\calM}{{\mathcal M}}

\renewcommand{\gg}{p}

\newcommand{\p}{p}
\newcommand{\dott}{\,\cdot\,}
\newcommand{\hatt}{\widehat}  % use \hat in subscripts and upperlimits of int.

\newcommand{\Div}{\operatorname{Div}}
\newcommand{\mini}{\wedge}

\newcommand{\no}{\nonumber}
\newcommand{\lb}{\label}
\newcommand{\f}{\frac}
\newcommand{\ul}{\underline}
\newcommand{\ol}{\overline}
\newcommand{\ti}{\tilde}
\newcommand{\wti}{\widetilde}
\newcommand{\N}{n}
\newcommand{\uc}{{\underline{c}}}

\newcommand{\Oh}{O}

\newcommand{\bi}{\bibitem}

\newcommand{\humu}{{ \hat{\underline{\mu} }}}

\newcommand{\hmu}{{\hat{\mu} }}

\newcommand{\umu}{{\underline{\mu}}}

\newcommand{\ome}{\omega}
\newcommand{\Ome}{\Omega}
\newcommand{\eps}{\varepsilon}

\newcommand{\ual}{{\underline{\alpha}}}

\newcommand{\Pinfp}{{P_{\infty_+}}}
\newcommand{\Pinfm}{{P_{\infty_-}}}

\newcommand{\Pinfpm}{{P_{\infty_\pm}}}

\newcommand{\Pinfmin}{{P_{\infty_{-}}}}
\newcommand{\Pinfplus}{{P_{\infty_{+}}}}
\newcommand{\uU}{{\underline{U}}}

\renewcommand{\Im}{\text{\rm Im}}

\DeclareMathOperator{\tl}{Tl}
\DeclareMathOperator{\sTl}{s-Tl}
\newcommand{\shTl}{{\mathop{\text{\rm s-}\widehat{\text{\rm Tl}}}}}
\DeclareMathOperator{\stl}{s-Tl}

\DeclareMathOperator{\sym}{Sym}

\newcommand{\symgg}{{\sym^\gg (\calK_\gg)}}

\allowdisplaybreaks
\numberwithin{equation}{section}
\newtheorem{theorem}{Theorem}[section]
\newtheorem{lemma}[theorem]{Lemma}

\newtheorem{hypothesis}[theorem]{Hypothesis}
\newtheorem{remark}[theorem]{Remark}
\theoremstyle{definition}

\newtheorem{example}[theorem]{Example}

\newcommand{\abs}[1]{\lvert#1\rvert}

\begin{document}

\title[The Toda Hierarchy Initial Value Problem]{The Algebro-Geometric Toda
Hierarchy Initial Value Problem for Complex-Valued Initial Data}
% Information for  author
\author[F.\ Gesztesy]{Fritz Gesztesy}
\address{Department of Mathematics,
University of Missouri,
Columbia, MO 65211, USA}
\email{fritz@math.missouri.edu}
\urladdr{http://www.math.missouri.edu/personnel/faculty/gesztesyf.html}
%\thanks{}
% Information for  author
\author[H.\ Holden]{Helge Holden}
\address{Department of Mathematical Sciences,
Norwegian University of
Science and Technology, NO--7491 Trondheim, Norway}
\email{holden@math.ntnu.no}
\urladdr{http://www.math.ntnu.no/\~{}holden/}
\thanks{Research supported in part by the Research Council of Norway,  
the US National Science Foundation under Grant No.\ DMS-0405526, and
the Austrian Science Fund (FWF) under Grant No.\ P17762.}
% Information for  author
\author[G.\ Teschl]{Gerald Teschl}
\address{Faculty of Mathematics\\University of Vienna, Nordbergstrasse
15, 1090 Wien,  Austria, and
International Erwin Schr\"odinger Institute for Mathematical Physics,
Boltzmanngasse 9, 1090 Wien, Austria}
\email{\href{mailto:Gerald.Teschl@univie.ac.at} 
{Gerald.Teschl@univie.ac.at}}
\urladdr{\href{http://www.mat.univie.ac.at/~gerald/}
{http://www.mat.univie.ac.at/\~{}gerald/}}
%\thanks{}
%----- end authors
%\dedicatory{}
%\date{August , 2005}
\date{\today}
\thanks{To appear in {\it Rev. Mat. Iberoamericana}.}

%%%%%%%%%%%%%%%%%%%%%%%%%%%%%%%%%%%%%%%%%%%
\begin{abstract}
We discuss the algebro-geometric initial value problem for the Toda hierarchy
with complex-valued initial data and prove unique solvability globally in
time for a set of initial (Dirichlet divisor) data of full measure. To this effect we develop a new algorithm for constructing stationary complex-valued algebro-geometric solutions of the Toda hierarchy, which is of independent interest as it solves the inverse 
algebro-geometric spectral problem for generally non-self-adjoint Jacobi operators, 
starting from a suitably chosen set of initial divisors of full measure. Combined with an appropriate first-order system of differential equations with respect to time (a substitute for the well-known Dubrovin equations), this yields the construction of global algebro-geometric solutions of the time-dependent Toda hierarchy. 

The inherent non-self-adjointness of the underlying Lax (i.e., Jacobi) operator associated with complex-valued coefficients for the Toda hierarchy poses a variety of difficulties that, to the best of our knowledge, are successfully overcome here for the first time. Our approach is not confined to the Toda hierarchy but applies generally to $1+1$-dimensional completely integrable (discrete and continuous) soliton equations. 
\end{abstract}
%%%%%%%%%%%%%%%%%%%%%%%%%%%%%%%%%%%%%%%%%%%%

\maketitle

%%%%%%%%%%%%%%%%%%%%%%%%%%%%%%%%%%%%%%%%%%
%%%%%%%%%%%%%%%%%%%%%%%%%%%%%%%%%%%%%%%%%% 
\section{Introduction} \lb{s1} 
%%%%%%%%%%%%%%%%%%%%%%%%%%%%%%%%%%%%%%%%%%
%%%%%%%%%%%%%%%%%%%%%%%%%%%%%%%%%%%%%%%%%%

The principal aim of this paper is an explicit construction of unique global solutions of the algebro-geometric initial value problem for the Toda hierarchy with complex-valued initial data. More precisely, we intend to describe a solution of the following problem: Given $p\in\bbN_0$, assume $a^{(0)}, b^{(0)}$ to be complex-valued solutions of the $p$th stationary Toda system $\stl_p(a,b)= 0$ associated with a prescribed nonsingular hyperelliptic curve $\calK_p$ of genus $p$ and let $r\in\bbN_0$; we want to construct unique global solutions $a=a(t_r),b=b(t_r)$ of the $r$th Tl  flow $\tl_r(a,b)=0$ with $a(t_{0,r})=a^{(0)}, b(t_{0,r})=b^{(0)}$ for some $t_{0,r}\in\bbR$. Thus, we seek a unique global solution of the initial value problem 
\begin{align}
\begin{split}
&\tl_r(a,b)=0, \label{1.1} \\  
&(a,b)\big|_{t_r=t_{0,r}}=\big(a^{(0)},b^{(0)}\big),  
\end{split} \\
&\sTl_p\big(a^{(0)},b^{(0)}\big)=0 \label{1.2}
\end{align}
for some $t_{0,r}\in\bbR$, $p,r\in\bbN_0$, where $a=a(n,t_r),b=b(n,t_r)$
satisfy 
\begin{align}
\begin{split}
& a \colon \bbZ\times\bbR\to\bbC\setminus\{0\}, \quad 
b \colon \bbZ\times\bbR\to\bbC,  \\
& a(\dott,t), b(\dott,t)\in \bbC^{\bbZ},\; t\in\bbR,
\quad a(n,\dott), b(n,\dott)\in C^1(\bbR), \; n\in\bbZ.   \lb{1.3A}
\end{split}
\end{align}

In the special case of a self-adjoint Lax (i.e., Jacobi) operator $L$, where $a$ and $b$ are real-valued and bounded, the actual solution of this algebro-geometric initial value problem consists of the following two-step procedure discussed in detail in 
\cite{BGHT98} (see also \cite[Sect.\ 1.3]{GH07}, \cite[Sect.\ 8.3]{Te00}):\footnote{We freely use the notation of divisors of degree $p$ as introduced in Appendix \ref{sA}.}

$(i)$ An algorithm that constructs finite nonspecial divisors 
$\calD_{\humu(n)}\in \symgg$ in real position for all $n\in\bbZ$ starting from an initial  Dirichlet divisor $\calD_{\humu(n_0)}\in \symgg$ in an appropriate real position (i.e., with Dirichlet eigenvalues in appropriate spectral gaps of $L$).  ``Trace formulas'' of the type 
\eqref{1.2.21b} and \eqref{1.2.21a} then construct  the stationary real-valued solutions $a^{(0)}, b^{(0)}$ of $\stl_p(a,b)= 0$.

$(ii)$ The first-order Dubrovin-type system of differential equations \eqref{3.6.16A}, augmented by the initial divisor $\calD_{\humu(n_0, t_{0,r})}=\calD_{\humu(n_0)}$ together with the analogous ``trace formulas'' \eqref{5.38}, \eqref{5.39} then yield unique global real-valued solutions $a=a(t_r),b=b(t_r)$ of the $r$th Tl  flow $\tl_r(a,b)=0$ satisfying $a(t_{0,r})=a^{(0)}, b(t_{0,r})=b^{(0)}$.

This approach works perfectly in the special self-adjoint case where the Dirichlet divisors 
$\humu(n,t_r)=(\hat\mu_1(n,t_r),\dots,\hat\mu_p(n,t_r))\in \symgg$, 
$(n,t_r)\in\bbZ\times\bbR$, yield Dirichlet eigenvalues $\mu_1(n,t_r),\dots,\mu_p(n,t_r)$ of the Lax operator $L$ situated in $p$ different spectral gaps of $L$ on the real axis. In particular, for fixed $(n,t_r)\in\bbZ\times\bbR$, the Dirichlet eigenvalues 
$\mu_j(n,t_r)$, $j=1,\dots,p$, are pairwise distinct and formulas \eqref{5.39} for $a$ and \eqref{3.6.16A} for $(d/dt_r)\mu_j(n,t_r)$, $j=1,\dots,p$, are well-defined.

This situation drastically changes if complex-valued initial data $a^{(0)}, b^{(0)}$ or 
$\calD_{\humu(n_0,t_{0,r})}$ are permitted. In this case the Dirichlet eigenvalues 
$\mu_j(n,t_r)$, $j=1,\dots,p$, are no longer confined to well separated spectral gaps of 
$L$ on the real axis and, in particular, they are in general no longer pairwise distinct and ``collisions'' between them can occur at certain values of $(n,t_r)\in\bbZ\times\bbR$. Thus, the stationary algorithm in step $(i)$ as well as the Dubrovin-type first-order system of differential equations \eqref{3.6.16A} in step $(ii)$ above, breaks down at such values of $(n,t_r)$. A priori, one has no control over such collisions, especially, it is not possible to identify initial conditions $\calD_{\humu(n_0, t_{0,r})}$ at some $(n_0,t_{0,r})\in\bbZ\times\bbR$ which avoid collisions for all $(n,t)\in\bbZ\times\bbR$. We solve this problem head on by explicitly permitting collisions in the stationary as well as time-dependent context from the outset. In the stationary context, we properly modify the algorithm described above in step $(i)$ in the self-adjoint case by alluding to a more general interpolation formalism (cf.\ Appendix \ref{sB}) for polynomials, going beyond the usual Lagrange interpolation formulas. In the time-dependent context we replace the first-order system of Dubrovin-type equations \eqref{3.6.16A}, augmented with the initial divisor $\calD_{\humu(n_0, t_{0,r})}$, by a different first-order system of differential equations \eqref{6.32} with initial conditions \eqref{6.33} which focuses on symmetric functions of $\mu_1(n,t_r),\dots,\mu_p(n,t_r)$ rather than individual Dirichlet eigenvalues 
$\mu_j(n,t_r)$, $j=1,\dots,p$. In this manner it will be shown that collisions of Dirichlet eigenvalues no longer pose a problem.  

In addition, there is a second nontrivial complication in the non-self-adjoint case: Since the Dirichlet eigenvalues $\mu_j(n,t_r)$, $j=1,\dots,p$, are no longer confined to spectral gaps of $L$ on the real axis as $(n,t_r)$ vary in $\bbZ\times\bbR$, it can no longer be guaranteed that $\mu_j(n,t_r)$, $j=1,\dots,p$, stay finite for all 
$(n,t_r)\in\bbZ\times\bbR$. As discussed in Section \ref{s4} in the stationary case, this phenomenon is related to certain deformations of the algebraic curve $\calK_p$ under which for some $n_0\in\bbZ$, $a(n_0)\to 0$ and $\mu_j(n_0+1)\to\infty$ for some 
$j\in\{1,\dots,p\}$. We solve this particular problem in the stationary as well as time-dependent case by properly restricting the initial Dirichlet divisors 
$\calD_{\humu(n_0, t_{0,r})}\in \symgg$ to a dense set of full measure.

Summing up, we offer a new algorithm to solve the inverse algebro-geometric spectral problem for generally non-self-adjoint Jacobi operators, starting from a properly chosen dense set of initial divisors of full measure. Combined with an appropriate first-order system of differential equations with respect to time (a substitute for the well-known Dubrovin equations), this yields the construction of global algebro-geometric solutions of the time-dependent Toda hierarchy. 

We emphasize that the approach described in this paper is not limited to the Toda  hierarchy but applies universally to constructing algebro-geometric solutions of $1+1$-dimensional integrable soliton equations. In particular, it applies to differential-difference (i.e., lattice) systems and we are now in the process of applying it to the Ablowitz--Ladik hierarchy. Moreover, the principal idea of replacing Dubrovin-type equations by a first-order system of the type \eqref{6.32} is also relevant in the context of general non-self-adjoint Lax operators for the continuous models in $1+1$-dimensions. (In particular, the models studied in detail in \cite{GH03} can be revisited from this point of view.) We also note that while the periodic case with complex-valued $a, b$ is of course included in our analysis, we throughout consider the more general algebro-geometric case (in which 
$a, b$ need not even be quasi-periodic).  

Finally we briefly describe the content of each section. Section \ref{s2} presents a quick summary of the basics of the Toda hierarchy, its recursive construction, Lax pairs, and zero-curvature equations. The stationary algebro-geometric Toda hierarchy solutions, the underlying hyperelliptic curve, trace formulas, etc., are the subject of Section \ref{s3}. A new algorithm solving the algebro-geometric inverse spectral problem for generally non-self-adjoint Jacobi operators is presented in Section \ref{s4}. In Section \ref{s5} we briefly summarize the properties of algebro-geometric time-dependent solutions of the Toda hierarchy and formulate the algebro-geometric initial value problem. Uniqueness and existence of global solutions of the algebro-geometric initial value problem as well as their explicit construction are then presented in our final and principal Section \ref{s6}. Appendix \ref{sA} reviews the basics of hyperelliptic Riemann surfaces of the Toda-type and sets the stage of much of the notation used in this paper. Various interpolation formulas of fundamental importance to our stationary inverse spectral algorithm developed in Section \ref{s4} are summarized in Appendix \ref{sB}. Finally, Appendix \ref{sC} summarizes asymptotic spectral parameter expansions of various quantities fundamental to the polynomial recursion formalism presented in Section \ref{s2}. These appendices support our intention to make this paper reasonably self-contained.
 
%%%%%%%%%%%%%%%%%%%%%%%%%%%%%%%%%%%%%%%%% 
%%%%%%%%%%%%%%%%%%%%%%%%%%%%%%%%%%%%%%%%%
\section{The Toda Hierarchy in a Nutshell} 
\label{s2}
%%%%%%%%%%%%%%%%%%%%%%%%%%%%%%%%%%%%%%%%%
%%%%%%%%%%%%%%%%%%%%%%%%%%%%%%%%%%%%%%%%%

In this section we briefly review the recursive construction of
the Toda hierarchy and associated Lax pairs and zero-curvature
equations following \cite{BGHT98}, \cite[Sect.\ 1.2]{GH07}, and
\cite[Ch.\ 12]{Te00}.

Throughout this section we make the following assumption: 
%%%%%%%%%%%%%%%%%%%%%%%%%%%%%%%%%%%%%%%%%%%
\begin{hypothesis} \lb{h2.1} 
Suppose
\begin{equation}
a, b \in \bbC^{\bbZ} \, \text{ and } \, a(n)\neq 0 \, \text{ for all }
\,   n\in\bbZ. \lb{1.2.1}
\end{equation}
\end{hypothesis}
%%%%%%%%%%%%%%%%%%%%%%%%%%%%%%%%%%%%%%%%%%%

Here $\bbC^J$ denotes the set of complex-valued sequences indexed by 
$J\subseteq\bbZ$.

We consider the second-order Jacobi difference expression
\begin{equation}
L=aS^{+}+a^{-}S^{-}+b, \lb{1.2.2}
\end{equation}
where $S^{\pm}$ denote the shift operators
\begin{equation}
(S^{\pm}f)(n)=f^{\pm}(n)=f(n{\pm}1), \quad n\in\bbZ, \; f \in
\bbC^{\bbZ}. \lb{1.2.3}
\end{equation}

To construct the stationary Toda hierarchy we need a second
difference expression of order $2\gg +2, \, \gg \in {\bbN_{0}},$
defined recursively in the following. We take the quickest route
to the construction of $P_{2\gg +2}$, and hence to the Toda hierarchy, by starting from the recursion relations
\eqref{1.2.4a}--\eqref{1.2.4c} below.

Define $\{{f_\ell}\}_{\ell \in \bbN_0}$ and $\{{g_\ell}\}_{\ell \in
\bbN_0}$ recursively by
\begin{align}
& f_0=1, \quad g_0=-c_1, \lb{1.2.4a} \\
& 2f_{\ell+1}+g_\ell+g_\ell^{-}-2bf_\ell=0, \quad \ell\in \bbN_0, \lb{1.2.4b} \\
&
g_{\ell+1}-g_{\ell+1}^{-}+2\big(a^2f_{\ell}^{+}-(a^{-})^2f_\ell^{-}\big)-b(g_\ell-g_\ell^{-})=0,
\quad \ell \in \bbN_{0}. \lb{1.2.4c}
\end{align}
Explicitly, one finds
\begin{align}
& f_0=1, \no \\
& f_1=b+c_1, \no \\
& f_2=a^2+(a^{-})^2+b^2+c_1b+c_2, \, \text{ etc.,} \lb{1.2.5}\\
& g_0=-c_1, \no\\
& g_1=-2a^2-c_2, \no \\
& g_2=-2a^2(b^+ +b)+c_1(-2 a^2)-c_3, \, \text{ etc.} \no 
\end{align}
Here $\{ c_\ell\}_{\ell\in \bbN}$ denote undetermined summation
constants which naturally arise when solving
\eqref{1.2.4a}--\eqref{1.2.4c}.

Subsequently, it will also be useful to work with the
corresponding homogeneous coefficients $\hat f_j$ and $\hat g_j$,
defined by vanishing of the constants $c_k,\, k \in \bbN$,
\begin{align}
\begin{split}
& \hat f_0=1, \quad \hat f_\ell=f_\ell \big|_{c_k=0, \, k=1,\dots,\ell},
\quad \ell\in\bbN,  \\
& \hat g_0 =0, \quad \hat g_\ell=g_\ell \big|_{c_k=0, \,
k=1,\dots,\ell+1}, \quad \ell\in\bbN_0. \lb{1.2.6}
\end{split}
\end{align}
Hence,
\begin{equation}
f_\ell=\sum_{k=0}^{\ell}c_{\ell-k}\hat{f}_{k},\quad
g_\ell=\sum_{k=1}^{\ell}c_{\ell-k}\hat{g}_{k}-c_{\ell+1},\quad
\ell\in\bbN_0, \lb{1.2.7}
\end{equation}
introducing
\begin{equation}
c_0=1. \lb{1.2.7a}
\end{equation}

Next we define difference expressions $P_{2\gg+2}$ of order
$2\gg+2$ by
\begin{equation}
P_{2\gg+2}=-L^{\gg+1}+\sum_{\ell=0}^{\gg} \Big(
g_\ell+2af_\ell S^{+}\Big)L^{\gg-\ell}+f_{\gg+1},\quad \gg\in\bbN_0.
\lb{1.2.8}
\end{equation}
Introducing the corresponding homogeneous difference expressions
$\hatt{P}_{2p+2}$ defined by
\begin{equation}
\hatt{P}_{2\ell+2}=P_{2\ell+2}\big|_{c_k=0,\,k=1,\dots,\ell}, \quad
\ell\in\bbN_0, \lb{3.2.9A}
\end{equation}
one finds
\begin{equation}
P_{2p+2} =\sum_{\ell=0}^p c_{p-\ell} \hatt{P}_{2\ell+2}. \lb{3.2.29}
\end{equation}
Using the recursion relations \eqref{1.2.4a}--\eqref{1.2.4c}, the
commutator of $P_{2\gg+2}$ and $L$ can be explicitly computed and
one obtains
\begin{align}
[P_{2\gg+2},L]=&-a\big(g_\gg^{+}+g_\gg+f_{\gg+1}^{+}+f_{\gg+1}-2b^{+}f_\gg^{+}
\big)S^{+} \no \\
&+2\big(-b(g_{\gg}+f_{\gg+1})+a^2f_\gg^{+}
-(a^{-})^2f_{\gg}^{-}+b^2f_\gg\big) \no \\
&-a^{-}\big(g_\gg+g_\gg^{-}+f_{\gg+1}+f_{\gg+1}^{-}-2bf_\gg
\big)S^{-}, \quad \gg\in\bbN_0. \lb{1.2.8a}
\end{align}
In particular, $(L,P_{2\gg+2})$ represents the celebrated
\textit{Lax pair} of the Toda hierarchy. Varying $\gg\in\bbN_0$,
the stationary Toda hierarchy is then defined in terms of the
vanishing of the commutator of $P_{2\gg+2}$ and $L$ in
\eqref{1.2.8a}, that is,
\begin{equation}
[P_{2\gg+2}, L] =\sTl_\gg(a,b)=0,\quad \gg\in\bbN_0. \lb{1.2.8aa}
\end{equation}
Thus one finds
\begin{align}
g_\gg+g_\gg^{-}+f_{\gg+1}+f_{\gg+1}^{-}-2bf_\gg&=0, \lb{1.2.8ab} \\
-b(g_\gg+f_{\gg+1})+a^2f_\gg^{+}-(a^{-})^2f_g^{-}+b^2f_\gg&=0.
\lb{1.2.8ac}
\end{align}
Using \eqref{1.2.4b} with $j=\gg$ one concludes that
\eqref{1.2.8ab} reduces to
\begin{equation}
f_{\gg+1}=f_{\gg+1}^{-}, \lb{1.2.8b}
\end{equation}
that is, $f_{\gg+1}$ is a lattice constant. Similarly, one infers
by subtracting $b$ times \eqref{1.2.8ab} from twice
\eqref{1.2.8ac} and using \eqref{1.2.4c} with $j=\gg$, that
$g_{\gg+1}$ is a lattice constant as well, that is,
\begin{equation}
g_{\gg+1}=g_{\gg+1}^{-}. \lb{1.2.8c}
\end{equation}
Equations \eqref{1.2.8b} and \eqref{1.2.8c} give rise to the  
stationary Toda hierarchy, which is introduced as
follows
\begin{equation}
\stl_p(a,b)= \begin{pmatrix} f_{p+1}^+-f_{p+1} \\
g_{p+1}-g_{p+1}^- \end{pmatrix}=0, \quad p\in\bbN_0. \lb{3.2.29d}
\end{equation}
Explicitly,
\begin{align}
\stl_0 (a,b) &=  \begin{pmatrix} b^+ -b\\
2\big((a^-)^2 -a^2\big)\end{pmatrix} =0, \no  \\%[3mm]
\stl_1 (a,b) &=  \begin{pmatrix} (a^+)^2 -(a^-)^2
+(b^+)^2 -b^2 \\ 2(a^-)^2 (b+b^-) -2a^2 (b^+ +b) 
\end{pmatrix}  \lb{3.2.16a} \\ 
& \quad \, + c_1 \begin{pmatrix} b^+ -b \\ 2 \big((a^-)^2 -a^2\big) 
\end{pmatrix} =0, \text{ etc.,} \no 
\end{align}
represent the first few equations of the stationary Toda hierarchy. By
definition, the set of solutions of \eqref{3.2.29d}, with $p$ ranging
in $\bbN_0$ and $c_\ell\in\bbC$, $\ell\in\bbN$, represents the
class of algebro-geometric Toda solutions. 

In the following we will frequently assume that $a, b$ satisfy the $p$th
stationary Toda system. By this we mean it satisfies one of the
$p$th stationary Toda equations after a particular choice of summation 
constants $c_\ell\in\bbC$, $\ell=1,\dots,p$, $p\in\bbN$, has been made.

In accordance with our notation introduced in \eqref{1.2.6} and
\eqref{3.2.9A}, the corresponding homogeneous stationary Toda equations
are defined by
\begin{equation}
\shTl_p (a,b)  =\stl_p (a,b) \big|_{c_\ell=0,\, \ell=1,\dots,p}=0,
\quad p\in\bbN_0. \label{1.2.8A}
\end{equation}

Next, we introduce polynomials $F_p$ and $G_{p+1}$ of
degree $p$ and $p+1$, with respect to the spectral parameter $z\in\bbC$ by
\begin{align}
F_{p} (z) &=\sum_{\ell=0}^p f_{p-\ell} z^\ell
=\sum_{\ell=0}^p c_{p-\ell}\hatt{F}_\ell(z),\lb{1.2.11a}\\
G_{p+1}(z) &=-z^{p+1}
+ \sum_{\ell=0}^p g_{p-\ell} z^\ell + f_{p+1}=\sum_{\ell=1}^{p+1}
c_{p+1-\ell}\hatt{G}_\ell(z) \lb{1.2.11b}
\end{align}
with $\hatt{F}_\ell$ and $\hatt{G}_\ell$ denoting the corresponding
homogeneous polynomials defined by
\begin{align}
\hatt{F}_0(z)&=F_0(z)=1, \no \\
\hatt{F}_\ell(z)&=F_\ell(z)\big|_{c_k=0,\,k=1,\dots,\ell}=\sum_{k=0}^\ell
\hat f_{\ell-k} z^k, \quad \ell\in\bbN_0,  \lb{1.2.11c}  \\
\hatt{G}_0(z)&=G_0(z)\big|_{c_1=0}=0, \quad \hatt{G}_1(z)=G_1(z)=-z-b, \no
\\
\hatt{G}_{\ell+1}(z)&=G_{\ell+1}(z)\big|_{c_k=0,\,k=1,\dots,\ell}
=-z^{\ell+1}+\sum_{k=0}^\ell \hat g_{\ell-k} z^k + \hat f_{\ell+1}, \quad
\ell\in\bbN.   \lb{1.2.11d}
\end{align}
Explicitly, one obtains
\begin{align}
F_0&=1, \no \\
F_1&=z+b+c_1, \no \\
F_2&=z^2+bz+a^2+(a^-)^2+b^2+c_1(z+b)+c_2, \, \text{ etc.,} \lb{1.2.12}
\\ G_1&=-z+b, \no \\
G_2&=-z^2+(a^-)^2-a^2+b^2+c_1(-z+b), \, \text{ etc.} \no 
\end{align}

Next, we study the restriction of the difference expression
$P_{2\gg+2}$ to the two-dimensional  kernel (i.e., the formal null
space in an algebraic sense as opposed to the functional analytic
one) of $(L-z)$. More precisely, let
\begin{align}
& \text{ker}(L-z)=\{\psi \colon \bbZ \to \bbC\cup\{\infty\} \mid
(L-z)\psi=0\}. \lb{1.2.13a}
\end{align}
Then \eqref{1.2.8} implies
\begin{equation}
P_{2\gg+2}\mid_{\text{ker}(L-z)}=\big(2aF_\gg(z)S^{+}+G_{\gg+1}(z)\big)\big|_{\text{ker}(L-z)}.
\lb{1.2.13}
\end{equation}
Therefore, the Lax relation \eqref{1.2.8aa} becomes
\begin{align}
2(z-b^{+})F_\gg^{+}-2(z-b)F_\gg+G_{\gg+1}^{+}-G_{\gg+1}^{-}&=0,
\lb{1.2.15a} \\
2a^2F_\gg^{+} -2(a^{-})^2F_\gg^{-} +(z-b)(G_{\gg+1}-G_{\gg+1}^-)&=0.
\lb{1.2.15b}
\end{align}
Additional manipulations yield
\begin{align}
2(z-b)F_\gg+G_{\gg+1}+G_{\gg+1}^{-}&=0, \lb{1.2.16a} \\
(z-b)^2F_\gg+(z-b)G_{\gg+1}+a^2F_\gg^{+}-(a^{-})^2F_\gg^{-}&=0.
 \lb{1.2.16b}
\end{align}
Indeed, adding $G_{p+1}-G_{p+1}$ to the left-hand side of \eqref{1.2.15a} 
(neglecting a trivial summation constant) yields
\eqref{1.2.16a} and inserting
\eqref{1.2.15a} into \eqref{1.2.15b} then implies \eqref{1.2.16b}. Varying
$p\in\bbN_0$, equations \eqref{1.2.16a}, \eqref{1.2.16b} provide an
alternative description of the stationary Toda hierarchy.

Combining equations \eqref{1.2.15b} and \eqref{1.2.16a} one
concludes that the quantity
\begin{equation}
R_{2\gg+2}(z) = G_{\gg+1}(z,n)^2-4a(n)^2F_\gg(z,n)F_\gg^{+}(z,n)
\lb{1.2.17}
\end{equation}
is a lattice constant, and hence depends on $z$ only. Thus, one can
write
\begin{equation}
R_{2\gg+2}(z) = \prod_{m=0}^{2\gg+1} (z-E_m), \quad \{
E_m\}_{m=0}^{2\gg+1} \subset \bbC. \lb{1.2.18}
\end{equation}
One can decouple \eqref{1.2.16a} and \eqref{1.2.16b} to obtain separate
equations for $F_p$ and $G_{p+1}$. For instance, computing $G_{p+1}$ from
\eqref{1.2.16b} and inserting the result into \eqref{1.2.16a} yields the
following linear difference equation for $F_p$
\begin{align}
&(z-b)^2(z-b^-)F_p-(z-b^-)^2(z-b)F_p^-
+\big((a^-)^2F_p^- -a^2F^+_p\big)(z-b^-)\no \\
& \quad +\big((a^{--})^2F_p^{--}-(a^-)^2F_p\big)(z-b)=0. \lb{3.2.35AA}
\end{align}
Similarly, insertion of \eqref{1.2.16b} into \eqref{1.2.17} permits one
to eliminate $G_{p+1}$ and results in the following nonlinear difference
equation for
$F_p$,
\begin{align}
&(z-b)^4 F_p^2-2a^2(z-b)^2F_pF_p^+ -2(a^-)^2(z-b)^2F_pF_p^-
+a^4(F_p^+)^2 \no \\
& \quad +(a^-)^4(F_p^-)^2 -2a^2(a^-)^2F_p^+ F_p^-= (z-b)^2R_{2p+2}(z).
\lb{3.2.35a}
\end{align}
On the other hand, computing $F_p$ in terms of $G_{p+1}$ and
$G_{p+1}^+$ using \eqref{1.2.16a} and inserting the result into
\eqref{1.2.16b} yields the following linear difference equation for
$G_{p+1}$
\begin{align}
&a^2(z-b^-)(G_{p+1}^++G_{p+1})- (a^-)^2(z-b^+)(G_{p+1}^-
+G_{p+1}^{--}) \no \\
& \quad  +(z-b^-)(z-b)(z-b^+)(G^-_{p+1}-G_{p+1})=0. \lb{3.2.35b}
\end{align}
Finally, inserting the result for $F_p$ into \eqref{1.2.17} yields the
following nonlinear difference equation for $G_{p+1}$
\begin{align}
&(z-b)(z-b^+)G_{p+1}^2 -a^2(G^-_{p+1}+G_{p+1})(G_{p+1}+G^+_{p+1}) \no \\ 
& \quad = (z-b)(z-b^+)R_{2p+2}. \lb{3.2.35c}
\end{align}
Equations \eqref{3.2.35a} and \eqref{3.2.35c} can be used to derive
nonlinear recursion relations for the homogeneous coefficients
$\hat f_\ell$ and $\hat g_\ell$ (i.e., the ones satisfying \eqref{1.2.6}
in the case of vanishing summation constants) as
proved in Theorem \ref{tE.8} in Appendix
\ref{sC}. This has interesting applications to the asymptotic
expansion of the Green's function of $L$ with respect to the spectral
parameter. In addition, as proven in Theorem \ref{tE.8},
\eqref{3.2.35a} leads to an explicit determination of the summation 
constants $c_1,\dots,c_p$ in
\begin{equation}
\stl_p(a,b)= 0, \quad p\in\bbN_0,  \lb{3.2.35A}
\end{equation}
in terms of the zeros $E_0,\dots,E_{2p+1}$ of the associated
polynomial $R_{2p+2}$ in \eqref{1.2.18}. In fact, one can prove (cf.\
Theorem \ref{tE.8}) that
\begin{equation}
c_k=c_k(\ul E), \quad k=1,\dots,\gg, \lb{1.2.12B}
\end{equation}
where
\begin{align}
c_k(\ul E) \no &=-\!\!\!\!\!\!\!\sum_{\substack{j_0,\dots,j_{2\gg+1}=0\\
        j_0+\cdots+j_{2\gg+1}=k}}^{k}\!\!
\f{(2j_0)!\cdots(2j_{2\gg+1})!} {2^{2k} (j_0!)^2\cdots
(j_{2\gg+1}!)^2 (2j_0-1)\cdots(2j_{2\gg+1}-1)}
E_0^{j_0}\cdots E_{2\gg+1}^{j_{2\gg+1}}, \no \\
& \hspace*{8cm} k=1,\dots,p, \label{1.2.12C}
\end{align}
are symmetric functions of $\ul E=(E_0,\dots,E_{2p+1})$.

We emphasize that the result \eqref{1.2.13} is valid independently
of whether or not $P_{2\gg+2}$ and $L$ commute.  However, the fact
that the two difference expressions $P_{2\gg+2}$ and $L$ commute
implies the existence of an algebraic relationship between them.
This gives rise to the Burchnall--Chaundy polynomial for the Toda
hierarchy first discussed in the discrete context by Na{\u\i}man
\cite{Na62}, \cite{Na64}.

%%%%%%%%%%%%%%%%%%%%%%%%%%%%%%%%%%%%%%%%%%%%
\begin{theorem} \label{t1.2.1}
Assume Hypothesis \ref{h2.1}, fix $\gg\in\bbN_0$ and suppose that
$P_{2\gg+2}$ and $L$ commute,
$[P_{2\gg+2},L]=0$, or equivalently, assume that $\sTl_{\gg}(a,b)=0$.
Then $L$ and $P_{2\gg+2}$ satisfy an algebraic relationship of the
type $($cf.\ \eqref{1.2.18}$)$
\begin{align}
\begin{split}
&\calF_\gg(L,P_{2\gg+2})= P_{2\gg+2}^2-R_{2\gg+2}(L)=0, \label{1.2.19} \\
& R_{2\gg+2}(z) = \prod_{m=0}^{2\gg+1} (z-E_m), \quad z\in \bbC. %\no
\end{split}
\end{align}
\end{theorem}
%%%%%%%%%%%%%%%%%%%%%%%%%%%%%%%%%%%%%%%%%%%%%

The expression $\calF_\gg(L,P_{2\gg+2})$ is called the
Burchnall--Chaundy polynomial of the Lax pair $(L,P_{2\gg+2})$ and it will
be used in Section \ref{s3} to introduce the underlying hyperelliptic
curve associated with the stationary Toda system $\sTl_{\gg}(a,b)=0$
(cf.\ \eqref{1.2.20}).

Next we turn to the time-dependent Toda hierarchy. For that
purpose the functions $a$ and $b$ are now considered as functions of both
the lattice point and time. For each equation in the  hierarchy, that is,
for  each $p$, we introduce a deformation (time) parameter
$t_p\in\bbR$ in $a, b$, replacing $a(n), b(n)$ by $a(n,t_{p}), b(n,t_p)$.
The second-order  difference expression $L$ (cf.\ \eqref{1.2.2}) now reads
\begin{equation}
L(t_p) = a(\dott,t_p) S^+ +a^-(\dott,t_p) S^- +b(\dott,t_p). \lb{3.2.11A}
\end{equation}
The quantities $\{f_\ell\}_{\ell\in\bbN_0}$,
$\{g_\ell\}_{\ell\in\bbN_0}$, and $P_{2p+2}$, $p\in\bbN_0$ are still
defined by \eqref{1.2.4a}--\eqref{1.2.4c} and \eqref{1.2.8},
respectively. The time-dependent Toda hierarchy is then obtained by
imposing the Lax commutator equations
\begin{equation}
L_{t_p}(t_p) -[P_{2p+2}(t_p), L(t_p)]=0, \quad t_p\in\bbR, \lb{3.2.11}
\end{equation}
varying $p\in\bbN_0$. Relation \eqref{3.2.11} implies
\begin{align}
&\left(a_{t_p} +a (g_p^+
+g_p + f_{p+1}^+ + f_{p+1} - 2b^+ f_p^+ )\right) S^+ \no\\
& \quad -\left(-b_{t_p} +2 \big(-b (g_p + f_{p+1}) +a^2 f_p^+
-(a^-)^2 f^-_p +b^2 f_p\big)\right) \lb{3.2.11aaa}\\
& \quad +\left(a_{t_p} +a (g_p^+
+g_p + f_{p+1}^+ + f_{p+1} - 2b^+ f_p^+ )\right)^- S^-=0.\no
\end{align}
Applying the same method we used to derive \eqref{1.2.8b}
and \eqref{1.2.8c} one concludes
\begin{align}
0&=L_{t_p} -[P_{2p+2}, L] \no \\
&=\left(a_{t_p}-a(f_{p+1}^+-f_{p+1})\right)S^+
-\left(-b_{t_p}-g_{p+1}+g_{p+1}^- \right) \no \\
& \quad +\left(a_{t_p}-a(f_{p+1}^+-f_{p+1})\right)^-S^-.
\lb{3.2.11a} %\\
%&=\tl_p (a,b)_1 S^+ -\tl_p (a,b)_2+\tl_p
%(a,b)^-_1 S^-, \quad p\in\bbN_{0}, \lb{3.2.11a}
\end{align}
Varying $p\in\bbN_0$, the collection of evolution equations
\begin{equation}
\tl_p(a,b) = \begin{pmatrix}  a_{t_p}-a(f_{p+1}^+-f_{p+1}) \\
b_{t_p}+g_{p+1}-g_{p+1}^- \end{pmatrix}=0,
\quad (n,t_p)\in\bbZ\times\bbR, \; p\in\bbN_0 \lb{3.2.13}
\end{equation}
then defines the time-dependent Toda hierarchy. Explicitly,
\begin{align}
\tl_0 (a,b) = & \begin{pmatrix}a_{t_0} -a(b^+ -b)\\ b_{t_0}
-2\big(a^2-(a^-)^2\big)\end{pmatrix} =0, \no  \\
\tl_1 (a,b) = & \begin{pmatrix}a_{t_1} -a\big((a^+)^2 -(a^-)^2
+(b^+)^2 -b^2\big)\\b_{t_1} +2(a^-)^2 (b+b^-) -2a^2 (b^+ +b) \end{pmatrix} 
\lb{3.2.16}  \\
&+ c_1 \begin{pmatrix} -a(b^+ -b)\\ -2\big(a^2-(a^-)^2\big)\end{pmatrix}
=0, \, \text{ etc.,} \no 
\end{align}
represent the first few equations of the time-dependent Toda hierarchy.
The system of equations, $\tl_0(a,b)=0$, is of course \textit{the} Toda
system.

The corresponding homogeneous Toda
equations obtained by taking all summation constants equal to zero,
$c_\ell=0$, $\ell=1,\dots, p$, are then denoted by
\begin{equation}
\widehat{\tl}_p (a,b)= \tl_p(a,b) \big|_{c_\ell =0,\,
\ell=1,\dots, p}. \lb{3.2.18}
\end{equation}

Restricting the Lax relation \eqref{3.2.11} to the
kernel $\ker(L-z)$ one finds that
\begin{align}
0 & = \big( L_{t_p} -[P_{2p+2}, L]\big) \big|_{\ker (L-z)}
=\big( L_{t_p}+(L-z) P_{2p+2}\big) \big|_{\ker (L-z)}\\
& = \bigg( a\Big(\frac{ a_{t_p}}{a} -\frac{a_{t_p}^-}{a^-}
+2 (z-b^+) F_p^+
-2(z-b) F_p +G_{p+1}^+ -G_{p+1}^- \Big) S^+ \no \\
& \quad \;\; + \Big( b_{t_p} +(z-b) \frac{a_{t_p}^-}{a^-}
+2(a^-)^2 F_p^- -2a^2 F_p^+ \no \\
&\qquad \quad \;\, +(z-b) (G_{p+1}^- -G_{p+1})\Big)\bigg)
\bigg|_{\ker(L-z)}. \lb{3.2.22a}
\end{align}
Hence one obtains
\begin{align}
& \frac{ a_{t_p}}{a} - \frac{ a_{t_p}^-}{a^-} = -2 (z-b^+) F_p^+
+2(z-b) F_p + G_{p+1}^- -G_{p+1}^+,
\lb{3.2.23time}\\
& b_{t_p} = -(z-b) \frac{a_{t_p}^-}{a^-} -2(a^-)^2 F_p^- +2a^2 F_p^+
-(z-b) (G_{p+1}^- -G_{p+1}). \lb{3.2.24time}
\end{align}
Further manipulations then yield,
\begin{align}
a_{t_p} &=-a\big(2(z-b^+) F_p^+
   +G_{p+1}^+ +G_{p+1}\big),  \lb{3.2.25time}  \\
b_{t_p} &=2 \big((z-b)^2 F_p + (z-b) G_{p+1} + a^2 F_p^+
- (a^-)^2 F_p^-\big).  \lb{3.2.26time} 
\end{align}
Indeed, \eqref{3.2.25time} follows by adding $G_{p+1} -G_{p+1}$ to
\eqref{3.2.23time} (neglecting a  trivial summation constant), and an
insertion of \eqref{3.2.25time} into \eqref{3.2.24time} implies
\eqref{3.2.26time}. Varying $p\in\bbN_0$, equations \eqref{3.2.25time} and
\eqref{3.2.26time} provide an alternative description of the
time-dependent Toda hierarchy.

%%%%%%%%%%%%%%%%%%%%%%%%%%%%%%%%%%%%%%%%%%%
\begin{remark}\lb{r3.2.7}
{}{}From \eqref{1.2.4a}--\eqref{1.2.4c} and \eqref{1.2.11a}, \eqref{1.2.11b}
one concludes that the coefficient $a$ enters quadratically in
$F_p$ and $G_{p+1}$, and hence the Toda hierarchy \eqref{3.2.13}
$($respectively \eqref{3.2.29d}$)$ is invariant under the substitution
\begin{equation}
a \to a_\eps =\{\eps(n) a(n)\}_{n\in\bbZ},
\quad \eps (n) \in \{1, -1\}, \; n\in\bbZ. \lb{3.2.28}
\end{equation}
\end{remark} 
%%%%%%%%%%%%%%%%%%%%%%%%%%%%%%%%%%%%%%%%%%%

We conclude this section by pointing out an alternative construction of
the Toda hierarchy using a zero-curvature approach instead of Lax pairs
$(L,P_{2p+2})$. To this end one defines the $2\times 2$ matrices
\begin{align}
   U(z) &= \begin{pmatrix} 0 & 1 \\ -a^-/a & (z-b)/a \end{pmatrix} ,
\lb{3.2.40} \\
V_{p+1}(z) &= \begin{pmatrix} G_{p+1}^-(z) & 2a^- F_p^-(z) \\
-2a^-F_{p}(z) & 2(z-b)F_p(z)+G_{p+1}(z) \end{pmatrix}, \quad p\in\bbN_0.
\label{3.2.41}
\end{align}
Then the stationary part of this section can equivalently
be based on the zero-curvature equation
\begin{align}
0&=UV_{p+1}-V_{p+1}^+U \label{3.2.42} \\
&=\f{2}{a}\begin{pmatrix} 0 & 0 \\[1mm]
a^-\big((z-b^+)F_p^+ -(z-b)F_p & a^2 F_p^+ -(a^-)^2 F_p^- \\
+2^{-1}(G_{p+1}^+-G_{p+1}^-)\big) &  +2^{-1}(z-b)(G_{p+1}-G_{p+1}^+)\\
& +(z-b)^2F_p-(z-b^+)(z-b)F_p^+ \end{pmatrix}. \no
\end{align}
Thus, one obtains \eqref{1.2.15a} from the $(2,1)$-entry in
\eqref{3.2.42}. Insertion of \eqref{1.2.15a} into the $(2,2)$-entry of
\eqref{3.2.42} then yields \eqref{1.2.15b}. Thus, one also obtains
\eqref{1.2.16a} and hence the $(2,2)$-entry of $V_{p+1}$ in \eqref{3.2.41}
simplifies to
\begin{equation}
V_{p+1,2,2}(z)= -G_{p+1}^-(z)  \lb{3.2.43}
\end{equation}
in the stationary case. Since $\det(U(z,n))=a^-(n)/a(n)\neq 0$,
$n\in\bbZ$, the zero-curvature equation \eqref{3.2.42} yields that
$\det(V_{p+1}(z,n))$ is a lattice constant $($i.e., independent of
$n\in\bbZ$$)$. The Burchnall--Chaundy polynomial $\calF_p(y,z)$
$($cf.\ \eqref{1.2.19} and especially, the hyperelliptic
curve \eqref{1.2.20}$)$ is then obtained from  the characteristic equation
of $V_{p+1}(z)$ by
\begin{align}
& \det(yI_2 - V_{p+1}(z,n)) \no \\
& \quad = y^2 + \det(V_{p+1}(z,n)) \no \\
& \quad = y^2 - G_{p-1}^-(z,n)^2 +
4a^-(n)^2F_p^-(z,n)F_{p}(z,n) \no \\
& \quad =y^2-R_{2p+2}(z)= 0, \label{3.2.44}
\end{align}
using \eqref{3.2.43}. (Here $I_2$ denotes the identity matrix in
$\bbC^2$.) Similarly, the time-dependent part
\eqref{3.2.11A}--\eqref{3.2.26time} can  equivalently be developed from
the zero-curvature equation
\begin{align}
0&=U_{t_{p}}+UV_{p+1}-V_{p+1}^+U  \label{3.2.45} \\ 
&=\f{1}{a}\begin{pmatrix} 0 & 0 \\[1mm]
a^-((a_{t_p}/a)-(a^-_{t_p}/a^-)) & -b_{t_p}-(z-b)(a_{t_p}/a) \\
+a^-\big(2(z-b^+)F_p^+ -2(z-b)F_p & +2a^2 F_p^+ -2(a^-)^2 F_p^- \\
+(G_{p+1}^+-G_{p+1}^-)\big) &  +(z-b)(G_{p+1}-G_{p+1}^+)\\
& +2(z-b)^2F_p-2(z-b^+)(z-b)F_p^+ \end{pmatrix}. \no
\end{align}
The $(2,1)$-entry in \eqref{3.2.45} yields \eqref{3.2.23time}, and inserting
\eqref{3.2.23time} into the $(2,2)$-entry of \eqref{3.2.45} yields
\eqref{3.2.24time} and hence also the basic equations defining the
time-dependent Toda hierarchy in \eqref{3.2.25time}, \eqref{3.2.26time}.

%%%%%%%%%%%%%%%%%%%%%%%%%%%%%%%%%%%%%%%%%%%%%%%%%%%%%%%%%%%%%%%%%%%%%%%%%%%%%%%%%%%%%%%%%%%%%%
\section{Properties of Stationary Algebro-Geometric Solutions \\
of the Toda Hierarchy} \lb{s3} 
%%%%%%%%%%%%%%%%%%%%%%%%%%%%%%%%%%%%%%%%%%%%%%%%%%%%%%%%%%%%%%%%%%%%%%%%%%%%%%%%%%%%%%%%%%%%%%
In this section we present a quick review of properties of
algebro-geometric solutions of the stationary Toda hierarchy. Since this
material is standard we omit all proofs and just refer to \cite{BGHT98} 
(cf.\ also \cite[Sect.\ 1.3]{GH07}, \cite[Chs.\ 8, 9]{Te00}) for detailed
presentations and an extensive list of references to the literature. 

For the notation employed in connection with elementary concepts
in algebraic geometry (more precisely, the theory of compact
Riemann surfaces), we refer to Appendix \ref{sA}.

Returning to Theorem \ref{t1.2.1}, we note that \eqref{1.2.19}
naturally leads to the hyperelliptic curve $\calK_\gg$ of genus
$\gg\in\bbN_0$, where
\begin{align}
\begin{split}
&\calK_\gg \colon \calF_\gg(z,y)=y^2-R_{2\gg+2}(z)=0,  \lb{1.2.20} \\
&R_{2\gg+2}(z) = \prod_{m=0}^{2\gg+1} (z-E_m), \quad \{
E_m\}_{m=0}^{2\gg+1} \subset \bbC.  
\end{split}
\end{align}

Throughout this section we make the following assumption:

%%%%%%%%%%%%%%%%%%%%%%%%%%%%%%%%%%%%%%%%%%% 
\begin{hypothesis} \lb{h3.1} 
Suppose that 
\begin{equation}
a, b \in \bbC^{\bbZ} \, \text{ and } \, a(n)\neq 0 \, \text{ for all } \, 
n\in\bbZ.  \lb{1.3}
\end{equation}
In addition, assume that the hyperelliptic curve $\calK_p$ in \eqref{1.2.20} is
nonsingular, that is, suppose that
\begin{equation}
E_m \neq E_{m'} \text{ for } m\neq m', \; m,m'=0,\dots,2\gg+1. \lb{1.3a}
\end{equation}
\end{hypothesis} 
%%%%%%%%%%%%%%%%%%%%%%%%%%%%%%%%%%%%%%%%%%%%%%%

The curve $\calK_\gg$ is compactified by joining two points
$P_{\infty_{\pm}}$, $P_{\infty_+}\neq P_{\infty_-},$ at
infinity. For notational simplicity, the resulting curve is still
denoted by $\calK_\gg$. Points $P$ on $\calK_\gg \setminus
\{P_{\infty_{+}}, P_{\infty_{-}}\}$ are  represented as pairs $P=(z,y)$, where
$y(\dott)$ is the meromorphic function on $\calK_\gg$ satisfying
$\calF_\gg(z,y)=0$. The complex structure on $\calK_\gg$ is then
defined in the usual way, see Appendix \ref{sA}. Hence,
$\calK_\gg$ becomes a two-sheeted hyperelliptic Riemann surface of
genus $\gg\in\bbN_0$ in a standard manner.

We also emphasize that by fixing the curve $\calK_\gg$ (i.e., by
fixing $E_0,\dots,E_{2\gg+1}$), the summation constants
$c_1,\dots,c_\gg$ in the corresponding stationary $\sTl_\gg$
equation are uniquely determined as is clear from \eqref{1.2.12B}
and \eqref{1.2.12C}, which establish the summation constants
$c_k$ as symmetric functions of $E_0,\dots,E_{2\gg+1}$.

For notational simplicity we will usually tacitly assume that
$\gg\in\bbN$. The trivial case $\gg=0$, which leads to
$a(n)^2=(E_1-E_0)^2/16$, $b(n)=(E_0+E_1)/2$, $n\in\bbZ$, is of no interest to
us in this paper.

In the following, the zeros\footnote{If $a,b\in
\ell^\infty(\bbZ)$, these zeros are the Dirichlet  eigenvalues of
a bounded operator on $\ell^2(\bbZ)$ associated with the difference
expression $L$ and a Dirichlet boundary condition  at $n\in\bbZ$.}
of the polynomial $F_\gg(\dott,n)$ (cf.\ \eqref{1.2.11a}) will
play a special role. We denote them by $\{\mu_j(n)\}_{j=1}^\gg$
and write
\begin{equation}
F_\gg(z,n) =\prod_{j=1}^\gg (z-\mu_j(n)). \lb{1.2.21}
\end{equation}
The next step is crucial; it permits us to ``lift'' the zeros
$\mu_j$ of $F_\gg$ from $\bbC$ to the curve $\calK_\gg$. {}From
\eqref{1.2.17} and \eqref{1.2.21} one infers
\begin{equation}
R_{2\gg+2}(z) - G_{\gg+1}(z)^2 = 0, \quad
z\in\{\mu_j, \mu_k^+\}_{j,k=1,\dots,\gg}. \lb{1.3.7}
\end{equation}
We now introduce $\{ \hat \mu_j(n) \}_{j=1,\dots,\gg}\subset
\calK_\gg$ by
%\begin{subequations}\lb{1.3.8}
\begin{equation}
 \hat{\mu}_j(n)=( \mu_j(n),-G_{\gg+1}(\mu_j(n),n))\in\calK_p,
\quad j=1,...,\gg, \; n\in\bbZ. \lb{1.2.24a}
\end{equation}
Next, we recall equation \eqref{1.2.17} and define the
fundamental meromorphic function $\phi(\dott,n)$ on $\calK_\gg$ by 
\begin{align}
\phi(P,n)&=\frac{y-G_{\gg+1}(z,n)}{2a(n)F_\gg (z, n)} \lb{1.2.22a}\\
& = \frac{-2a(n)F_{\gg} (z,n+1)}{y+G_{\gg+1}(z,n)},\quad\lb{1.2.22b}\\
& \hspace*{-.05cm} P = (z,y)\in\calK_\gg, \; n\in\bbZ,  \no
\end{align}
with divisor $(\phi(\dott, n))$ of $\phi(\dott, n)$ given by
\begin{equation}
\big( \phi(\dott,n) \big)=\calD_{P_{\infty_+} \hat{\underline{
\mu}}(n+1)}-\calD_{P_{\infty_-}\hat{\underline{\mu}}(n)}, 
\lb{1.2.23}
\end{equation}
using \eqref{1.2.21} and \eqref{1.2.24a}. Here we abbreviated
\begin{equation}
\hat{\underline \mu} = \{\hat \mu_1, \dots, \hat \mu_\gg\} \in
\symgg \lb{1.2.24b}
\end{equation}
(cf.\ the notation introduced in Appendix \ref{sA}). We note that several 
$\mu_j(n)$ may be equal for a given lattice point $n\in\bbZ$. Moreover, since
$-G_{p+1}(\mu_j(n),n)$ takes on the same value for all coinciding zeros
$\mu_j(n)$, no finite special divisors $\calD_{\humu(n)}$ can ever arise in
$\phi$ (cf.\ also Lemma \ref{l1.3.9ba}).

The stationary Baker--Akhiezer function $\psi(\dott,n,n_0)$ on
$\calK_\gg\setminus\{P_{\infty_{\pm}} \}$ is then defined in
terms of $\phi(\dott,n)$ by
\begin{align}
& \psi(P,n,n_0)= \begin{cases} \prod_{m=n_0}^{n-1}\phi(P,m) &
\text{for}\quad n\geq n_0+1, \\
1 & \text{for}\quad n=n_0, \\
\prod_{m=n}^{n_0-1} {\phi(P,m)}^{-1} \quad &\text{for}\quad n\leq
n_0-1,  \end{cases} \lb{1.2.25a}  \\
& \hspace*{3.56cm} P\in\calK_p\setminus\{P_{\infty_{\pm}}\}, \;
(n,n_0)\in\bbZ^2,  
\no 
\end{align}
with divisor $\big(\psi(\dott,n,n_0)\big)$ of $\psi(P,n,n_0)$
given by
\begin{equation}
\big(\psi(\dott,n,n_0)\big)=\calD_{\underline{\hat{\mu}}(n)}
-\calD_{\underline{\hat{\mu}}(n_0)}+(n-n_0)(\calD_{P_{\infty
_ +}}-\calD_{P_{\infty_ -}}). \lb{1.2.25b}
\end{equation}
For future purposes we also introduce the following Baker--Akhiezer
vector,
\begin{equation}
\Psi(P,n,n_0)=\begin{pmatrix} \psi^-(P,n,n_0) \\ 
\psi(P,n,n_0)\end{pmatrix}, \quad P\in\calK_p\setminus\{P_{\infty_{\pm}}\},
\;  (n,n_0)\in\bbZ^2.
\end{equation}

Basic properties of $\phi$, $\psi$, and $\Psi$ are summarized in the
following result. We abbreviate by 
\begin{equation}
W(f,g)=a(fg^+ - f^+ g)  
\end{equation}
the Wronskian of two complex-valued sequences $f$ and $g$, and denote 
$P^*=(z,-y)$ for $P=(z,y)\in\calK_p$.

%%%%%%%%%%%%%%%%%%%%%%%%%%%%%%%%%%%%%%%%%%%%%
\begin{lemma} \lb{l1.3.1}
Assume Hypothesis \ref{h3.1} and suppose that $a, b$ satisfy the
$\gg$th stationary Toda system \eqref{3.2.29d}. Moreover, let $P= (z,y)
\in \calK_\gg \setminus \{P_{\infty_{\pm}}\}$ and $(n,n_0) \in
\bbZ^2$.  Then $\phi$ satisfies the Riccati-type equation
\begin{align}
& a\phi(P) + a^{-}\phi^{-}(P)^{-1} = z-b, \lb{1.2.26}
\intertext{as well as}
& \phi(P) \phi (P^*)= \f{F^{+}_{\gg}(z)}{F_\gg (z)},\lb{1.2.27}\\
&\phi (P) + \phi (P^*)= -\f{G_{\gg+1} (z)}{aF_\gg (z)},\lb{1.2.28}\\
& \phi(P)-\phi (P^*)=\f{y(P)}{aF_\gg (z)}. \lb{1.2.29}
\end{align}
Moreover, $\psi$ and $\Psi$ satisfy
\begin{align}
& \big(L-z(P)\big) \psi(P) =0, \quad \big(P_{2\gg+2}-y(P)\big)
\psi(P) =0, \lb{1.2.30} \\
& \Psi^+(P) = U(z)\Psi(P), \quad y\Psi(P) = V_{p+1}\Psi(P), \lb{3.19} \\
& \psi (P,n,n_0) \psi (P^*, n, n_0)= \f{F_\gg
(z,n)}{F_\gg(z,n_0)},\lb{1.2.31}\\
& a(n)\big(\psi(P,n,n_0)\psi(P^*,n+1,n_0)
+\psi(P^*,n,n_0)\psi(P,n+1,n_0)\big)  \no \\
&\quad=-\f{G_{\gg+1}(z,n)}{F_\gg(z,n_0)},
\lb{1.2.32a} \\
&W(\psi (P,\dott,n_0), \psi (P^*, \dott,
n_0))=-\f{y(P)}{F_{\gg}(z,n_0)}. \lb{1.2.32ab}
\end{align}
\end{lemma}
%%%%%%%%%%%%%%%%%%%%%%%%%%%%%%%%%%%%%%%%%%%%%%%

Combining the polynomial recursion approach with \eqref{1.2.21}
readily yields trace formulas for the Toda invariants, which are
expressions of $a$ and $b$ in terms of the zeros $\mu_j$ of
$F_\gg$. We introduce the abbreviation,
\begin{equation}
b^{(k)}(n)=\f{1}{2}\sum_{m=0}^{2\gg+1} E_m^k - \sum_{j=1}^\gg 
\mu_j^k(n),\quad k\in \bbN.    \lb{3.21}
\end{equation}

%%%%%%%%%%%%%%%%%%%%%%%%%%%%%%%%%%%%%%%%%%% 
\begin{lemma} \lb{l1.3.9a} 
Assume Hypothesis \ref{h3.1} and suppose that $a, b$ satisfy the
$\gg$th stationary Toda system \eqref{3.2.29d}. Then,
\begin{equation}
b(n)=\f{1}{2}\sum_{m=0}^{2\gg+1} E_m - \sum_{j=1}^\gg \mu_j(n), \quad n\in\bbZ.
\lb{1.2.21b} 
\end{equation}
In addition, if for all $n\in\bbZ$, $\mu_j(n)\neq \mu_k(n)$ for $j\neq k$,
$j,k=1,\dots,p$, then,
\begin{equation}
a(n)^2=\f{1}{2}\sum_{j=1}^{\gg}
y(\hat{\mu}_j(n))\prod_{\substack{k=1\\ k\neq j}}^\gg
(\mu_j(n)-\mu_k(n))^{-1}+\f{1}{4}\big(b^{(2)}(n) - b(n)^2\big), \quad 
n\in\bbZ. \lb{1.2.21a} 
\end{equation}
\end{lemma}  
%%%%%%%%%%%%%%%%%%%%%%%%%%%%%%%%%%%%%%%%%%%

The case where some of the $\mu_j$ coincide in \eqref{1.2.21a} requires a
more elaborate argument that will be presented in Section \ref{s4}.

Since nonspecial Dirichlet divisors $\calD_{\humu}$ and the linearization
property of the Abel map when applied to $\calD_{\humu}$ will play a
fundamental role later on, we also recall the following facts.

%%%%%%%%%%%%%%%%%%%%%%%%%%%%%%%%%%%%%%%%%% 
\begin{lemma} \lb{l1.3.9ba}
Assume Hypothesis \ref{h3.1} and suppose that $a, b$ satisfy the
$\gg$th stationary Toda system \eqref{3.2.29d}. Let $\calD_{\humu}$,
$\humu=\{\hmu_1,\dots,\hmu_\gg\}\in\symgg$, be the Dirichlet divisor of
degree
$\gg$ associated with $a$, $b$ defined according to
\eqref{1.2.24a}, that is,
\begin{equation}
\hmu_j(n)=\big(\mu_j(n),-G_{\gg+1}(\mu_j(n),n)\big)\in\calK_p, \quad
j=1,\dots,\gg, \; n\in\bbZ. \lb{1.3.59AA}
\end{equation}
Then $\calD_{\humu(n)}$ is nonspecial for all $n\in\bbZ$.  
Moreover, the Abel map linearizes the auxiliary divisor $\calD_{\humu}$
in the sense that
\begin{equation}
\ual_{Q_0} (\calD_{\hat\umu(n)}) = \ual_{Q_0} (\calD_{\hat\umu (n_0)})
-(n-n_0) \underline{A}_{P_{\infty_-}} (P_{\infty_+}), \lb{3.3.32A}
\end{equation}
where $Q_0\in\calK_p$ is a given base point. \\
If in addition, $a, b \in \ell^{\infty}(\bbZ)$, then there exists a
constant $C_{\mu}>0$ such that
\begin{equation}
|\mu_j(n)|\leq C_{\mu}, \quad j=1,\dots,\gg, \; n\in\bbZ. \lb{1.3C}
\end{equation}
\end{lemma} 
%%%%%%%%%%%%%%%%%%%%%%%%%%%%%%%%%%%%%%%%%%%

%%%%%%%%%%%%%%%%%%%%%%%%%%%%%%%%%%%%%%%%%%% 
\begin{remark} \lb{r3.5}
We note that by construction, the divisors $\calD_{\humu(n)}$, $n\in\bbZ$,
as introduced in \eqref{1.2.24a} are all finite and hence nonspecial by
Lemma \ref{l1.3.9ba}. On the other hand, as we will see in the next
Section \ref{s4}, given a nonspecial divisor $\calD_{\hat\umu (n_0)}$, the
solution $\calD_{\humu(n)}$ of equation \eqref{3.3.32A} may cease to be a
finite divisor at some $n\in\bbZ$. 
\end{remark}
%%%%%%%%%%%%%%%%%%%%%%%%%%%%%%%%%%%%%%%%%%%%%%%

%%%%%%%%%%%%%%%%%%%%%%%%%%%%%%%%%%%%%%%%%
%%%%%%%%%%%%%%%%%%%%%%%%%%%%%%%%%%%%%%%%% 
\section{An Algorithm for Solving the Inverse Algebro-Geometric Spectral Problem for (Non-self-adjoint) Jacobi Operators} 
\lb{s4}  
%%%%%%%%%%%%%%%%%%%%%%%%%%%%%%%%%%%%%%%%%
%%%%%%%%%%%%%%%%%%%%%%%%%%%%%%%%%%%%%%%%%
The aim of this section is to derive an algorithm that enables
one to construct algebro-geometric solutions for the
stationary Toda hierarchy for complex-valued initial data. Equivalently, we offer 
a solution of the inverse algebro-geometric spectral problem for general (non-self-adjoint) Jacobi operators, starting with initial divisors in general complex position.

Up to the end of Section \ref{s3} the material was standard (see
\cite{BGHT98}  and \cite[Sect.\ 1.3]{GH07}, \cite[Chs.\ 8, 9]{Te00} for
details) and based on the assumption that $a, b \in \bbC^{\bbZ}$
satisfy the $\gg$th stationary Toda system \eqref{3.2.29d}. Now we
embark on the corresponding inverse problem consisting of constructing
a solution of \eqref{3.2.29d} given certain initial data. More precisely,
we seek to construct solutions $a, b \in\bbC^{\bbZ}$  satisfying the
$\gg$th stationary Toda system
\eqref{3.2.29d} starting from a properly restricted set $\calM_0$ of
finite nonspecial Dirichlet divisor initial data
$\calD_{\humu(n_0)}$ at some fixed $n_0\in\bbZ$,
\begin{align}
\begin{split}
\humu(n_0)&=\{\hmu_1(n_0),\dots,\hmu_\gg(n_0)\}\in \calM_0, \quad 
\calM_0\subset\symgg, \\
\hmu_j(n_0)&=\big(\mu_j(n_0),-G_{\gg+1}(\mu_j(n_0),n_0)\big), \quad
j=1,\dots,\gg.  \lb{4.1}
\end{split}
\end{align}

Of course we would like to ensure that the sequences obtained via our 
algorithm do not blow up. To investigate when this happens, we study 
the image of our divisors under the Abel map. The key
ingredient in our analysis will be  \eqref{3.3.32A} which yields a linear
discrete dynamical system on the Jacobi variety
$J(\calK_p)$. In particular, we will be led to investigate solutions 
$\calD_{\hat\umu}$ of the discrete initial value problem
\begin{align}
\begin{split}
&\ual_{Q_0} (\calD_{\hat\umu(n)}) = \ual_{Q_0} (\calD_{\hat\umu (n_0)})
-(n-n_0) \underline{A}_{P_{\infty_-}} (P_{\infty_+}), \lb{4.2} \\
&\humu(n_0)=\{\hmu_1(n_0),\dots,\hmu_\gg(n_0)\}\in \symgg,
\end{split}
\end{align}
where $Q_0\in\calK_p$ is a given base point. Eventually, we will be
interested in solutions $\calD_{\hat\umu}$ of \eqref{4.2} with initial
data $\calD_{\humu(n_0)}$ satisfying \eqref{4.1} and $\calM_0$ to be
specified as in (the proof of) Lemma \ref{l4.1}.

Before proceeding to develop the stationary Toda algorithm, we briefly
analyze the dynamics of \eqref{4.2}.

%%%%%%%%%%%%%%%%%%%%%%%%%%%%%%%%%%%%%%%%%% 
\begin{lemma}
Let $\calD_{\humu(n)}$ be defined via \eqref{4.2} for some divisor 
$\calD_{\humu(n_0)}\in\symgg$. \\
$(i)$ If $\calD_{\humu(n)}$ is finite and nonspecial and $\calD_{\humu(n+1)}$
is  infinite, then $\calD_{\humu(n+1)}$ contains $\Pinfp$ but not
$P_{\infty_-}$. \\
$(ii)$ If $\calD_{\humu(n)}$ is nonspecial and $\calD_{\humu(n+1)}$ is
special, then $\calD_{\humu(n)}$ contains $P_{\infty_+}$ at least twice. \\
Items $(i)$ and $(ii)$ hold if $n+1$ is replaced by $n-1$ and $P_{\infty_+}$
by $P_{\infty_-}$.  
\end{lemma} 
%%%%%%%%%%%%%%%%%%%%%%%%%%%%%%%%%%%%%%%%%%% 
\begin{proof}
$(i)$ Suppose one point in $\calD_{\humu(n+1)}$ equals $P_{\infty_-}$ and 
denote the remaining ones by $\calD_{\tilde{\underline{\mu}}(n+1)}$. Then
\eqref{4.2} tells us
$\underline{\alpha}_{P_0}(\calD_{\tilde{\underline{\mu}}(n+1)}) 
+ \underline{A}_{P_0}(P_{\infty_+})
= \underline{\alpha}_{P_0}(\calD_{\humu(n)})$. Since we assumed 
$\calD_{\humu(n)}$ to be nonspecial, we have $\calD_{\humu(n)}=
\calD_{\tilde{\underline{\mu}}(n+1)} + \calD_{P_{\infty_+}}$ contradicting
finiteness of $\calD_{\humu(n)}$.

\noindent $(ii)$ We choose $P_0$ to be a branch point such that 
$\underline{A}_{P_0}(P^*)=-\underline{A}_{P_0}(P)$. In particular, if
$\calD_{\humu(n+1)}$ is special, then it contains a pair of points $(Q,Q^*)$
whose contribution will cancel under the Abel map, that is,
$\underline{\alpha}_{P_0}(\calD_{\humu(n+1)})= 
\underline{\alpha}_{P_0}(\calD_{{\underline{\hat \nu}}(n+1)})$
for some $\calD_{{\underline{\hat\nu}}(n+1)}\in \sym^{p-2} (\calK_\gg)$. 
But invoking \eqref{4.2} shows that 
$\underline{\alpha}_{P_0}(\calD_{\humu(n)})=
\underline{\alpha}_{P_0}(\calD_{{\underline{\hat\nu}}(n+1)}) + 2
\underline{A}_{P_0}(P_{\infty_+})$. As $\calD_{\humu(n)}$ was assumed
nonspecial, this shows that $\calD_{\humu(n)} =
\calD_{{\underline{\hat \nu}}(n+1)} + 2  \calD_{P_{\infty_+}}$, as claimed.
\end{proof} 
%%%%%%%%%%%%%%%%%%%%%%%%%%%%%%%%%%%%%%%%%%%

This yields the following behavior of $\calD_{\humu(n)}$ if we start 
with some nonspecial finite initial divisor $\calD_{\humu(n_0)}$: As $n$
increases, $\calD_{\humu(n)}$ stays nonspecial as long as it remains
finite. If it becomes infinite, then it is still nonspecial and contains
$P_{\infty_+}$ at least once (but not $P_{\infty_-}$). Further increasing
$n$, all instances of $P_{\infty_+}$ will be rendered into $P_{\infty_-}$
step by step, until we have again a nonspecial divisor that has the same
number of $P_{\infty_-}$ as the first infinite one had $P_{\infty_+}$.
Generically, we expect the subsequent divisor to be finite and
nonspecial again.

Next we show that most initial divisors are nice in the 
sense that their iterates stay away from infinity. Since we want to
show that this set is of full measure, it will be convenient for us to
identify $\symgg$ with the Jacobi variety $J(\calK_\gg)$
via the Abel map and take the Haar measure on $J(\calK_\gg)$.
Of course, the Abel map is only injective when restricted to the
set of nonspecial divisors, but these are the only ones we are interested
in. 

%%%%%%%%%%%%%%%%%%%%%%%%%%%%%%%%%%%%%%%%%% 
\begin{lemma} \lb{l4.1} 
The set $\calM_0\subset\symgg$ of initial divisors $\calD_{\humu(n_0)}$
for which $\calD_{\humu(n)}$, defined via
\eqref{4.2}, is finite and hence nonspecial for all $n\in\bbZ$, forms a
dense set of full measure  in the set $\symgg$ of nonnegative
divisors of degree $p$.  
%the Jacobi variety $J(\calK_\gg)$.  
\end{lemma} 
%%%%%%%%%%%%%%%%%%%%%%%%%%%%%%%%%%%%%%%%%% 
\begin{proof}
Let $\calM_\infty$ be the set of divisors in $\symgg$ for which 
(at least) one point is equal to $P_{\infty_+}$. The image
$\underline{\alpha}_{P_0}(\calM_\infty)$ of $\calM_\infty$  is given
by 
\begin{equation}
\underline{\alpha}_{P_0}(\calM_\infty)=\underline{A}_{P_0}(P_{\infty_+}) + 
\underline{\alpha}_{P_0}(\sym^{p-1} (\calK_\gg))\subset J(\calK_\gg). 
\end{equation}
Since the (complex) dimension of $\sym^{p-1} (\calK_\gg)$ is 
$p-1$, its image must be of measure zero by Sard's theorem (see, e.g.,
\cite[Sect.\ 3.6]{AMR88}). Similarly, let $\calM_{\rm sp}$ be the set of
special  divisors, then its image is given by 
\begin{equation}
\underline{\alpha}_{P_0}(\calM_{\rm sp}) =
\underline{\alpha}_{P_0}(\sym^{p-2} (\calK_\gg)), 
\end{equation}
assuming $P_0$ to be a branch point. In particular, we conclude that 
$\underline{\alpha}_{P_0}(\calM_{\rm sp})\subset
\underline{\alpha}_{P_0}(\calM_\infty)$ and thus
$\underline{\alpha}_{P_0}(\calM_{\rm sing})=
\underline{\alpha}_{P_0}(\calM_\infty)$ has measure zero, where
\begin{equation}
\calM_{\rm sing}=\calM_\infty\cup \calM_{\rm sp}. 
\end{equation}
Hence,
\begin{equation}
\bigcup_{n\in\bbZ} \big(\underline{\alpha}_{P_0}(\calM_{\rm sing}) 
+ n\underline{A}_{P_{\infty_-}}(P_{\infty_+}) \big)  \lb{4.5}
\end{equation}
is of measure zero as well.
But this last set contains all initial divisors which will hit
$P_{\infty_+}$ or become special at some $n$. We denote by  $\calM_0$ the
inverse image of the complement of the set \eqref{4.5} under the Abel map,
\begin{equation}
\calM_0= 
\underline{\alpha}_{P_0}^{-1} \bigg(\symgg \Big\backslash 
\bigcup_{n\in\bbZ} \big(\underline{\alpha}_{P_0}(\calM_{\rm sing}) 
+ n\underline{A}_{P_{\infty_-}}(P_{\infty_+}) \big)\bigg).
\end{equation}
Since $\calM_0$ is of full measure, it is automatically dense in $\symgg$.
\end{proof} 
%%%%%%%%%%%%%%%%%%%%%%%%%%%%%%%%%%%%%%%%%% 

We briefly illustrate some aspects of this analysis in the special case
$p=1$ (i.e., the case case where \eqref{1.2.20} represents an elliptic
Riemann surface) in more detail. 

%%%%%%%%%%%%%%%%%%%%%%%%%%%%%%%%%%%%%%%%%% 
\begin{example} \lb{e4.3}  The case $p=1$. \\ 
In this case we have
\begin{align}
F_1(z,n)&= z - \mu_1(n), \no \\
G_2(z,n)&= R_4(\hmu_1(n))^{1/2} + (z-b(n)) F_1(z,n),  \\
R_4(z)&=\prod_{m=0}^3 (z-E_m), \no
\end{align}
and hence a quick calculations shows that
\begin{align}
\begin{split}
G_2(z,n)^2 - R_4(z) & = 4a(n)^2 (z-\mu_1(n))(z-\mu_1(n+1))   \\
&= (z-\mu_1(n))(4a(n)^2 z - 4 a(n)^2 b(n) + \tilde{E}), 
\end{split} 
\end{align}
where
\begin{equation}
\tilde{E}= \frac{1}{8}(E_0+E_1-E_2-E_3) (E_0-E_1+E_2-E_3) (E_0-E_1-E_2+E_3).
\end{equation}
Solving for $\mu_1(n+1)$, one obtains
\begin{equation}
\mu_1(n+1) = b(n) - \frac{\tilde{E}}{4a(n)^2}.
\end{equation}
This shows that $\mu_1(n_0+1)\to\infty$, in fact,
$\mu_1(n_0+1)=\Oh(a(n_0)^{-2})$ as $a(n_0)\to 0$ during an appropriate
deformation of the parameters $E_m$, $m=0,\dots,3$. In particular, as $a(n_0)\to 0$, one thus infers $b(n_0+1)\to\infty$ during such a deformation since 
\begin{equation}
b(n)=\f{1}{2}\sum_{m=0}^{3} E_m - \mu_1(n), \quad n\in\bbZ, 
\end{equation}
specializing to $p=1$ in the trace formula \eqref{1.2.21b}. Next, we illustrate
the set $\calM_\infty$ in the case $p=1$. (We recall that 
$\calM_{\rm sp}=\emptyset$ and hence $\calM_{\rm sing}=\calM_\infty$ if
$p=1$.) By \eqref{4.2} one infers 
\begin{equation}
A_{\Pinfp}(\hat \mu_1(n))=A_{\Pinfp}(\hat \mu_1(n_0))+
(n-n_0) A_{\Pinfp}(\Pinfm), \quad n, n_0 \in\bbZ.  \lb{4.12a}
\end{equation}
We note that $\hat \mu_1 \in \calM_\infty$ is equivalent to 
\begin{equation}
\text{there is an $n\in\bbZ$ such that } \, \hat\mu_1(n) = \Pinfp 
\text{ (or $\Pinfm$).}   \lb{4.13a}
\end{equation}
By \eqref{4.12a}, relation \eqref{4.13a} is equivalent to
\begin{equation}
A_{\Pinfp}(\hat \mu_1(n_0))+A_{\Pinfp}(\Pinfm) \, \bbZ = 0 \pmod{L_1}. 
\lb{4.14a}
\end{equation}
Thus, $\calD_{\hat\mu_1(n_0)}\in\calM_0\subset\calK_1$ if and only if 
\begin{equation}
A_{\Pinfp}(\hat \mu_1(n_0))+A_{\Pinfp}(\Pinfm) \, \bbZ \neq 0 \pmod{L_1} 
\lb{4.15a}
\end{equation}
or equivalently, if and only if 
\begin{equation}
A_{\Pinfm}(\hat \mu_1(n_0))+A_{\Pinfm}(\Pinfp) \, \bbZ \neq 0 \pmod{L_1}. 
\lb{4.16aa}
\end{equation}
\end{example} 
%%%%%%%%%%%%%%%%%%%%%%%%%%%%%%%%%%%%%%%%%%%

\medskip

Next, we describe the stationary Toda algorithm. Since this is a somewhat
lengthy affair, we will break it up into several steps.  

\smallskip

%%%%%%%%%%%%%%%%%%%%%%%%%%%%%%%%%%%%%%%%%%% 
\noindent {\bf The Stationary (Complex) Toda Algorithm:}
%%%%%%%%%%%%%%%%%%%%%%%%%%%%%%%%%%%%%%%%%%%

We prescribe the following data: \\
$(i)$ The set
\begin{equation}
\{E_m\}_{m=0}^{2p+1} \subset\bbC, \quad E_m\neq E_{m'} \, \text{ for } \, 
m\neq m', \; m,m'=0,\dots,2p+1   \lb{4.8}
\end{equation}
for some fixed $p\in\bbN$. Given $\{E_m\}_{m=0}^{2p+1}$, we introduce 
the function $R_{2p+2}$ and the (nonsingular) hyperelliptic curve
$\calK_p$ as in \eqref{1.2.20}. \\
$(ii)$ The nonspecial divisor 
\begin{equation}
\calD_{\humu(n_0)}\in \symgg,   \lb{4.9}
\end{equation}
where $\humu(n_0)$ is of the form
\begin{align}
\begin{split}
\humu(n_0)&=\{\hat\mu_1(n_0),\dots,\hat\mu_p(n_0)\}  \\
&=\{\underbrace{\hat\mu_1(n_0),\dots,\hat\mu_1(n_0)}_{p_1(n_0) \text{
times}},
\dots,\underbrace{\hat\mu_{q(n_0)},\dots,
\hat\mu_{q(n_0)}}_{p_{q(n_0)}(n_0) \text{ times}}\}    \lb{4.10}
\end{split}
\end{align}
with 
\begin{equation}
\hat\mu_k(n_0)=(\mu_k(n_0),y(\hat\mu_k(n_0))), \quad  
\mu_k(n_0)\neq \mu_{k'}(n_0) \, \text{ for } \, k\neq k', \; 
k,k'=1,\dots,q(n_0),     \lb{4.11}
\end{equation}
and
\begin{equation}
p_k(n_0)\in\bbN, \; k=1,\dots,q(n_0), \quad 
\sum_{k=1}^{q(n_0)} p_k(n_0) = p.   \lb{4.12}
\end{equation}

With $\{E_m\}_{m=0}^{2p+1}$ and $\calD_{\humu(n_0)}$ prescribed, we next
introduce the following quantities (for $z\in\bbC$):
\begin{align} 
F_p(z,n_0)&=\prod_{k=1}^{q(n_0)} (z-\mu_k(n_0))^{p_k(n_0)},  \lb{4.13} \\
T_{p-1}(z,n_0)&=-F_p(z,n_0)\sum_{k=1}^{q(n_0)}\sum_{\ell=0}^{p_k(n_0)-1} 
\f{\big(d^\ell
\big(R_{2p+2}(\zeta)^{1/2}\big)/d\zeta^\ell\big)\big|_{\zeta
=\mu_k(n_0)}}{\ell!(p_k(n_0)-\ell-1)!} \lb{4.14} \\ 
& \hspace*{-.5cm} \times 
\Bigg(\f{d^{p_k(n_0)-\ell-1}}{d \zeta^{p_k(n_0)-\ell-1}}\Bigg(
(z-\zeta)^{-1}\prod_{k'=1, \, k'\neq k}^{q(n_0)} 
(\zeta-\mu_{k'}(n_0))^{-p_{k'}(n_0)} 
\Bigg)\Bigg)\Bigg|_{\zeta=\mu_k(n_0)},  \no  \\
b(n_0)&=\f{1}{2}\sum_{m=0}^{2p+1}E_m -\sum_{k=1}^{q(n_0)} p_k(n_0)
\mu_k(n_0),  \lb{4.15} \\ 
G_{p+1}(z,n_0)&=-(z-b(n_0)) F_p(z,n_0)+T_{p-1}(z,n_0).  \lb{4.16}
\end{align}
Here the sign of the square root in \eqref{4.14} is chosen according to
\eqref{4.11},
\begin{equation}
\hat\mu_k(n_0)=(\mu_k(n_0),y(\hat\mu_k(n_0)))
=\big(\mu_k(n_0), R_{2p+2}(\mu_k(n_0))^{1/2}\big), \quad  
k=1,\dots,q(n_0).   \lb{4.16a}
\end{equation}

Next we record a series of facts: \\
\smallskip
$\textbf{(I)}$ By construction (cf.\ Lemma \ref{B.1}),
\begin{align}
\begin{split}
T_{p-1}^{(\ell)}(\mu_k(n_0),n_0)=- 
\f{d^\ell \big(R_{2p+2}(z)^{1/2}\big)}{dz^\ell}\bigg|_{z=\mu_k(n_0)}
=G^{(\ell)}_{p+1}(\mu_k(n_0),n_0), \\
\ell=0,\dots,p_k(n_0)-1, \; k=1,\dots,q(n_0),    \lb{4.17}
\end{split}
\end{align}
(here the superscript $(\ell)$ denotes $\ell$ derivatives w.r.t. $z$) and hence
\begin{equation}
\hat\mu_k(n_0)=(\mu_k(n_0),-G_{p+1}(\mu_k(n_0),n_0)), \quad 
k=1,\dots,q(n_0).   \lb{4.17a}
\end{equation}
$\textbf{(II)}$ Since $\calD_{\humu(n_0)}$ is nonspecial by hypothesis,
one concludes that
\begin{equation}
p_k(n_0)\geq 2 \, \text{ implies } \, R_{2p+2}(\mu_k(n_0))\neq 0, \quad 
k=1,\dots,q(n_0).    \lb{4.18}
\end{equation}
$\textbf{(III)}$ By $\textbf{(I)}$ and $\textbf{(II)}$ one computes
\begin{align}
\begin{split}
\f{d^\ell \big(G_{p+1}(z,n_0)^2\big)}{dz^\ell}\bigg|_{z=\mu_k(n_0)}= 
\f{d^\ell R_{2p+2}(z)}{dz^\ell}\bigg|_{z=\mu_k(n_0)}, \\
z\in\bbC, \quad \ell=0,\dots,p_k(n_0)-1, \; k=1,\dots,q(n_0).    \lb{4.19}
\end{split}
\end{align}
$\textbf{(IV)}$ By \eqref{4.16} and \eqref{4.19} one infers that $F_p$
divides
$R_{2p+2}-G_{p+1}^2$. \\
$\textbf{(V)}$ By \eqref{4.15} and \eqref{4.16} one verifies
that
\begin{equation}
R_{2p+2}(z)-G_{p+1}(z,n_0)^2 \underset{z\to \infty}{=} \Oh(z^{2p}).  
\lb{4.20}
\end{equation}
By $\textbf{(IV)}$ and \eqref{4.20} we may write 
\begin{equation}
R_{2p+2}(z)-G_{p+1}(z,n_0)^2=F_p(z,n_0) \check F_{p-r}(z,n_0+1), 
\quad z\in\bbC,   \lb{4.20a}
\end{equation}
for some $r\in\{0,\dots,p\}$, 
where the polynomial $\check F_{p-r}$ has degree $p-r$. 
If in fact $\check F_0 =0$, then $R_{2p+2}(z)=G_{p+1}(z,n_0)^2$ would
yield double zeros of $R_{2p+2}$, contradicting our basic hypothesis
\eqref{4.8}. Thus we conclude that in the case $r=p$, $\check F_{0}$ 
cannot vanish identically and hence we may break up \eqref{4.20a} in the
following manner 
\begin{equation}
\check \phi(P,n_0)=\f{y-G_{p+1}(z,n_0)}{F_p(z,n_0)}
=\f{\check F_{p-r}(z,n_0+1)}{y+G_{p+1}(z,n_0)}, \quad P=(z,y)\in\calK_p. 
\lb{4.20b}
\end{equation}
Next we decompose
\begin{equation}
\check F_{p-r}(z,n_0+1)=\check C \prod_{j=1}^{p-r} (z-\mu_j(n_0+1)),
\quad z\in\bbC,   \lb{4.20c}
\end{equation}
where $\check C \in\bbC\setminus\{0\}$ and
$\{\mu_j(n_0+1)\}_{j=1}^{p-r}\subset 
\bbC$ (if $r=p$ we follow the usual convention and replace the product in
\eqref{4.20c} by $1$). By inspection of the local zeros and poles as well
as the behavior near $\Pinfpm$ of the function $\check\phi (\dott,n_0)$,
its divisor, $\big(\check \phi(\dott,n_0) \big)$, is given by
\begin{equation}
\big(\check \phi(\dott,n_0) \big)=\calD_{P_{\infty_+} \hat{\underline{
\mu}}(n_0+1)}-\calD_{P_{\infty_-}\hat{\underline{\mu}}(n_0)},
\lb{4.20d}
\end{equation}
where
\begin{equation}
\hat{\ul{\mu}}(n_0+1)=\{\hat{\mu}_1(n_0+1),\dots,
\hat{\mu}_{p-r}(n_0+1),\underbrace{\Pinfp,\dots,\Pinfp}_{r
\text{ times}}\}.  
\lb{4.20e}
\end{equation}
In particular, 
\begin{equation}
\calD_{\hat{\ul{\mu}}(n_0+1)} \, \text{ is a finite divisor if
and only if } \, r=0.  \lb{4.20f}
\end{equation}
We note that 
\begin{equation}
\ual_{Q_0} (\calD_{\hat{\ul{\mu}}(n_0+1)}) = \ual_{Q_0}
(\calD_{\hat\umu (n_0)}) - \underline{A}_{P_{\infty_-}}
(P_{\infty_+}), \lb{4.20g} 
\end{equation}
in accordance with \eqref{4.2}. \\
$\textbf{(VI)}$ {\it Assuming} that \eqref{4.20} is precisely of order
$z^{2p}$ as $z\to\infty$, that is, assuming $r=0$ in \eqref{4.20a}, we
rewrite
\eqref{4.20a} in the more appropriate manner
\begin{equation}
R_{2p+2}(z)-G_{p+1}(z,n_0)^2=-4a(n_0)^2 F_p(z,n_0)F_p(z,n_0+1), 
\quad z\in\bbC,   \lb{4.21}
\end{equation}
where we introduced the coefficient $a(n_0)^2$ to make $F_p(\dott,n_0+1)$
a monic polynomial  of degree $p$. (We will later discuss
conditions which indeed guarantee that $r=0$, cf.\ \eqref{4.20f} and the
discussion in step $\textbf{(XI)}$ below.) By construction,
$F_p(\dott,n_0+1)$ is then of the type
\begin{align}
\begin{split}
F_p(z,n_0+1)=\prod_{k=1}^{q(n_0+1)} (z-\mu_k(n_0+1))^{p_k(n_0+1)}, 
\quad \sum_{k=1}^{q(n_0+1)} p_k(n_0+1) =p,   \\
\mu_k(n_0+1) \neq \mu_{k'}(n_0+1) \, \text{ for } \, k\neq k', \; 
k, k' =1,\dots,q(n_0+1), \;\; z\in\bbC,  \lb{4.22}
\end{split}
\end{align}
and we {\it define}
\begin{equation}
\hat\mu_k(n_0+1)=(\mu_k(n_0+1),G_{p+1}(\mu_k(n_0+1),n_0)), \quad 
k=1,\dots,q(n_0+1).  \lb{4.23}
\end{equation}
Moreover, we introduce the divisor 
\begin{equation}
\calD_{\humu(n_0+1)}\in \symgg   \lb{4.23a}
\end{equation}
by
\begin{align}
\begin{split}
\humu(n_0+1)&=\{\hat\mu_1(n_0+1),\dots,\hat\mu_p(n_0+1)\}   \\
&=\{\underbrace{\hat\mu_1(n_0+1),\dots,
\hat\mu_1(n_0+1)}_{p_1(n_0+1) \text{ times}},
\dots,\underbrace{\hat\mu_{q(n_0+1)},\dots,
\hat\mu_{q(n_0+1)}}_{p_{q(n_0+1)}(n_0+1) \text{ times}}\}.    \lb{4.23b}
\end{split}
\end{align}
In particular, because of the definition \eqref{4.23}, 
$\calD_{\humu(n_0+1)}$ is nonspecial and hence 
\begin{equation}
p_k(n_0+1)\geq 2 \, \text{ implies } \, R_{2p+2}(\mu_k(n_0+1))\neq 0,
\quad  k=1,\dots,q(n_0+1).    \lb{4.24}
\end{equation}
Again we note that 
\begin{equation}
\ual_{Q_0} (\calD_{\hat\umu(n_0+1)}) = \ual_{Q_0} (\calD_{\hat\umu (n_0)})
- \underline{A}_{P_{\infty_-}} (P_{\infty_+}), \lb{4.24a}
\end{equation}
in accordance with \eqref{4.2}.  \\
$\textbf{(VII)}$ Introducing
\begin{equation}
b(n_0+1)=\f{1}{2}\sum_{m=0}^{2p+1}E_m -\sum_{k=1}^{q(n_0+1)} p_k(n_0+1)
\mu_k(n_0+1),  \lb{4.25} 
\end{equation}
and interpolating $G_{p+1}(\dott,n_0)$ with $F_p(\dott,n_0+1)$ rather
than $F_p(\dott,n_0)$ yields
\begin{equation}
G_{p+1}(z,n_0)=-(z-b(n_0+1))F_p(z,n_0+1)-T_{p-1}(z,n_0+1), \quad
z\in\bbC,   \lb{4.26}
\end{equation}
where
\begin{align}
&T_{p-1}(z,n_0+1)=F_p(z,n_0+1)  \no \\
& \times \sum_{k=1}^{q(n_0+1)}\sum_{\ell=0}^{p_k(n_0+1)-1} 
\f{\big(d^\ell
\big(R_{2p+2}(\zeta)^{1/2}\big)/d\zeta^\ell\big)\big|_{\zeta
=\mu_k(n_0+1)}}{\ell!(p_k(n_0+1)-\ell-1)!}   \lb{4.27}  \\ 
& \times \Bigg(\f{d^{p_k(n_0+1)-\ell-1}}{d
\zeta^{p_k(n_0+1)-\ell-1}}\Bigg( (z-\zeta)^{-1}\prod_{k'=1, \, k'\neq
k}^{q(n_0+1)}  (\zeta-\mu_{k'}(n_0+1))^{-p_{k'}(n_0+1)} 
\Bigg)\Bigg)\Bigg|_{\zeta=\mu_k(n_0+1)}.  \no 
\end{align}
Here the sign of the square root in \eqref{4.27} is chosen in accordance
with \eqref{4.23}, that is,
\begin{align}
\hat\mu_k(n_0+1)&=(\mu_k(n_0+1),y(\hat\mu_k(n_0+1))) \no \\
&=(\mu_k(n_0+1),G_{p+1}(\mu_k(n_0+1),n_0))   \no \\
&=\big(\mu_k(n_0+1), R_{2p+2}(\mu_k(n_0+1))^{1/2}\big), \quad  
k=1,\dots,q(n_0+1).   \lb{4.27a}
\end{align}
$\textbf{(VIII)}$ An explicit computation of $a(n_0)^2$ then yields 
\begin{align}
a(n_0)^2&=\f{1}{2}\sum_{k=1}^{q(n_0)}
\f{\big(d^\ell \big(R_{2p+2}(z)^{1/2}\big)/d z^\ell\big)\big|_{z
=\mu_k(n_0)}}{(p_k(n_0)-1)!}  \no \\
& \quad \times \prod_{k'=1, \, k'\neq k}^{q(n_0)}
(\mu_k(n_0)-\mu_{k'}(n_0))^{-p_k(n_0)} 
+\f{1}{4}\big(b^{(2)}(n_0)-b(n_0)^2 \big).  \lb{4.29}
\end{align}
Here and in the following we abbreviate
\begin{equation}
b^{(2)}(n)=\f{1}{2}\sum_{m=0}^{2p+1}E_m^2 
-\sum_{k=1}^{q(n)} p_k(n) \mu_k(n)^2  \lb{4.30}
\end{equation}
for an appropriate range of $n\in\bbN$.

The result \eqref{4.29} is obtained as follows: One starts from the
identity \eqref{4.21}, inserts the expressions \eqref{4.13} and 
\eqref{4.16} for $F_{p}(\dott,n_0)$ and  $G_{p+1}(\dott,n_0)$,
respectively, then inserts the explicit form \eqref{4.14} of
$T_{p-1}(\dott,n_0)$, and finally collects all terms of order $z^{2p}$ as
$z\to\infty$. An entirely elementary but fairly tedious calculation then
produces \eqref{4.29}. 

In the special case $q(n_0)=p$, $p_k(n_0)=1$, $k=1,\dots,p$, \eqref{4.29}
and \eqref{4.30} reduce to \eqref{1.2.21a} and \eqref{3.21} (for $k=2$).
\\
$\textbf{(IX)}$ Introducing
\begin{equation}
G_{p+1}(z,n_0+1)=-(z-b(n_0+1)) F_p(z,n_0+1)+T_{p-1}(z,n_0+1)  \lb{4.31}
\end{equation}
one then obtains
\begin{equation}
G_{p+1}(z,n_0+1)=-G_{p+1}(z,n_0)-2(z-b(n_0+1))F_p(z,n_0+1).  \lb{4.32}
\end{equation}
$\textbf{(X)}$ At this point one can iterate the procedure step by step to
construct 
$F_p(\dott,n)$, $G_{p+1}(\dott,n)$, $T_{p-1}(\dott,n)$, $a(n)$, $b(n)$,
$\mu_k(n)$, $k=1,\dots,q(n)$, etc., for $n\in [n_0,\infty)\cap\bbZ$,
subject to the following assumption (cf.\ \eqref{4.20f}) at each step:
\begin{equation}
\calD_{\humu(n+1)} \, \text{ is a finite divisor (and hence } \, 
a(n) \neq 0) \text{ for all } \, n\in[n_0,\infty) \cap \bbZ.   \lb{4.33}
\end{equation}
The formalism is symmetric with respect to $n_0$ and can equally well be
developed for $n\in (-\infty,n_0]\cap\bbZ$ subject to the analogous
assumption
\begin{equation}
\calD_{\humu(n-1)} \, \text{ is a finite divisor (and hence } \, 
a(n) \neq 0) \text{ for all } \, n\in(-\infty,n_0] \cap \bbZ.  \lb{4.34}
\end{equation}
Indeed, one first interpolates $G_{p+1}(\dott,n_0-1)$ with the help of 
$F_p(\dott,n_0)$, then with $F_p(\dott,n_0-1)$, etc. 

Moreover, we once again remark for consistency reasons that
\begin{equation}
\ual_{Q_0} (\calD_{\hat\umu(n)}) = \ual_{Q_0} (\calD_{\hat\umu (n_0)})
-(n-n_0) \underline{A}_{P_{\infty_-}} (P_{\infty_+}), 
\quad n\in\bbZ,  \lb{4.34a}
\end{equation}
in agreement with our starting point \eqref{4.2}.  \\
$\textbf{(XI)}$ Choosing the initial data $\calD_{\humu(n_0)}$ such that 
\begin{equation}
\calD_{\humu(n_0)}\in \calM_0,  \lb{4.35}
\end{equation}
where $\calM_0\subset\symgg$ is the set of finite divisors introduced in
Lemma \ref{l4.1}, then guarantees that assumptions \eqref{4.33} and
\eqref{4.34} are satisfied for all $n\in\bbZ$. \\
$\textbf{(XII)}$ Performing these iterations for all
$n\in\bbZ$, one then arrives at the following set of equations for
$F_p$ and $G_{p+1}$ after the following elementary manipulations: 
Utilizing 
\begin{equation}
G_{p+1}^2-4a^2F_pF_p^+ = R_{2p+2} = (G_{p+1}^-)^2-4(a^-)^2F_p^-F_p,
\lb{4.36}
\end{equation}
and inserting
\begin{equation}
G_{p+1}^+=-G_{p+1}-2(z-b^+)F_p^+   \lb{4.37}
\end{equation}
into 
\begin{equation}
G_{p+1}^2- (G_{p+1}^-)^2 -4a^2F_pF_p^+ +4(a^-)^2F_p^-F_p=0
\lb{4.38}
\end{equation}
then yields
\begin{equation}
2 a^2 F_p^+ - 2(a^-)^2 F_p^-  +(z-b)(G_{p+1}-G_{p+1}^-) =0. \lb{4.39}
\end{equation}
Subtracting \eqref{4.37} from its shifted version $G_{p+1}=-G_{p+1}^- -2
(z-b)F_p$ then also yields
\begin{equation}
2(z-b^+)F_p^+ -2(z-b)F_p +G_{p+1}^+ - G_{p+1}^- =0. \lb{4.40}
\end{equation} 
As discussed in Section \ref{s2}, \eqref{4.39} and \eqref{4.40} are
equivalent to the stationary Lax and zero-curvature equations
\eqref{1.2.8aa} and \eqref{3.2.42}. At this stage we have verified the
basic hypotheses of Section \ref{s3} (i.e., Hypothesis \ref{h3.1} and the
assumption that $a, b$ satisfy the $p$th stationary Toda system 
\eqref{3.2.29d}) and hence all results of Section \ref{s3} apply. 

Finally, we briefly summarize these considerations:

%%%%%%%%%%%%%%%%%%%%%%%%%%%%%%%%%%%%%
\begin{theorem} \lb{t4.3}
Suppose the set $\{E_m\}_{m=0}^{2p+1} \subset\bbC$ satisfies 
$E_m\neq E_{m'}$ for $m\neq m'$, $m,m'=0,\dots,2p+1$, and introduce the
function $R_{2p+2}$ and the hyperelliptic curve $\calK_p$ as in
\eqref{1.2.20}. Choose a nonspecial divisor $\calD_{\humu(n_0)}\in
\calM_0$, where $\calM_0\subset\symgg$ is the set of finite divisors 
introduced in Lemma \ref{l4.1}. Then the stationary $($complex$)$ Toda
algorithm as outlined in steps \textbf{$($I\,$)$}--\textbf{$($XII\,$)$}
produces solutions $a, b$ of the $\gg$th stationary Toda system,
\begin{equation}
\stl_p(a,b)= \begin{pmatrix} f_{p+1}^+-f_{p+1} \\
g_{p+1}-g_{p+1}^- \end{pmatrix}=0, \quad p\in\bbN_0, \lb{4.54}
\end{equation}
satisfying \eqref{1.3} and  
\begin{align}
a(n)^2&=\f{1}{2}\sum_{k=1}^{q(n)}
\f{\big(d^\ell \big(R_{2p+2}(z)^{1/2}\big)/d z^\ell\big)\big|_{z
=\mu_k(n)}}{(p_k(n)-1)!}  \no \\
& \quad \times \prod_{k'=1, \, k'\neq k}^{q(n)}
(\mu_k(n)-\mu_{k'}(n))^{-p_k(n)} 
+\f{1}{4}\big(b^{(2)}(n)-b(n)^2 \big),  \lb{4.55} \\
b(n)&=\f{1}{2}\sum_{m=0}^{2p+1}E_m -\sum_{k=1}^{q(n)} p_k(n)
\mu_k(n), \quad n\in\bbZ. \lb{4.56} 
\end{align}
Moreover, Lemmas \ref{l1.3.1}--\ref{l1.3.9ba} apply.  
\end{theorem}
%%%%%%%%%%%%%%%%%%%%%%%%%%%%%%%%%%%%%

%%%%%%%%%%%%%%%%%%%%%%%%%%%%%%%%%%%%%
\begin{remark}  \lb{r4.5}
Suppose the hypotheses of the previous theorem are satisfied and
$a(n_0)$, $b(n_0)$, $b(n_0+1)$, $F_p(z,n_0)$, $F_p(z,n_0+1)$,
$G_{p+1}(z,n_0)$, and $G_{p+1}(z,n_0+1)$ have already been computed
using step \textbf{$($I\,$)$}--\textbf{$($IX\,$)$}. Then, alternatively, one can use
\begin{align}
(a^-)^2 F_p^- &= a^2 F_p^+ + 2^{-1}(z-b)(G_{p+1}-G_{p+1}^+) 
+(z-b)^2F_p  \no \\
& \quad -(z-b^+)(z-b)F_p^+,\\
G^- &= 2((z-b^+)F_p^+ -(z-b)F_p) + G_{p+1}^+
\end{align}
$($derived from \eqref{3.2.42}$)$ to compute $a(n)$, $b(n)$, $F_p(z,n)$, $G_{p+1}(z,n)$ for $n<n_0$ and
\begin{align}
a^+ F^{++} &= a F_p - 2^{-1}(z-b)(G_{p+1}^+ -G_{p+1}),\\
G^{++} &= G - 2((z-b^{++})F_p^{++} - (z-b^+)F_p^+)
\end{align}
to compute $a(n-1)$, $b(n)$, $F_p(z,n)$, $G_{p+1}(z,n)$ for $n>n_0+1$.
\end{remark}
%%%%%%%%%%%%%%%%%%%%%%%%%%%%%%%%%%%%%

Theta function representations of $a$ and $b$ can now be derived in complete analogy to the self-adjoint case. Since the final results are formally the same as in the self-adjoint case we just refer, for instance, to \cite{BGHT98}, \cite{DT76}, \cite{DKN90}, 
\cite{DMN76}, \cite[Sect.\ 1.3]{GH07}, \cite{Kr78}, \cite{Kr82}, \cite{Kr82a}, \cite{Kr83} (cf.\ also the appendix written in \cite{Du81}), \cite{vMM79}, 
\cite[Appendix, Sect.\ 9]{NMPZ84}, \cite[Sect.\ 9.2]{Te00}, \cite[Sect.\ 4.5]{To89}.

The stationary $($complex$)$ Toda algorithm as outlined in steps \textbf{$($I\,$)$}--\textbf{$($XII\,$)$}, starting from a nonspecial divisor $\calD_{\humu(n_0)}\in \calM_0$, represents a solution of the inverse algebro-geometric spectral problem for generally non-self-adjoint Jacobi operators. While we do not assume periodicity (or even quasi-periodicity), let alone real-valuedness of the coefficients of the underlying Jacobi operator, once can view this algorithm a continuation of the inverse periodic spectral problem started around 1975 (in the self-adjoint context) by Kac and van Moerbeke 
\cite{KvM75}, \cite{KvM75a} and Flaschka \cite{Fl75}, continued in the seminal papers by van Moerbeke \cite{vM76}, Date and Tanaka \cite{DT76}, and Dubrovin, Matveev, and Novikov \cite{DMN76}, and further developed by  Krichever \cite{Kr78}, McKean 
\cite{McK79}, van Moerbeke and Mumford \cite{vMM79}, Mumford \cite{Mu78}, and others, in part in the more general quasi-periodic algebro-geometric case.

We note that in general (i.e., unless one is, e.g., in the special
periodic or self-adjoint case), $\calD_{\humu(n)}$ will get
arbitrarily close to $P_{\infty_+}$ since straight motions on the torus
are generically dense (see e.g. \cite[Sect.\ 51]{Ar89} or \cite[Sects.\ 1.4, 1.5]{KH95}). 
Thus, no uniform bound on the sequences $a(n), b(n)$ exists as $n$ varies in $\bbZ$. In particular, these complex-valued algebro-geometric solutions of some of the equations of the stationary Toda hierarchy, generally, will not be quasi-periodic (cf.\ the usual definition of quasi-periodic functions, e.g., in \cite[p.\ 31]{PF92}) with respect to $n$. For the special case of complex-valued and quasi-periodic Jacobi matrices where all quasi-periods are real-valued, we refer to \cite{BG05a} (cf.\ also \cite{BG05}).

%%%%%%%%%%%%%%%%%%%%%%%%%%%%%%%%%%%%%%%%%%% 
%%%%%%%%%%%%%%%%%%%%%%%%%%%%%%%%%%%%%%%%%%%   
\section{Properties of Algebro-Geometric Solutions \\ of the  
Time-Dependent Toda Hierarchy} \lb{s5}  
%%%%%%%%%%%%%%%%%%%%%%%%%%%%%%%%%%%%%%%%%%%
%%%%%%%%%%%%%%%%%%%%%%%%%%%%%%%%%%%%%%%%%%%

In this section we present a quick review of properties of
algebro-geometric solutions of the time-dependent Toda hierarchy. Since
this material is standard we omit all proofs and just refer to
\cite{BGHT98}  (cf.\ also \cite[Sect.\ 1.4]{GH07}, \cite[Chs.\ 12,
13]{Te00}) for detailed presentations and an extensive list references to
the literature. 

Throughout this section we will make the following assumption:

%%%%%%%%%%%%%%%%%%%%%%%%%%%%%%%%%
\begin{hypothesis} \lb{h3.4.1}
Suppose that $a, b$ satisfy
\begin{align}
\begin{split}
& a(\dott,t), b(\dott,t)\in \bbC^\bbZ,\; t\in\bbR,
\quad a(n,\dott), b(n,\dott)\in C^1(\bbR), \; n\in\bbZ,    \lb{5.1} \\
& a(n,t) \neq 0, \; (n,t)\in \bbZ\times\bbR
\end{split}
\end{align}
and assume that the hyperelliptic curve $\calK_p$, $p\in\bbN_0$, is nonsingular.
\end{hypothesis}
%%%%%%%%%%%%%%%%%%%%%%%%%%%%%%%%%

In order to briefly analyze algebro-geometric solutions of the
time-dependent Toda hierarchy we proceed as follows. Given $p\in\bbN_0$,
consider a complex-valued solution $a^{(0)}, b^{(0)}$ of the $p$th stationary Toda
equation $\stl_p(a,b)= 0$, associated with $\calK_p$ and a given set of
summation constants $\{c_\ell\}_{\ell=1,\dots,p}\subset\bbC$. Next, let
$r\in\bbN_0$; we intend to consider solutions $a=a(t_r),b=b(t_r)$ of the
$r$th Tl  flow $\tl_r(a,b)=0$ with $a(t_{0,r})=a^{(0)},
b(t_{0,r})=b^{(0)}$ for some $t_{0,r}\in\bbR$. To emphasize that the
summation constants in the definitions of the stationary and the
time-dependent Tl equations are independent of each other, we indicate
this by adding a tilde on all the time-dependent quantities. Hence we
shall employ the notation 
$\wti P_{2r+2}$, $\wti V_{r+1}$, $\wti F_r$, $\wti G_{r+1}$, $\tilde f_{s}$,
$\tilde g_{s}$, $\tilde c_{s}$, in order to distinguish them from $P_{2p+2}$,
$V_{p+1}$, $F_p$, $G_{p+1}$, $f_{\ell}$, $g_{\ell}$, $c_{\ell}$,
in the following. In addition, we will follow a more elaborate notation
inspired by Hirota's $\tau$-function approach and indicate the individual
$r$th $\tl$ flow by  a separate time variable $t_r \in \bbR$. More precisely,
we will review properties of solutions $a,b$ of the time-dependent
algebro-geometric initial value problem
\begin{align}
\begin{split}
&\wti\tl_r(a,b)=\begin{pmatrix}a_{t_r}-a\big(\tilde{f}^+_{p+1}(a,b)-
\tilde{f}_{p+1}(a,b)\big) \\ 
 b_{t_r}+\tilde{g}_{p+1}(a,b)-\tilde{g}^-_{p+1}(a,b) \end{pmatrix}=0, 
\label{43} \\  
&(a,b)\big|_{t_r=t_{0,r}}=\big(a^{(0)},b^{(0)}\big),  
\end{split} \\
&\sTl_p\big(a^{(0)},b^{(0)}\big)=
\begin{pmatrix}-a\big({f^+}_{p+1}\big(p^{(0)},q^{(0)}\big)
-{f}_{p+1}\big(p^{(0)},q^{(0)}\big)\big) \\
 {g}_{p+1}\big(a^{(0)},b^{(0)}\big)
-g^-_{p+1}\big(a^{(0)},b^{(0)}\big) \end{pmatrix}=0
\label{42}
\end{align}
for some $t_{0,r}\in\bbR$, $p,r\in\bbN_0$, where $a=a(n,t_r),b=b(n,t_r)$
satisfy \eqref{5.1} and a fixed curve $\calK_p$ is associated
with the stationary solutions $a^{(0)}, b^{(0)}$ in \eqref{42}. In terms of
Lax pairs this amounts to solving
\begin{align}
\f{d}{dt_r}L(t_r)-[\wti{P}_{2r+2}(t_r), L(t_r)]&=0,\quad
t_r \in \bbR, \label{44} \\
 [P_{2p+2}(t_{0,r}),L(t_{0,r})]&= 0. \label{44a}
\end{align}
Anticipating that the Tl flows are isospectral deformations of
$L(t_{0,r})$, we are going a step further replacing \eqref{44a} by
\begin{equation}
[P_{2p+2}(t_r),L(t_r)]=0,\quad t_r \in \bbR. \label{45} 
\end{equation}
This then implies
\begin{equation}
P_{2p+2}(t_r)^2= R_{2p+2}(L(t_r))= \prod_{m=0}^ {2p+1} 
(L(t_r)-E_m), \quad t_r \in \bbR. \label{46}
\end{equation}
Actually, instead of working with \eqref{44}, \eqref{44a}, and 
\eqref{45}, one can equivalently take the zero-curvature equations
\eqref{3.2.45} as one's point of  departure, that is, one can also start from
\begin{align}
U_{t_{r}}+U\wti{V}_{r+1}-\wti V_{r+1}^+U&=0, \label{47} \\
UV_{p+1}-V^+_{p+1}U&=0, \label{48}
\end{align}
where (cf.\ \eqref{1.2.11a}, \eqref{1.2.11b}, \eqref{3.2.40}, \eqref{3.2.41})
\begin{align}
  U(z) &= \begin{pmatrix} 0 & 1 \\ -a^-/a & (z-b)/a \end{pmatrix}, \no \\
 V_{p+1}(z) &= \begin{pmatrix}
  G_{p+1}^-(z) & 2a^-F_p^-(z) \\ -2a^-F_p(z) & 
2(z-b)F_p+G_{p+1}(z) \end{pmatrix},  
\label{49} \\ 
 \wti{V}_{r+1}(z) &= \bigg(\begin{matrix}
 \wti{G}_{r+1}^-(z) & 2a^-\wti{F}_r^-(z) \\-2a^-\wti{F}_r(z) &
2(z-b)\wti F_r(z)+\wti{G}_{r+1}(z)
 \end{matrix}\bigg), \no 
\end{align}
and 
\begin{align} 
F_p(z)&= \sum_{{\ell}=0}^{p}f_{p-{\ell}}z^{\ell}
=\prod_{j=1}^p (z-\mu_j), \quad f_0=1, \label{410} \\
G_{p+1}(z)&= -z^{p+1} + 
\sum_{{\ell}=0}^{p}g_{p-{\ell}}z^{\ell}+f_{p+1},
      \; \; g_0=-c_1,  \label{411} \\
\wti F_r(z)&= \sum_{s=0}^{r}\tilde f_{r-s}z^s, \quad \tilde
f_0=1,  \label{412aa} \\ 
\wti G_{r+1}(z)&= -z^{r+1}+
\sum_{s=0}^{r}\tilde g_{r-s}z^s+ \tilde f_{r+1},
      \; \; \tilde g_0=-\tilde c_1,  \label{412ab} 
\end{align}
for fixed  $p, r \in \bbN_0$. Here $f_{\ell}$, $\tilde
f_s$, $g_{\ell}$, and $\tilde g_s$, $\ell=0,\dots,p$, $s=0,\dots,r$, are
defined as in \eqref{1.2.4a}--\eqref{1.2.4c} with appropriate sets of
summation constants $c_\ell$, $\ell\in\bbN$, and $\tilde c_k$, $k\in\bbN$.  Explicitly, \eqref{47} and \eqref{48} are equivalent to (cf.\ \eqref{3.2.25time}, \eqref{3.2.26time}, \eqref{1.2.16a}, \eqref{1.2.16b})
\begin{align}
a_{t_r} &=-a\big(2(z-b^+) \wti F_r^+
   +\wti G_{r+1}^+ + \wti G_{r+1}\big),  \lb{5.15}  \\
b_{t_r} &=2 \big((z-b)^2 \wti F_r
+ (z-b) \wti G_{r+1} + a^2 \wti F_r^+
- (a^-)^2 \wti F_r^-\big),  \lb{5.16}  \\
0&=2(z-b^{+})F_\gg^{+}+G_{\gg+1}^{+}+G_{\gg+1}, \lb{5.17} \\
0&=(z-b)^2F_\gg+(z-b)G_{\gg+1}+a^2F_\gg^{+}-(a^{-})^2F_\gg^{-}, \lb{5.18}
\end{align}
respectively. In particular, \eqref{1.2.17} holds in the present
$t_r$-dependent setting, that is,
\begin{equation}
G^{2}_{p+1} - 4 a^2 F_pF^+_p=R_{2p+2}. \label{412g}
\end{equation}

As in \eqref{1.2.24a} one introduces
\begin{align}
\hmu_j(n,t_r)&=(\mu_j(n,t_r),-G_{p+1}(\mu_j(n,t_r),n,t_r))\in\calK_p,
\quad j=1,\dots,p, \; (n,t_r)\in\bbZ\times\bbR, \lb{3.6.7AA}  \\
\hmu_j^+(n,t_r)&=(\mu_j^+(n,t_r),G_{p+1}(\mu_j^+(n,t_r),n,t_r))\in\calK_p,
\quad j=1,\dots,p, \; (n,t_r)\in\bbZ\times\bbR, \lb{3.6.7BB}
\end{align}
and notes that the regularity assumptions \eqref{5.1} on $a, b$ imply
continuity of $\mu_j$ with respect to $t_r\in\bbR$ (away from collisions of
zeros, $\mu_j$ are of course $C^\infty$).  

In analogy to \eqref{1.2.22a}, \eqref{1.2.22b}, one defines the 
meromorphic function $\phi (\dott,n,t_r)$ on $\calK_p$,
\begin{align}
\phi (P, n, t_r) & =
\frac{y-G_{p+1} (z, n,t_r)}{2a (n,t_r) F_p (z, n,t_r)} \lb{3.6.7} \\
& = \frac{-2a (n,t_r) F_p (z, n+1, t_r)} {y+G_{p+1} (z, n,t_r)},
\lb{3.6.7a} \\
& \hspace*{-.1cm} P(z,y)\in\calK_p, \; (n,t_r)\in\bbZ\times\bbR,  \no 
\end{align} 
with divisor $(\phi(\dott,n,t_r))$ of $\phi(\dott,n,t_r)$ given by
\begin{equation}
(\phi(\dott,n,t_r))=\calD_{\Pinfp\humu(n+1,t_r)}
-\calD_{\Pinfm\humu(n,t_r)}, \lb{3.6.7AB}
\end{equation}
using \eqref{410} and  \eqref{3.6.7AA}. 

The time-dependent Baker--Akhiezer function $\psi (P,n,n_0,t_r,t_{0,r})$, 
meromorphic on $\calK_p\setminus \{\Pinfp,\Pinfm\}$, is then defined in
terms of $\phi$ by
\begin{align}
&\psi (P,n,n_0, t_r,t_{0,r})\no\\
&\quad=\exp \left( \int_{t_{0,r}}^{t_r} \, ds \big(2a (n_0, s)
\wti{F}_r (z,n_0,s) \phi (P, n_0, s)
+\wti{G}_{r+1} (z,n_0, s)\big)\right) \no\\
&\qquad\times \begin{cases}
\prod_{m=n_0}^{n-1} \phi (P,m,t_r)& \text{for
$n \geq n_0+1$},\\ 1 & \text{for $n=n_0$},\\
\prod_{m=n}^{n_0-1} \phi (P,m,t_r)^{-1}& \text{for
$n\leq n_0 -1$}, \end{cases}  \lb{3.6.8} \\
& \hspace*{.5cm} P\in\calK_p\setminus 
\{P_{\infty_{\pm}}\}, \; (n,n_0,t_r,t_{0,r})\in\bbZ^2\times\bbR^2. 
\no 
\end{align}
For subsequent purposes we also introduce the following Baker--Akhiezer
vector,
\begin{align}
\begin{split}
&\Psi(P,n,n_0,t_r,t_{0,r})=\begin{pmatrix} \psi^-(P,n,n_0,t_r,t_{0,r}) \\ 
\psi(P,n,n_0,t_r,t_{0,r})\end{pmatrix}, \\
& \hspace*{.25cm} P\in\calK_p\setminus\{P_{\infty_{\pm}}\},
\;  (n,n_0,t_r,t_{0,r})\in\bbZ^2\times\bbR^2.
\end{split}
\end{align}

Basic properties of $\phi$, $\psi$, and $\Psi$ are summarized in the following
lemma.

%%%%%%%%%%%%%%%%%%%%%%%%%%%%
\begin{lemma} \lb{l3.3.2} 
Assume Hypothesis \ref{h3.4.1} and suppose that $a, b$ satisfy 
\eqref{5.15}--\eqref{5.18}. In addition, let
$P=(z,y)\in\calK_p\setminus\{P_{\infty_\pm}\}$, 
$(n,n_0,t_r,t_{0,r})\in\bbZ^2\times\bbR^2$, and $r\in\bbN_0$. Then
$\phi$ satisfies
\begin{align}
& a \phi(P) +a^- (\phi^- (P))^{-1} =z-b,  \lb{3.6.10} \\
&\phi_{t_r}(P)
=-2 a \big(\wti{F}_r (z) \phi (P)^2+\wti{F}^+_r (z)\big) 
 +2 (z-b^+) \wti{F}^+_r (z) \phi (P)   \no \\ 
& \hspace*{1.43cm} 
+ \big(\wti{G}^+_{r+1} (z) -\wti{G}_{r+1}(z)\big) \phi(P),  \lb{3.6.11} \\
&\phi(P) \phi(P^*) = \f{F^+_p (z)}{F_p (z)}, \lb{3.6.13a} \\
&\phi (P) -\phi(P^*)= \f{y(P)}{a F_p (z)}, \lb{3.6.13c} \\
&\phi(P) + \phi(P^*) = -\f{G_{p+1}(z)}{a F_p (z)}. \lb{3.6.13d}
\end{align}
Moreover, $\psi$ and $\Psi$ satisfy
\begin{align} 
& (L-z(P))\psi(P) =0, \;\; (P_{2p+2}-y(P))\psi(P) =0,  \lb{3.6.13A} \\
&\psi_{t_r} (P)=\wti P_{2r+2}\psi(P) \lb{3.6.12A} \\
&\hspace*{1.03cm} = 2a \wti{F}_r (z) \psi^+ (P) + \wti{G}_{r+1} (z) \psi(P),
\lb{3.6.13} \\ 
& \Psi^+(P) = U(z)\Psi(P), \quad y\Psi(P) = V_{p+1}\Psi(P), \lb{5.34} \\
& \Psi_{t_r}(P)= \wti V_{r+1}(z)\Psi(P),  \lb{5.35} \\
&\psi(P, n, n_0,t_r,t_{0,r}) \psi(P^*, n, n_0,t_r,t_{0,r})
= \f{F_p (z, n,t_r)}{F_p (z, n_0,t_{0,r})}, \lb{3.6.13b}  \\
& a(n,t_r)\big(\psi(P,n,n_0,t_r,t_{0,r})\psi(P^*,n+1,n_0,t_r,t_{0,r}) \no \\
& \quad +\psi(P^*,n,n_0,t_r,t_{0,r})\psi(P,n+1,n_0,t_r,t_{0,r})\big) 
=-\f{G_{\gg+1}(z,n,t_r)}{F_\gg(z,n_0,t_{0,r})},
\lb{5.37} \\
&W(\psi (P,\dott,n_0,t_r,t_{0,r}), \psi (P^*, \dott,
n_0,t_r,t_{0,r}))=-\f{y(P)}{F_{\gg}(z,n_0,t_{0,r})}. \lb{5.38a}
\end{align}
\end{lemma}
%%%%%%%%%%%%%%%%%%%%%%%%

In complete analogy to the case of stationary trace formulas one obtains
trace formulas in the time-dependent setting (cf.\ the abbreviation
\eqref{3.21}).

%%%%%%%%%%%%%%%%%%%%%%%%%%%%
\begin{lemma} \lb{l3.4.4}
Assume Hypothesis \ref{h3.4.1} and suppose that $a, b$ satisfy 
\eqref{5.15}--\eqref{5.16}. Then,
\begin{equation}
b=\f{1}{2}\sum_{m=0}^{2\gg+1} E_m - \sum_{j=1}^\gg \mu_j.
\lb{5.38} 
\end{equation}
In addition, if for all $(n,t_r)\in\bbZ\times\bbR$, $\mu_j(n,t_r)\neq
\mu_k(n,t_r)$ for $j\neq k$, $j,k=1,\dots,p$, then,
\begin{equation}
a^2=\f{1}{2}\sum_{j=1}^{\gg}
y(\hat{\mu}_j)\prod_{\substack{k=1\\ k\neq j}}^\gg
(\mu_j-\mu_k)^{-1}+\f{1}{4}\big(b^{(2)} - b^2\big). \lb{5.39} 
\end{equation}
\end{lemma}
%%%%%%%%%%%%%%%%%%%%%%%%%%%%%%%%%%%

For completeness we next mention the Dubrovin equations for the time
variation of the Dirichlet eigenvalues of the Toda lattice.

%%%%%%%%%%%%%%%%%%%%%%%%%%%%%%%%%%%%%%
\begin{lemma} \lb{l3.4.3} 
Assume Hypothesis \ref{h3.4.1} and suppose that $a, b$ satisfy 
\eqref{5.15}--\eqref{5.16}. In addition, assume that the
zeros $\mu_j (n,t_r)$, $j=1,\dots,p$, of $F_p(\dott,n,t_r)$ remain distinct
for all $(n,t_r)\in\bbZ\times\bbR$. Then,
\begin{align}
\begin{split}
&\frac{d}{dt_r} \mu_j (n,t_r) 
=-2\ti{F}_r (\mu_j (n,t_r), n,t_r) \frac{y(\hat\mu_j (n,t_r))}
{\prod^p_{\substack{ \ell =1 \\ \ell\neq j}}
(\mu_j(n,t_r) -\mu_\ell (n,t_r))}, \lb{3.6.16A} \\
& \hspace*{5.35cm} j=1,\dots,p, \; (n,t_r) \in\bbZ\times\bbR. 
\end{split}
\end{align}
\end{lemma}
%%%%%%%%%%%%%%%%%%%%%%%%%%%%%%%%%%%%%%

When attempting to solve the Dubrovin system \eqref{3.6.16A}, it must be 
augmented with appropriate divisors $\calD_{\humu(n_0,t_{0,r})}\in\sym^{p} \calK_p$ 
as initial conditions.

For the $t_r$-dependence of $F_p$ and $G_{p+1}$ one obtains the following result.

%%%%%%%%%%%%%%%%%%%%%%%%%%%%
\begin{lemma} \lb{l3.4.2}
Assume Hypothesis \ref{h3.4.1} and suppose that $a, b$ satisfy 
\eqref{5.15}--\eqref{5.16}. In addition, let
$(z,n,t_r)\in\bbC\times\bbZ\times\bbR$. Then,
\begin{align}
F_{p,t_r} &=2\big(F_p \wti{G}_{r+1} - G_{p+1} \wti{F}_r \big),
\lb{3.6.15}\\
G_{p+1,t_r} &=4a^2 \big(F_p \wti{F}^+_r
- F^+_p \wti{F}_r \big). \lb{3.6.16}
\end{align}
In particular, \eqref{3.6.15} and \eqref{3.6.16} are equivalent to
\begin{equation}
V_{p+1, t_r} = [\wti V_{r+1}, V_{p+1}].   \lb{5.41}
\end{equation}
\end{lemma}
%%%%% proof of lemma %%%%%%

It will be shown in Section \ref{s6} that Lemma \ref{l3.4.2} in
conjunction with the fundamental identity \eqref{412g} yields a
first-order system of differential equations for $f_\ell$, $g_\ell$,
$\ell=1,\dots,p$, that serves as a pertinent substitute for the Dubrovin
equations \eqref{3.6.16A} even (in fact, especially) when some of the $\mu_j$
coincide.

As in the case of trace formulas, also Lemma \ref{l1.3.9ba} on nonspecial
Dirichlet divisors $\calD_{\humu}$ and the linearization property of the Abel
map when applied to $\calD_{\humu}$ extends to the present time-dependent
setting. For the latter fact we need to introduce a particular differential
of the second kind, $\wti{\Ome}_r^{(2)}$, defined as follows. Let
$\ome_{P_{\infty_\pm}, q}^{(2)}$ be the normalized Abelian differential of
the second kind (i.e., with vanishing $a$-periods)
with a single pole at $P_{\infty_\pm}$ of the form
\begin{equation}
\ome_{P_{\infty_\pm},q}^{(2)} =\big(\zeta^{-2-q}
+\Oh(1)\big) \, d\zeta \text{ near }
P_{\infty_\pm}, \quad q \in\bbN_0.
\lb{3.6.23}
\end{equation}
Given the summation constants $\tilde{c}_1, \ldots, \tilde{c}_r$ in
$\wti{F}_r$ (cf.\ \eqref{412aa}), we then define
\begin{equation}
\wti{\Ome}_r^{(2)} =\sum_{q=0}^r (q+1) \tilde{c}_{r-q}
\big(\ome_{P_{\infty_+},q}^{(2)} - \ome_{P_{\infty_-},q}^{(2)}\big), \quad
\tilde{c}_0 =1.
\lb{3.6.24}
\end{equation}
Since the differentials $\ome_{P_{\infty_\pm},q}^{(2)}$ were supposed to be
normalized we have
\begin{equation}
\int_{a_j} \wti{\Ome}_r^{(2)} =0, \quad j=1,\dots,p. \lb{3.6.25}
\end{equation}
Moreover, writing
\begin{equation}
\ome_j =\bigg( \sum_{m=0}^\infty d_{j,m}
(P_{\infty_\pm}) \zeta^m \bigg) \,
d\zeta
=\pm \bigg( \sum_{m=0}^\infty d_{j,m} (P_{\infty_+})
\zeta^m \bigg) \, d
\zeta \text{ near } P_{\infty_\pm},
\lb{3.6.26}
\end{equation}
relation \eqref{a35} yields for the vector of $b$-periods
$\wti{\uU}^{(2)}_r$ of $\wti{\Ome}_r^{(2)}$,
\begin{align}
& \wti{\uU}^{(2)}_r=\big(\wti{U}_{r,1}^{(2)},\dots,\wti{U}_{r,p}^{(2)}\big),
\\ 
&\wti{U}_{r,j}^{(2)} = \frac{1}{2\pi i }\int_{b_j} \wti{\Ome}_r^{(2)}
= 2 \sum_{q=0}^r \tilde{c}_{r-q} d_{j,q}(P_{\infty_+}), \quad j=1,\dots,p.
\lb{3.6.27}
\end{align}

The time-dependent analog of Lemma \ref{l1.3.9ba} then reads as follows.

%%%%%%%%%%%%%%%%%%%%%%%%%%%%%%%%%%%%%%%%%%%%%%
\begin{lemma} \lb{l3.6.2}
Assume Hypothesis \ref{h3.4.1} and suppose that $a, b$ satisfy 
\eqref{5.15}--\eqref{5.16}. Let $\calD_{\humu}$,
$\humu=\{\hmu_1,\dots,\hmu_\gg\}\in\symgg$, be the Dirichlet divisor of
degree $\gg$ associated with $a, b$ defined according to \eqref{3.6.7AA}, that
is,
\begin{equation}
\hmu_j(n,t_r)=(\mu_j(n,t_r),-G_{\gg+1}(\mu_j(n,t_r),n,t_r))\in\calK_p, \quad
j=1,\dots,\gg, \; (n,t_r)\in\bbZ\times\bbR. \lb{5.46}
\end{equation}
Then $\calD_{\humu(n,t_r)}$ is nonspecial for all $(n,t_r)\in\bbZ\times\bbR$.
Moreover, the Abel map linearizes the auxiliary divisor $\calD_{\humu}$
in the sense that
\begin{equation}
\ual_{Q_0} (\calD_{\hat\umu(n,t_r)}) = 
\ual_{Q_0} (\calD_{\hat\umu (n_0,t_{0,r})})
-(n-n_0) \underline{A}_{P_{\infty_-}} (P_{\infty_+}) 
-(t_r-t_{0,r})\wti{\uU}^{(2)}_r, \lb{3.6.55}
\end{equation}
where $Q_0\in\calK_p$ is a given base point and $\wti{\uU}^{(2)}_r$ is the
vector of $b$-periods of the differential of the second kind
$\wti{\Ome}_r^{(2)}$ introduced in \eqref{3.6.27}.  \\
In addition, if $a, b \in L^\infty(\bbZ\times\bbR)$,  then there exists a
constant
$C_{\mu}>0$ such that
\begin{equation}
|\mu_j(n,t_r)|\leq C_{\mu}, \quad j=1,\dots,\gg, \; (n,t_r)\in\bbZ\times\bbR.
\lb{5.47}
\end{equation}
\end{lemma}
%%%%%%%%%%%%%%%%%%%%%%%%%%%%%%%%%%%%%%%%%%%%%%
\begin{proof}
We will prove that
\begin{align}
\psi (P,n,n_0, t_r,t_{0,r}) = & C(n,t_r)
\frac{\theta(\ul{z}(P,n, t_r))}{\theta(\ul{z}(P,n_0, t_{0,r}))}  \\
& \times \exp\left( (n-n_0) \int_{Q_0}^P \ome_{P_{\infty_+},P_{\infty_-}}^{(3)} +
(t_r-t_{0,r}) \int_{Q_0}^P \wti{\Ome}_r^{(2)} \right),  \no
\end{align}
where
\begin{equation}
\ul{z}(P,n,t_r)=  \underline{A}_{Q_0} (P) - \ual_{Q_0} (\calD_{\hat\umu(n,t_r)})
+\ul \Xi_{Q_0}. 
\end{equation}
By Lemma 13.4 of \cite{Te00} it suffices to show that the essential singularities
at $P_{\infty_\pm}$ are equal. That is, by \eqref{3.6.8} we need to show that
\begin{align}
&\psi (P,n_0,n_0, t_r,t_{0,r})\no\\
&\quad=\exp \left( \int_{t_{0,r}}^{t_r} \, ds \big(2a (n_0, s)
\wti{F}_r (z,n_0,t_r) \phi (P, n_0, t_{0,r})
+\wti{G}_{r+1} (z,n_0, s)\big)\right) \no\\
&\quad= \exp \left( \pm(t_r-t_{0,r}) \sum_{k=0}^r c_{r-k} \zeta^{-k-1} + O(1)\right) 
\, \text{ for $P$ near $P_{\infty_\pm}$.} 
\end{align}
Using \eqref{3.6.7} and \eqref{3.6.15} one obtains
\begin{equation}
\psi (P,n_0,n_0, t_r,t_{0,r})
= \left(\frac{F_p (z,n_0,t_r)}{F_p (z,n_0,t_{0,r})} \right)^{1/2} \exp \left( y \int_{t_{0,r}}^{t_r} ds \, 
\frac{\wti{F}_r (z,n_0,s)}{F_p (z,n_0,s)} \right) 
\end{equation}
and the desired asymptotics follow from Theorem \ref{tE.8}, which tells us that 
\begin{equation}
\frac{y}{F_p (z,n_0,s)} \hatt{F}_k (z,n_0,s) = \pm\zeta^{-k-1} + O(1) 
\, \text{ for $P$ near $P_{\infty_\pm}$,} 
\end{equation}
together with \eqref{1.2.11a}.
\end{proof}
%%%%%%%%%%%%%%%%%%%%%%%%%%%%%%%%%%%%%%%%%%%%%%

Again the analog of Remark \ref{r3.5} applies in the present
time-dependent context.

%%%%%%%%%%%%%%%%%%%%%%%%%%%%%%%%%%%%%%%%%%%%%%%
%%%%%%%%%%%%%%%%%%%%%%%%%%%%%%%%%%%%%%%%%%%%%%%
\section{The Algebro-Geometric Toda Hierarchy Initial Value 
Problem} \lb{s6}
%%%%%%%%%%%%%%%%%%%%%%%%%%%%%%%%%%%%%%%%%%%%%%%
%%%%%%%%%%%%%%%%%%%%%%%%%%%%%%%%%%%%%%%%%%%%%%%

In this section we consider the algebro-geometric Toda hierarchy initial value
problem \eqref{43}, \eqref{42} with complex-valued initial data. For a
generic set of initial data we will prove unique solvability of the
initial value problem globally in time. 

While it is naturally in the special self-adjoint case to base the
solution of the algebro-geometric initial value problem on the Dubrovin
equations \eqref{3.6.16A} (and the trace formula \eqref{5.38} for $b$ and
formula \eqref{5.39} for $a^2$), this strategy meets with difficulties in
the non-self-adjoint case as Dirichlet eigenvalues $\hat \mu_j$ may now
collide on $\calK_p$ and hence the denominator of \eqref{3.6.16A} can
blow up. Hence, we will develop an alternative strategy based on the use
of elementary symmetric functions of the variables
$\{\mu_j\}_{j=1,\dots,p}$ in this section, which can accommodate
collisions of $\hat\mu_j$. In short, our strategy will be as follows:

$(i)$ Replace the first-order autonomous Dubrovin system \eqref{3.6.16A}
of differential equations in $t_r$ for the Dirichlet eigenvalues
$\mu_j(n,t_r)$, $j=1,\dots,p$, augmented by appropriate initial
conditions, by the first-order autonomous system 
\eqref{6.32}, \eqref{6.33} for the coefficients $f_j$, $j=1,\dots,p$,
$g_j$, $j=1,\dots,p-1$, and $g_p+f_{p+1}$ with respect to $t_r$. (We
note that $f_j$, $j=1,\dots,p$, are symmetric functions of
$\mu_1,\dots,\mu_p$.) Solve this first-order autonomous system 
in some time interval $(t_{0,r}-T_0,t_{0,r}+T_0)$ under
appropriate initial conditions at $(n_0,t_{0,r})$ derived from an
initial (nonspecial) Dirichlet divisor
$\calD_{\humu(n_0,t_{0,r})}$.

$(ii)$ Use the stationary algorithm derived in Section \ref{s4} to
extend the solution of step $(i)$ from
$\{n_0\}\times(t_{0,r}-T_0,t_{0,r}+T_0)$ to 
$\bbZ \times (t_{0,r}-T_0,t_{0,r}+T_0)$ (cf.\ Lemma \ref{l6.2}).

$(iii)$ Prove consistency of this approach, that is, show that the
discrete algorithm of Section \ref{s4} is compatible with the
time-dependent Lax and zero-curvature equations in the sense that first
solving the autonomous system \eqref{6.32}, \eqref{6.33} and then
applying the discrete algorithm, or first applying the discrete
algorithm and then solving the autonomous system \eqref{6.32},
\eqref{6.33} yields the same result whenever the same endpoint 
$(n,t_r)$ is reached (cf.\ the discussion following Lemma \ref{l6.2} and
the subsequent Lemma \ref{l6.3} and Theorem \ref{t6.4}). 

$(iv)$ Prove that there is a dense set of initial conditions of full
measure for which this strategy yields global solutions of the
algebro-geometric Toda hierarchy initial value problem (cf.\ Lemma
\ref{l6.5} and Theorem \ref{t6.6}). 

\smallskip

To set up this formalism we need some preparations. From the
outset we make the following assumption.  

%%%%%%%%%%%%%%%%%%%%%%%%%%%%%%%%%%%%%%%%% 
\begin{hypothesis} \lb{h6.1} 
Suppose that 
\begin{equation}
a, b \in \bbC^{\bbZ} \, \text{ and } \, a(n)\neq 0 \, \text{ for all }
\,  n\in\bbZ,   \lb{6.0}
\end{equation}
and assume that $a, b$ satisfy the $p$th stationary Toda system 
\eqref{3.2.29d}. In addition, suppose that the hyperelliptic curve $\calK_p$ 
in \eqref{1.2.20} is nonsingular. 
\end{hypothesis} 
%%%%%%%%%%%%%%%%%%%%%%%%%%%%%%%%%%%%%%%%%

Assuming Hypothesis \ref{h6.1}, we consider the polynomials $F_p$,
$G_{p+1}$, $\wti F_r$, and
$\wti G_{r+1}$ given by \eqref{410}--\eqref{412ab} for fixed  $p, r \in
\bbN_0$. Here $f_{\ell}$, $\tilde f_s$, $g_{\ell}$, and $\tilde g_s$,
$\ell=0,\dots,p$, $s=0,\dots,r$, are defined as in
\eqref{1.2.4a}--\eqref{1.2.4c} with appropriate sets of summation
constants. 

Our aim will be to find an autonomous first-order system of
ordinary differential equations with respect to $t_r$ of $f_{\ell}$ and
$g_{\ell}$ rather than $\mu_j$. Indeed, we will take the coupled system
of differential equations \eqref{3.6.15}, \eqref{3.6.16}, properly
rewritten next, as our point of departure. In order to turn
\eqref{3.6.15}, \eqref{3.6.16} into a system of first-order ordinary
differential equations for $f_\ell$ and $g_\ell$, we first need to
eliminate $f_\ell^+$, $\ti f_s$, $\ti g_s$, and $\ti f_s^+$ in terms of
$f_\ell$ and $g_\ell$ as follows. 

Using \eqref{1.2.7}, \eqref{1.2.11a}, \eqref{1.2.11c}, and Theorem
\ref{tE.8} one infers 
\begin{align}
\wti F_r (z) &= \sum_{s=0}^r \ti f_{r-s} z^s = \sum_{s=0}^r \ti c_{r-s}
\hatt F_s (z),   \lb{6.5} \\ 
\hatt F_\ell (z) &=\sum_{k=0}^\ell \hat f_{\ell-k} z^k, \quad 
\hat f_0=1, \quad 
\hat f_\ell = \sum_{k=0}^{\ell \wedge p} \hat c_{\ell-k} (\ul E) f_k,  
\quad \ell\in\bbN_0,  \lb{6.6}
\end{align}
where $m\mini n=\min\{m,n\}$ and $\hat c_\ell(\ul E)$ has been introduced
in \eqref{T2.26f}. Hence one obtains
\begin{equation}
\ti f_0=1, \quad \ti f_s = \calF_{1,s} (f_1,\dots,f_p), \quad
s=1,\dots,r,    \lb{6.7}
\end{equation}
where $\calF_{1,s}$, $s=1,\dots,r$, are polynomials in $p$ variables.

Next, using \eqref{1.2.7}, \eqref{1.2.11b}, \eqref{1.2.11d}, and Theorem
\ref{tE.8} one concludes
\begin{align}
\wti G_{r+1}(z) &=-z^{r+1}
+ \sum_{s=0}^r \ti g_{r-s} z^s + \ti f_{r+1}=\sum_{s=1}^{r+1}
\ti c_{r+1-s}\hatt{G}_s (z),  \no \\
\hatt{G}_0(z)&=G_0(z)\big|_{c_1=0}=0, \quad \hatt{G}_1(z)=G_1(z)=-z-b, \no
\\
\hatt{G}_{\ell+1}(z)&=G_{\ell+1}(z)\big|_{c_k=0,\,k=1,\dots,\ell}
=-z^{\ell+1}+\sum_{k=0}^\ell \hat g_{\ell-k} z^k + \hat f_{\ell+1}, \quad
\ell\in\bbN,   \lb{6.8}  \\
\hat g_0 &= 0, \quad 
\hat g_\ell =\sum_{k=0}^{\ell\wedge p} \hat
c_{\ell-k}(\ul E) (g_k  + f_{p+1}\delta_{p,k}) - \hat c_{\ell+1}(\ul E),
\quad \ell\in\bbN. \lb{6.9}
\end{align}
Hence one concludes
\begin{equation}
\ti g_0 =-\ti c_1, \quad 
\ti g_s = \calF_{2,s} (f_1,\dots,f_p,g_1,\dots,g_{p-1},(g_p+f_{p+1})),
\quad s=1,\dots,r,  
\lb{6.10}
\end{equation}
where $\calF_{2,s}$, $s=1,\dots,r$, are polynomials in $2p$ variables. We
also recall (cf.\ \eqref{1.2.8b}) that $f_{p+1}$ is a lattice constant,
that is,
\begin{equation}
f_{\gg+1}=f_{\gg+1}^{-}. \lb{6.12}
\end{equation}

Next we invoke the fundamental identity \eqref{1.2.17} in the form
\begin{equation}
-4a^2 F_p^+ =\f{R_{2p+2}-G_{p+1}^2}{F_p}.   \lb{6.13}
\end{equation}
While \eqref{6.13} at this point only holds in the stationary context, we
will use it later on also in the $t_r$-dependent context and
verify after the time-dependent solutions of \eqref{43}, \eqref{42} have
been obtained that \eqref{6.13} indeed is valid for all
$(n,t_r)\in\bbZ\times\bbR$. A comparison of powers of $z$ in \eqref{6.13}
then yields
\begin{align}
\begin{split}
4 a^2 f_0^+ &= -2g_1 -2c_2,  \\
4 a^2 f_\ell^+ &= \calF_{3,\ell}
(f_1,\dots,f_p,g_1,\dots,g_{p-1},(g_p+f_{p+1})), \quad \ell=1,\dots,p,  
\lb{6.14}
\end{split}
\end{align}
where $\calF_{3,\ell}$, $\ell=1,\dots,p$, are polynomials in $2p$
variables. 

Finally, combining \eqref{6.5}, \eqref{6.6}, \eqref{6.13}, and
\eqref{6.14}, one obtains 
\begin{align}
\begin{split}
4 a^2 \ti f_0^+ &= -2g_1 -2c_2,  \\
4a^2 \ti f^+_s &= \calF_{4,s}
(f_1,\dots,f_p,g_1,\dots,g_{p-1},(g_p+f_{p+1})),
\quad s=1,\dots,r,    \lb{6.15}
\end{split}
\end{align}
where $\calF_{4,s}$, $s=1,\dots,3$, are polynomials in $2p$
variables. 

We emphasize that also the Dubrovin equations \eqref{3.6.16A} require an
analogous rewriting of $\wti F_r$ in terms of (symmetric functions of)
$\mu_j$ in order to represent a first-order system of differential
equations for
$\mu_j$, $j=1,\dots,p$.

Next, we make the transition to the algebro-geometric initial
value problem \eqref{43}, \eqref{42}. We introduce a deformation (time)
parameter $t_r\in\bbR$ in $a=a(t_r)$ and $b=b(t_r)$, and hence obtain
$t_r$-dependent quantities $f_\ell=f_\ell(t_r)$, $g_\ell=g_\ell(t_r)$,
$F_p(z)=F_p(z,t_r)$, $G_{p+1}(z)=G_{p+1}(z,t_r)$, etc. At a fixed
initial time $t_{0,r}\in\bbR$ we require that
\begin{equation}
(a,b)|_{t_r=t_{0,r}} = \big(a^{(0)},b^{(0)}\big),   \lb{6.16}
\end{equation}
where $a^{(0)}=a(\dott,t_{0,r}),b^{(0)}=b(\dott,t_{0,r})$ satisfy the
$p$th stationary Toda  equation \eqref{3.2.29d} as in
\eqref{6.0}--\eqref{6.15}. As discussed in Section \ref{s4}, in order to
guarantee that the stationary solutions \eqref{6.16} can be constructed
for all $n\in\bbZ$ one starts  from a particular divisor 
\begin{equation}
\calD_{\humu(n_0,t_{0,r})}\in\calM_0   \lb{6.17}
\end{equation} 
where $\humu(n_0,t_{0,r})$ is of the form
\begin{align}
&\humu(n_0,t_{0,r})=\{\hat\mu_1(n_0,t_{0,r},\dots,\hat\mu_p(n_0,t_{0,r})\}
\lb{6.18}  \\ 
& \quad =\{\underbrace{\hat\mu_1(n_0,t_{0,r}),\dots,
\hat\mu_1(n_0,t_{0,r})}_{p_1(n_0,t_{0,r}) \text{ times}},
\dots,\underbrace{\hat\mu_{q(n_0,t_{0,r})}(n_0,t_{0,r}),\dots,
\hat\mu_{q(n_0,t_{0,r})}(n_0,t_{0,r})}_{p_{q(n_0,t_{0,r})}(n_0,t_{0,r})
\text{ times}}\}   
\no
\end{align}
with 
\begin{align}
\begin{split}
& \hat\mu_k(n_0,t_{0,r})=(\mu_k(n_0,t_{0,r}),y(\hat\mu_k(n_0,t_{0,r}))),
\\ 
& \mu_k(n_0,t_{0,r})\neq \mu_{k'}(n_0,t_{0,r}) \, \text{ for } \, k\neq
k', \;  k,k'=1,\dots,q(n_0,t_{0,r}),    \lb{6.19}
\end{split}
\end{align}
and
\begin{equation}
p_k(n_0,t_{0,r})\in\bbN, \; k=1,\dots,q(n_0,t_{0,r}), \quad 
\sum_{k=1}^{q(n_0,t_{0,r})} p_k(n_0,t_{0,r}) = p.   \lb{6.20}
\end{equation}
Next we recall
\begin{align} 
F_p(z,n_0,t_{0,r})&=\sum_{{\ell}=0}^{p}f_{p-{\ell}}(n_0,t_{0,r})z^{\ell}
=\prod_{k=1}^{q(n_0,t_{0,r})}
(z-\mu_k(n_0,t_{0,r}))^{p_k(n_0,t_{0,r})},  \lb{6.21} \\
T_{p-1}(z,n_0,t_{0,r})&=-F_p(z,n_0,t_{0,r})  \no \\
& \hspace*{-1cm} \times \sum_{k=1}^{q(n_0,t_{0,r})}
\sum_{\ell=0}^{p_k(n_0,t_{0,r})-1} 
\f{\big(d^\ell
\big(R_{2p+2}(\zeta)^{1/2}\big)/d\zeta^\ell\big)\big|_{\zeta
=\mu_k(n_0,t_{0,r})}}{\ell!(p_k(n_0,t_{0,r})-\ell-1)!} \no \\ 
& \hspace*{-1cm} \times 
\Bigg(\f{d^{p_k(n_0,t_{0,r})-\ell-1}}{d
\zeta^{p_k(n_0,t_{0,r})-\ell-1}}\Bigg( (z-\zeta)^{-1}  \lb{6.22} \\
& \hspace*{-.3cm} \times \prod_{k'=1, \,
k'\neq k}^{q(n_0,t_{0,r})} 
(\zeta-\mu_{k'}(n_0,t_{0,r}))^{-p_{k'}(n_0,t_{0,r})} 
\Bigg)\Bigg)\Bigg|_{\zeta=\mu_k(n_0,t_{0,r})},  \no  \\
b(n_0,t_{0,r})&=\f{1}{2}\sum_{m=0}^{2p+1}E_m -\sum_{k=1}^{q(n_0,t_{0,r})}
p_k(n_0,t_{0,r}) \mu_k(n_0,t_{0,r}),  \lb{6.23} \\ 
G_{p+1}(z,n_0,t_{0,r})&= -z^{p+1} + 
\sum_{{\ell}=0}^{p}g_{p-{\ell}}(n_0,t_{0,r}) z^{\ell}+f_{p+1}(t_{0,r}),
 \no \\
&=-(z-b(n_0,t_{0,r}))
F_p(z,n_0,t_{0,r})+T_{p-1}(z,n_0,t_{0,r}).   \lb{6.24}
\end{align}
Here the sign of the square root in \eqref{6.22} is chosen as usual by
\begin{align}
\hat\mu_k(n_0,t_{0,r})&=(\mu_k(n_0,t_{0,r}),y(\hat\mu_k(n_0,t_{0,r}))) 
\no \\
&=\big(\mu_k(n_0,t_{0,r}), R_{2p+2}(\mu_k(n_0,t_{0,r}))^{1/2}\big) 
 \lb{6.25} \\
&=(\mu_k(n_0,t_{0,r}), -G_{p+1}(\mu_k(n_0,t_{0,r}),n_0,t_{0,r})), 
\quad   k=1,\dots,q(n_0,t_{0,r}).  \no 
\end{align}
By \eqref{6.21} one concludes that \eqref{6.18} uniquely determines
$F_p(z,n_0,t_{0,r})$ and hence
\begin{equation}
f_{1}(n_0,t_{0,r}),\dots,f_{p}(n_0,t_{0,r}).   \lb{6.27}
\end{equation}
By \eqref{6.22}--\eqref{6.27} one concludes that also
$G_{p+1}(z,n_0,t_{0,r})$ and hence 
\begin{equation}
g_{1}(n_0,t_{0,r}),\dots,g_{p-1}(n_0,t_{0,r}),
g_p(n_0,t_{0,r})+f_{p+1}(t_{0,r})   \lb{6.28}
\end{equation}
are uniquely determined by the initial divisor
$\calD_{\humu(n_0,t_{0,r})}$ in \eqref{6.17}.

Summing up the discussion in \eqref{6.5}--\eqref{6.28}, we can transform
the differential equations 
\begin{align}
F_{p,t_r}(z,n_0,t_r) &=2\big(F_p(z,n_0,t_r) \wti{G}_{r+1}(z,n_0,t_r) \no\\
& \hspace*{.75cm} - G_{p+1}(z,n_0,t_r) \wti{F}_r(z,n_0,t_r) \big), 
\lb{6.29}\\ G_{p+1,t_r}(z,n_0,t_r) &=4a(n_0,t_r)^2 \big(F_p(z,n_0,t_r)
\wti{F}^+_r(z,n_0,t_r) \no \\
& \hspace*{2.2cm} - F^+_p(z,n_0,t_r) \wti{F}_r(z,n_0,t_r) \big)
\lb{6.30}
\end{align}
subject to the constraint
\begin{equation}
-4a^2 F_p^+(z,n_0,t_r)
=\f{R_{2p+2}(z)-G_{p+1}(z,n_0,t_r)^2}{F_p(z,n_0,t_r)}, 
\lb{6.31}
\end{equation}
and associated with an initial divisor $\calD_{\humu(n_0,t_{0,r})}$ in
\eqref{6.17} into the following autonomous first-order system of ordinary
differential equations (for fixed $n=n_0$),
\begin{align}
&f_{j,t_r}=\calF_j(f_1,\dots,f_p,g_1,\dots,g_{p-1},g_p+f_{p+1}), \quad 
j=1,\dots,p, \no \\
&g_{j,t_r}=\calG_j(f_1,\dots,f_p,g_1,\dots,g_{p-1},g_p+f_{p+1}), \quad 
j=1,\dots,p-1, \lb{6.32} \\
&(g_p+f_{p+1})_{t_r}=\calG_p
(f_1,\dots,f_p,g_1,\dots,g_{p-1},g_p+f_{p+1})  \no
\end{align}
with initial condition
\begin{align}
&f_j(n_0,t_{0,r}), \quad j=1,\dots,p, \no \\ 
&g_j(n_0,t_{0,r}),\quad j=1,\dots,p-1,  \lb{6.33} \\
&g_p(n_0,t_{0,r})+f_{p+1}(t_{0,r}),   \no 
\end{align}
where $\calF_j$, $\calG_j$, $j=1,\dots,p$, are polynomials in $2p$
variables. As just discussed, the initial conditions \eqref{6.33} are uniquely determined by the initial divisor $\calD_{\humu(n_0,t_{0,r})}$ in \eqref{6.17}.

Being autonomous with polynomial right-hand sides, there
exists a $T_0>0$, such that the first-order initial value problem
\eqref{6.32}, \eqref{6.33} has a unique solution 
\begin{align}
\begin{split}
&f_j=f_j(n_0,t_r), \quad j=1,\dots,p,  \\
&g_j=g_j(n_0,t_r),  \quad j=1,\dots,p-1, \lb{6.34} \\
&g_p+f_{p+1}=g_p(n_0,t_{r})+f_{p+1}(t_{r})   \\
& \text{for all } \, t_r\in (t_{0,r}-T_0,t_{0,r}+T_0)   
\end{split}
\end{align} 
(cf., e.g., \cite[Sect.\ III.10]{Wa98}). Given the solution \eqref{6.34},
we next introduce the following quantities (where $t_r\in
(t_{0,r}-T_0,t_{0,r}+T_0)$):  
\begin{align} 
F_p(z,n_0,t_{r})&=\sum_{{\ell}=0}^{p}f_{p-{\ell}}(n_0,t_{r})z^{\ell}
=\prod_{k=1}^{q(n_0,t_{r})}
(z-\mu_k(n_0,t_{r}))^{p_k(n_0,t_{r})},  \lb{6.35} \\
T_{p-1}(z,n_0,t_{r})&=-F_p(z,n_0,t_{r})  \no \\
& \hspace*{-1cm} \times \sum_{k=1}^{q(n_0,t_{r})}
\sum_{\ell=0}^{p_k(n_0,t_{r})-1} 
\f{\big(d^\ell
\big(R_{2p+2}(\zeta)^{1/2}\big)/d\zeta^\ell\big)\big|_{\zeta
=\mu_k(n_0,t_{r})}}{\ell!(p_k(n_0,t_{r})-\ell-1)!} \no \\ 
& \hspace*{-1cm} \times 
\Bigg(\f{d^{p_k(n_0,t_{r})-\ell-1}}{d
\zeta^{p_k(n_0,t_{r})-\ell-1}}\Bigg( (z-\zeta)^{-1}  \lb{6.36} \\
& \hspace*{-.3cm} \times \prod_{k'=1, \,
k'\neq k}^{q(n_0,t_{r})} 
(\zeta-\mu_{k'}(n_0,t_{r}))^{-p_{k'}(n_0,t_{r})} 
\Bigg)\Bigg)\Bigg|_{\zeta=\mu_k(n_0,t_{r})},  \no  \\
b(n_0,t_{r})&=\f{1}{2}\sum_{m=0}^{2p+1}E_m -\sum_{k=1}^{q(n_0,t_{r})}
p_k(n_0,t_{r}) \mu_k(n_0,t_{r}),  \lb{6.37} \\ 
G_{p+1}(z,n_0,t_{r})&= -z^{p+1} + 
\sum_{{\ell}=0}^{p}g_{p-{\ell}}(n_0,t_{r}) z^{\ell}+f_{p+1}(t_{r})
 \no \\
&=-(z-b(n_0,t_{r}))
F_p(z,n_0,t_{r})+T_{p-1}(z,n_0,t_{r}).   \lb{6.38}
\end{align}
In particular, this leads to the divisor 
\begin{equation}
\calD_{\humu(n_0,t_{r})}\in\symgg   \lb{6.38a}
\end{equation} 
and the sign of the square root in \eqref{6.36} is chosen as usual by
\begin{align}
\begin{split}
\hat\mu_k(n_0,t_{r})&=(\mu_k(n_0,t_{r}),
-G_{p+1}(\mu_k(n_0,t_{r}),n_0,t_r))  \\ 
&=\big(\mu_k(n_0,t_{r}), R_{2p+2}(\mu_k(n_0,t_{r}))^{1/2}\big),
\quad   k=1,\dots,q(n_0,t_{r})   \lb{6.39}
\end{split}
\end{align}
and
\begin{align}
\begin{split}
&\humu(n_0,t_r)=\{\mu_1(n_0,t_{r},\dots,\mu_p(n_0,t_{r})\}  \\ 
& \quad =\{\underbrace{\mu_1(n_0,t_{r}),\dots,
\mu_1(n_0,t_{r})}_{p_1(n_0,t_{r}) \text{ times}},
\dots,\underbrace{\mu_{q(n_0,t_{r})}(n_0,t_{r}),\dots,
\mu_{q(n_0,t_{r})}(n_0,t_{r})}_{p_{q(n_0,t_{r})}(n_0,t_{r}) 
\text{ times}}\}  \lb{6.40}
\end{split}
\end{align}
with 
\begin{equation}
\mu_k(n_0,t_{r})\neq \mu_{k'}(n_0,t_{r}) \, \text{ for } \, k\neq
k', \;  k,k'=1,\dots,q(n_0,t_{r})    \lb{6.41}
\end{equation}
and
\begin{equation}
p_k(n_0,t_{r})\in\bbN, \; k=1,\dots,q(n_0,t_{r}), \quad 
\sum_{k=1}^{q(n_0,t_{r})} p_k(n_0,t_{r}) = p.   \lb{6.42}
\end{equation}
By construction (cf.\ \eqref{6.39}), the divisor
$\calD_{\humu(n_0,t_{r})}$ is nonspecial for all $t_r\in
(t_{0,r}-T_0,t_{0,r}+T_0)$.

In exactly the same manner as in \eqref{4.17}--\eqref{4.19} one then
infers that $F_p(\dott,n_0,t_r)$ divides $R_{2p+2}-G_{p+1}^2$ (since
$t_r$ is just a fixed additional parameter). Moreover, arguing as in
\eqref{4.20}--\eqref{4.20f} we now assume that the polynomial 
\begin{equation}
R_{2p+2}(z)-G_{p+1}(z,n_0,t_r)^2 \underset{z\to\infty}{=}\Oh(z^{2p}) 
\lb{6.43}
\end{equation}
is precisely of maximal order $2p$ for all $t_r\in
(t_{0,r}-T_0,t_{0,r}+T_0)$. One then obtains
\begin{align}
\begin{split}
R_{2p+2}(z)-G_{p+1}(z,n_0,t_r)^2=-4a(n_0,t_r)^2
F_p(z,n_0,t_r)F_p(z,n_0+1,t_r), \\ 
(z,t_r)\in \bbC\times (t_{0,r}-T_0,t_{0,r}+T_0), \lb{6.44}
\end{split}
\end{align}
where we introduced the coefficient $a(n_0,t_r)^2$ to make
$F_p(\dott,n_0+1,t_r)$ a monic polynomial of degree $p$. As in Section
\ref{s4}, the assumption that the polynomial $F_p(\dott,n_0+1,t_r)$ is
precisely of order $p$ is implied by the hypothesis that
\begin{equation}
\calD_{\humu(n_0,t_{r})}\in\calM_0 \, \text{ for all } \, t_r\in
(t_{0,r}-T_0,t_{0,r}+T_0),   \lb{6.45}
\end{equation}  
a point we will revisit later (cf.\ Lemma \ref{l6.5}). Given \eqref{6.44},
we obtain consistency with \eqref{6.13} for $n=n_0$ and $t_r\in
(t_{0,r}-T_0,t_{0,r}+T_0)$. 

The explicit formula for $a(n_0,t_r)^2$ then reads (for $t_r\in
(t_{0,r}-T_0,t_{0,r}+T_0)$)
\begin{align}
a(n_0,t_r)^2&=\f{1}{2}\sum_{k=1}^{q(n_0,t_r)}
\f{\big(d^\ell \big(R_{2p+2}(z)^{1/2}\big)/d z^\ell\big)\big|_{z
=\mu_k(n_0,t_r)}}{(p_k(n_0,t_r)-1)!}  \no \\
& \qquad \times \prod_{k'=1, \, k'\neq k}^{q(n_0,t_r)}
(\mu_k(n_0,t_r)-\mu_{k'}(n_0,t_r))^{-p_k(n_0,t_r)}   \lb{6.45A} \\
& \quad +\f{1}{4}\big(b^{(2)}(n_0,t_r)-b(n_0,t_r)^2 \big).  \no
\end{align}
Here and in the following we use the abbreviation
\begin{equation}
b^{(2)}(n,t_r)=\f{1}{2}\sum_{m=0}^{2p+1}E_m^2 
-\sum_{k=1}^{q(n,t_r)} p_k(n,t_r) \mu_k(n,t_r)^2   \lb{6.45B}
\end{equation}
for appropriate ranges of $(n,t_r)\in \bbN\times\bbR$.

With \eqref{6.35}--\eqref{6.45B} in place, we can now apply the stationary
formalism as summarized in Theorem \ref{t4.3}, subject to the
additional hypothesis \eqref{6.45}, for each fixed $t_r\in
(t_{0,r}-T_0,t_{0,r}+T_0)$. This yields, in particular, the quantities
\begin{equation} 
F_p, \, G_{p+1}, \, a, \, b, \, \text{ and } \, \humu \, \text{ for } \, 
(n,t_r) \in \bbZ\times(t_{0,r}-T_0,t_{0,r}+T_0),  \lb{6.45a}
\end{equation}
which are of the form \eqref{6.35}--\eqref{6.45B}, replacing the fixed
$n_0\in\bbZ$ by an arbitrary $n\in\bbZ$. In addition, one has the
following fundamental identities (cf.\ \eqref{4.32}, \eqref{4.36},
\eqref{4.39}, and \eqref{4.40}), which we summarize in the following
result.

%%%%%%%%%%%%%%%%%%%%%%%%%%%%%%%%%%%%%%%%%%%%%%
\begin{lemma}  \lb{l6.2}
Assume Hypothesis \ref{h6.1} and condition \eqref{6.45}. Then the
following relations are valid:
\begin{align}
R_{2p+2}-G_{p+1}^2 + 4a^2 F_pF_p^+ & =0,  \lb{6.46} \\ 
2(z-b^+)F_p^+ + G_{p+1}^+ + G_{p+1} &=0,  \lb{6.47} \\
2 a^2 F_p^+ - 2(a^-)^2 F_p^-  +(z-b)(G_{p+1}-G_{p+1}^-) &=0, \lb{6.48} \\
2(z-b^+)F_p^+ -2(z-b)F_p +G_{p+1}^+ - G_{p+1}^- &=0 \lb{6.49} \\
& \hspace*{-4.25cm} \text{on } \, \bbC\times \bbZ \times
(t_{0,r}-T_0,t_{0,r}+T_0)       \no
\end{align}
and hence the stationary part, \eqref{48}, of the algebro-geometric
initial value problem holds
\begin{equation}
UV_{p+1} - V^+_{p+1}U = 0 \, \text{ on } \, \bbC\times \bbZ \times
(t_{0,r}-T_0,t_{0,r}+T_0).   \lb{6.50}   
\end{equation}
In particular, Lemmas \ref{l1.3.1}--\ref{l1.3.9ba} apply.
\end{lemma}
%%%%%%%%%%%%%%%%%%%%%%%%%%%%%%%%%%%%%%%%%%%%%%

Lemma \ref{l6.2} now raises the following important consistency issue:
On one hand, one can solve the initial value problem \eqref{6.32},
\eqref{6.33} at $n=n_0$ in some interval $t_r \in
(t_{0,r}-T_0,t_{0,r}+T_0)$, and then extend the quantities
$F_p, G_{p+1}$ to all $\bbC\times\bbZ\times(t_{0,r}-T_0,t_{0,r}+T_0)$
using the stationary algorithm summarized in Theorem \ref{t4.3} as just 
recorded in Lemma \ref{l6.2}. On the other hand, one can solve the
initial value problem \eqref{6.32}, \eqref{6.33} at $n=n_1$, $n_1\neq
n_0$, in some interval $t_r \in (t_{0,r}-T_1,t_{0,r}+T_1)$ with the
initial condition obtained by applying the discrete algorithm to the
quantities $F_p, G_{p+1}$ starting at $(n_0,t_{0,r})$ and ending at
$(n_1,t_{0,r})$. Consistency then requires that the two approaches yield
the same result at $n=n_1$ for $t_r$ in some open neighborhood of
$t_{0,r}$. 

Equivalently, and pictorially speaking, envisage a vertical $t_r$-axis and
a horizontal $n$-axis. Then, consistency demands that first solving the
initial value problem \eqref{6.32}, \eqref{6.33} at $n=n_0$ in some
$t_r$-interval around $t_{0,r}$ and using the stationary algorithm to
extend $F_p, G_{p+1}$ horizontally to $n=n_1$ and the same $t_r$-interval
around $t_{0,r}$, or first applying the stationary algorithm starting at
$(n_0,t_{0,r})$ to extend $F_p, G_{p+1}$ horizontally to $(n_1,t_{0,r})$
and then solving  the initial value problem \eqref{6.32}, \eqref{6.33} at
$n=n_1$ in some $t_r$-interval around $t_{0,r}$ should produce the same
result at $n=n_1$ in a sufficiently small open $t_r$ interval around
$t_{0,r}$.

To settle this consistency issue, we will prove the following result. To this end 
we find it convenient to replace the
initial value problem \eqref{6.32}, \eqref{6.33} by the original
$t_r$-dependent zero-curvature equation \eqref{47},
$U_{t_r}+U\wti V_{r+1}- \wti V^+_{r+1}U=0$ on
$\bbC\times\bbZ\times(t_{0,r}-T_0,t_{0,r}+T_0)$.  

%%%%%%%%%%%%%%%%%%%%%%%%%%%%%%%%%%%%%%%%% 
\begin{lemma}  \lb{l6.3}
Assume Hypothesis \ref{h6.1} and condition \eqref{6.45}. Moreover,
suppose that \eqref{6.29}--\eqref{6.31} hold on $\bbC\times\{n_0\}\times
(t_{0,r}-T_0,t_{0,r}+T_0)$. Then \eqref{6.29}--\eqref{6.31} hold on $\bbC\times\bbZ\times (t_{0,r}-T_0,t_{0,r}+T_0)$, that is,
\begin{align}
& F_{p,t_r}(z,n,t_r) =2\big(F_p(z,n,t_r) \wti{G}_{r+1}(z,n,t_r)
\no \\ 
& \hspace*{2.8cm}   
-G_{p+1}(z,n,t_r) \wti{F}_r(z,n,t_r) \big),  \lb{6.51} \\ 
& G_{p+1,t_r}(z,n,t_r) = 4a(n,t_r)^2 \big(F_p(z,n,t_r)
\wti{F}^+_r(z,n,t_r)  \no \\ 
& \hspace*{4.5cm} - F^+_p(z,n,t_r) \wti{F}_r(z,n,t_r) \big), 
\lb{6.52} \\  
& R_{2p+2}(z) = G_{p+1}(z,n,t_r)^2 -4a(n,t_r)^2 F_p(z,n,t_r)
F_p^+(z,n,t_r),  \lb{6.53}  \\ 
& \hspace*{3.3cm} 
(z,n,t_r)\in \bbC\times\bbZ\times (t_{0,r}-T_0,t_{0,r}+T_0). \no
\end{align}
Moreover,
\begin{align}
&\phi_{t_r}(P,n,t_r)
 =-2 a(n,t_r) \big(\wti{F}_r (z,n,t_r) \phi(P,n,t_r)^2
+\wti{F}^+_r(z,n,t_r)\big) \no \\
& \hspace*{2.2cm} +2 (z-b^+(n,t_r)) \wti{F}^+_r(z,n,t_r)
\phi(P,n,t_r)  \lb{6.54} \\  
& \hspace*{2.2cm} 
+ \big(\wti{G}^+_{r+1}(z,n,t_r) -\wti{G}_{r+1}(z,n,t_r)\big)
\phi(P,n,t_r),  \no \\
&a_{t_r}(n,t_r)=-a(n,t_r)\big(2(z-b^+(n,t_r)) \wti F_r^+(z,n,t_r) 
\no \\
& \hspace*{3.35cm}   
+ \wti G_{r+1}^+(z,n,t_r) + \wti G_{r+1}(z,n,t_r)\big), 
\lb{6.55} \\
& b_{t_r}(n,t_r) =2 \big((z-b(n,t_r))^2 \wti F_r(z,n,t_r) +
(z-b(n,t_r)) \wti G_{r+1}(z,n,t_r)  \no \\
& \hspace*{1.7cm} + a^2(n,t_r) \wti F_r^+(z,n,t_r) -
(a^-(n,t_r))^2 \wti F_r^-(z,n,t_r)\big), 
\lb{6.56}  \\ 
& \hspace*{3.25cm} 
(z,n,t_r)\in \bbC\times\bbZ\times (t_{0,r}-T_0,t_{0,r}+T_0). \no
\end{align}
\end{lemma} 
%%%%%%%%%%%%%%%%%%%%%%%%%%%%%%%%%%%%%%%%%
\begin{proof} 
By Lemma \ref{l6.2} we have \eqref{3.6.7}, \eqref{3.6.7a}, 
\eqref{3.6.10}, \eqref{3.6.13a}--\eqref{3.6.13d}, and \eqref{6.46}--\eqref{6.49} for
$(n,t_r)\in\bbZ\times(t_{0,r}-T_0,t_{0,r}+T_0)$ at our disposal. 
Differentiating \eqref{6.53} at $n=n_0$ with respect to $t_r$, inserting \eqref{6.51}
and \eqref{6.52} at $n=n_0$, yields 
\begin{align}
\begin{split}
&2F_p^+ a_{t_r} + a F^+_{p,t_r}=2a\big(G_{p+1}\wti F_r^+- F_p^+ \wti
G_{r+1}\big)    \\
& \quad =2F^+_p a \big(-2(z-b^+)\wti F_r^+ -\wti G_{r+1}^+ - \wti
G_{r+1}\big)  + 2 a\big(F_p^+ \wti G^+_{r+1}-G^+_{p+1} \wti F_r^+\big)    \lb{6.57} 
\end{split}
\end{align}
at $n=n_0$. By inspection,
\begin{equation}
F_p^+(z) \wti G^+_{r+1}(z)-G^+_{p+1}(z) \wti F_r^+(z)
\underset{|z|\to\infty}{=} 
\Oh(z^{p-1}).   \lb{6.58}
\end{equation}
This can be shown directly using formulas such as 
\eqref{1.2.11a}--\eqref{1.2.11d}, \eqref{6.5}, \eqref{6.6}, \eqref{6.8},
and \eqref{6.9}. It also follows from \eqref{3.6.15} and the fact that
$F_p$ is a monic polynomial of degree $p$. Thus one concludes that
\begin{equation}
2F_p^+ a_{t_r}=2F^+_p a \big(-2(z-b^+)\wti F_r^+ -\wti G_{r+1}^+ - \wti
G_{r+1}\big)  
\end{equation} 
at $n=n_0$, and upon cancelling $2F_p^+$ that \eqref{6.55} holds at $n=n_0$. This and \eqref{6.57} then also proves that \eqref{6.51} holds at $n=n_0+1$.

Next, differentiating $2aF_p\phi=y-G_{p+1}$ at $n=n_0$ with respect to $t_r$
inserting \eqref{6.51}, \eqref{6.52}, and \eqref{6.55} at $n=n_0$, and using 
\eqref{3.6.7a} to replace $2aF_p^+$ by $-(y+G_{p+1})\phi$ and
\eqref{3.6.7} to replace $(G_{p+1}-y)$ by $-2aF_p\phi$, yields
\eqref{6.54} at $n=n_0$ upon cancelling the factor $2aF_p$.

Differentiating \eqref{6.47} with respect to $t_r$ (fixing
$n=n_0$), inserting \eqref{6.47} (to replace $G^+_{p+1}$), \eqref{6.52} at $n=n_0$, 
and \eqref{6.51} at $n=n_0+1$ yields
\begin{align}
0 & =  -2 F^+_p\big(b^+_{t_r}-2(z-b^+)^2 \wti F^+_r
 + 2 a^2 \wti F_r -2(z-b^+) \wti G^+_{r+1} \big) \no \\
& \quad +4 (z-b^+)^2 F^+_p \wti F^+_r + 4(z-b^+) G_{p+1} \wti F^+_r
+ 4 (a)^2 F_p \wti F^+_r +G^+_{p+1,t_r}  \no \\
& = - 2 F^+_p\big(b^+_{t_r} - 2 (z-b^+)^2 \wti F^+_r - 2 (z-b^+) \wti
G^+_{r+1} +2 a^2 \wti F_r - 2 (a^+)^2 \wti F^{++}_r\big) \no \\
& \quad -4 (a^+)^2 F^+_p \wti F_r^{++} +4 (z-b^+)^2 F^+_p \wti F^+_r 
+ 4 (z-b^+) G_{p+1} \wti F^+_r \no \\
& \quad + 4a^2 F_p \wti F^+_r + G^+_{p+1, t_r} \no \\
& = - 2 F^+_p\big(b^+_{t_r} - 2 (z-b^+)^2 \wti F^+_r - 2 (z-b^+) \wti
G^+_{r+1} +2 a^2 \wti F_r - 2 (a^+)^2 \wti F^{++}_r\big) \lb{6.59}   \\
& \quad + G^+_{p+1, t_r} - 4 (a^+)^2 F^+_p \wti F^{++}_r +\big(4 a^2
F_p +4(z-b^+)^2 F^+_p + 4(z-b^+) G_{p+1}\big) 
\wti F^+_r  \no
\end{align}
at $n=n_0$. Combining \eqref{6.47} and \eqref{6.48} at $n=n_0$ one computes 
\begin{equation}
4 (a^+)^2 F^{++}_p = 4a^2 F_p + 4 (z-b^+)^2 F^+_p + 4(z-b^+)G_{p+1}  
\lb{6.59a} 
\end{equation}
at $n=n_0$. Insertion of \eqref{6.59a} into \eqref{6.59} then yields 
\begin{align}
0 & = - 2 F^+_p\big(b^+_{t_r} - 2 (z-b^+)^2 \wti F^+_r - 2 (z-b^+) \wti
G^+_{r+1} +2 a^2 \wti F_r - 2 (a^+)^2 \wti F^{++}_r\big) \no \\
& \quad + G^+_{p+1, t_r} - 4 (a^+)^2 F^+_p \wti F^{++}_r + 
4 (a^+)^2 F_p^{++} \wti F^+_r  \lb{6.59b}
\end{align}
at $n=n_0$. In close analogy to \eqref{6.58} one observes that
\begin{equation}
F_p^{+}(z) \wti F^{++}_{r}(z)-F^{++}_{p}(z) \wti F_r^+(z)
\underset{|z|\to\infty}{=} 
\Oh(z^{p-1}) \, \text{ for } \, p\in\bbN.  \lb{6.59c}
\end{equation}
Thus, since $F^+_p$ is a monic polynomial of degree $p$, \eqref{6.59b}
proves that
\begin{equation}
b^+_{t_r} - 2 (z-b^+)^2 \wti F^+_r - 2 (z-b^+) \wti
G^+_{r+1} +2 a^2 \wti F_r - 2 (a^+)^2 \wti F^{++}_r=0  \lb{6.59d}
\end{equation}
at $n=n_0$, upon cancelling $F^+_p$. Thus, \eqref{6.56} holds at $n=n_0+1$. 
Simultaneously, this proves \eqref{6.52} at $n=n_0+1$.

Iterating the arguments just presented (and performing the analogous considerations for $n<n_0$) then extends these results to all lattice points $n\in\bbZ$ and hence proves \eqref{6.51}--\eqref{6.56} for  
$(z,n,t_r)\in \bbC\times\bbZ\times (t_{0,r}-T_0,t_{0,r}+T_0)$. 
\end{proof} 
%%%%%%%%%%%%%%%%%%%%%%%%%%%%%%%%%%%%%%%%%%

We summarize Lemmas \ref{l6.2} and \ref{l6.3} next.

%%%%%%%%%%%%%%%%%%%%%%%%%%%%%%%%%%%%%%%% 
\begin{theorem}  \lb{t6.4}
Assume Hypothesis \ref{h6.1} and condition \eqref{6.45}. Moreover,
suppose that 
\begin{align}
\begin{split}
&f_j=f_j(n_0,t_r), \quad j=1,\dots,p,  \\
&g_j=g_j(n_0,t_r),  \quad j=1,\dots,p-1, \lb{6.60} \\
&g_p+f_{p+1}=g_p(n_0,t_{r})+f_{p+1}(t_{r})  \\
& \hspace*{2mm} \text{for all } \, t_r\in (t_{0,r}-T_0,t_{0,r}+T_0),  
\end{split}
\end{align} 
satisfies the autonomous first-order system of ordinary differential
equations \eqref{6.32} $($for fixed $n=n_0$$)$,
\begin{align}
&f_{j,t_r}=\calF_j(f_1,\dots,f_p,g_1,\dots,g_{p-1},g_p+f_{p+1}), \quad 
j=1,\dots,p, \no \\
&g_{j,t_r}=\calG_j(f_1,\dots,f_p,g_1,\dots,g_{p-1},g_p+f_{p+1}), \quad 
j=1,\dots,p-1, \lb{6.61} \\
&(g_p+f_{p+1})_{t_r}=\calG_p
(f_1,\dots,f_p,g_1,\dots,g_{p-1},g_p+f_{p+1})  \no
\end{align}
with initial condition
\begin{align}
&f_j(n_0,t_{0,r}), \quad j=1,\dots,p, \no \\ 
&g_j(n_0,t_{0,r}),\quad j=1,\dots,p-1,  \lb{6.62} \\
&g_p(n_0,t_{0,r})+f_{p+1}(t_{0,r}).   \no 
\end{align}
Then $F_p$ and $G_{p+1}$ as constructed in \eqref{6.5}--\eqref{6.45a} on 
$\bbC\times\bbZ\times(t_{0,r}-T_0,t_{0,r}+T_0)$ satisfy the
zero-curvature equations \eqref{47}, \eqref{48}, and \eqref{5.41},
\begin{align}
U_{t_{r}}+U\wti{V}_{r+1}-\wti V_{r+1}^+U&=0, \label{6.63} \\
UV_{p+1}-V^+_{p+1}U&=0, \label{6.64} \\
V_{p+1, t_r} - [\wti V_{r+1}, V_{p+1}]&= 0 \label{6.64a} \\
& \hspace*{-4.25cm} \text{on } \, \bbC\times \bbZ \times
(t_{0,r}-T_0,t_{0,r}+T_0),      \no
\end{align}
with $U$, $V_{p+1}$, and $\wti V_{r+1}$ given by \eqref{49}. In
particular, $a, b$ satisfy the algebro-geometric initial value problem
\eqref{43}, \eqref{42}
\begin{align}
\begin{split}
&\wti\tl_r(a,b)=\begin{pmatrix}a_{t_r}-a\big(\tilde{f}^+_{p+1}(a,b)-
\tilde{f}_{p+1}(a,b)\big) \\ 
 b_{t_r}+\tilde{g}_{p+1}(a,b)-\tilde{g}^-_{p+1}(a,b) \end{pmatrix}=0, 
\label{6.65} \\  
&(a,b)\big|_{t_r=t_{0,r}}=\big(a^{(0)},b^{(0)}\big),  
\end{split} \\
&\sTl_p\big(a^{(0)},b^{(0)}\big)=
\begin{pmatrix}-a\big({f^+}_{p+1}\big(p^{(0)},q^{(0)}\big)
-{f}_{p+1}\big(p^{(0)},q^{(0)}\big)\big) \\
 {g}_{p+1}\big(a^{(0)},b^{(0)}\big)
-g^-_{p+1}\big(a^{(0)},b^{(0)}\big) \end{pmatrix}=0 
\label{6.66}  \\
& \hspace*{5.85cm} \text{on } \, \bbZ \times
(t_{0,r}-T_0,t_{0,r}+T_0),       \no
\end{align}
and are given by
\begin{align}
a(n,t_r)^2&=\f{1}{2}\sum_{k=1}^{q(n,t_r)}
\f{\big(d^\ell \big(R_{2p+2}(z)^{1/2}\big)/d z^\ell\big)\big|_{z
=\mu_k(n,t_r)}}{(p_k(n,t_r)-1)!}  \no \\
& \qquad \times \prod_{k'=1, \, k'\neq k}^{q(n,t_r)}s
(\mu_k(n,t_r)-\mu_{k'}(n,t_r))^{-p_k(n,t_r)}   \lb{6.75} \\
& \quad +\f{1}{4}\big(b^{(2)}(n,t_r)-b(n,t_r)^2 \big),  \no \\
b(n,t_{r})&=\f{1}{2}\sum_{m=0}^{2p+1}E_m -\sum_{k=1}^{q(n,t_{r})}
p_k(n,t_{r}) \mu_k(n,t_{r}),  \lb{6.76} \\
& \hspace*{.6cm} (z,n,t_r)\in \bbZ \times
(t_{0,r}-T_0,t_{0,r}+T_0).       \no
\end{align}
Moreover, Lemmas \ref{l1.3.1}--\ref{l1.3.9ba} and
\ref{l3.3.2}--\ref{l3.4.2} apply. 
\end{theorem}
%%%%%%%%%%%%%%%%%%%%%%%%%%%%%%%%%%%%%%%%%%%%

As in the stationary case, the theta function representations of $a$ and $b$ in the time-dependent context can be derived in complete analogy to the self-adjoint case. Since the final results are formally the same as in the self-adjoint case we again just refer, for instance, to \cite{BGHT98}, \cite{DT76}, \cite{DKN90}, \cite{DMN76}, 
\cite[Sect.\ 1.4]{GH07},  \cite{Kr78} (cf.\ also the appendix written in \cite{Du81}), 
\cite{vMM79}, \cite[Appendix, Sect.\ 9]{NMPZ84}, \cite[Sect.\ 13.2]{Te00}, 
\cite[Sect.\ 4.6]{To89}, \cite[Ch.\ 28]{To89a}.  

As in Lemma \ref{l4.1} we now show that also in the time-dependent case, 
most initial divisors are nice in the sense that the corresponding divisor
trajectory stays away from infinity for all $(n,t_{r})\in\bbZ\times\bbR$.

%%%%%%%%%%%%%%%%%%%%%%%%%%%%%%%%%%%%%%%%%%%%
\begin{lemma}  \lb{l6.5}
The set $\calM_1$ of initial divisors $\calD_{\humu(n_0,t_{0,r})}$
for which $\calD_{\humu(n,t_r)}$, defined via \eqref{3.6.55}, is nonspecial
and finite for all $(n,t_r)\in\bbZ\times\bbR$, forms a dense set of full
measure in the set $\symgg$ of nonnegative divisors of degree $p$.  
%Jacobi variety $J(\calK_\gg)$.  
\end{lemma}
%%%%%%%%%%%%%%%%%%%%%%%%%%%%%%%%%%%%%%%%%%%%
\begin{proof}
Let $\calM_{\rm sing}$ be as introduced in the proof of Lemma \ref{l4.1}.
Then
\begin{align}
& \bigcup_{t_r\in\bbR} \Big(\underline{\alpha}_{P_0}(\calM_{\rm sing}) 
+ t_r \wti{\uU}^{(2)}_r\Big) \no \\
& \quad = \bigcup_{t_r\in\bbR} \Big(\underline{A}_{P_0}(P_{\infty_+}) + 
\underline{\alpha}_{P_0}(\sym^{p-1} (\calK_\gg)) 
+ t_r \wti{\uU}^{(2)}_r \Big)
\end{align}
is of  measure zero\ as well, since it is the image of 
$\bbR\times \sym^{p-1} (\calK_\gg)$ which misses one real dimension in
comparison to the $2p$ real dimensions of $J(\calK_\gg)$.  But then
\begin{equation}
\bigcup_{(n,t_r)\in\bbZ\times\bbR}
\Big(\underline{\alpha}_{P_0}(\calM)  +
n\underline{A}_{P_{\infty_-}}(P_{\infty_+}) 
+ t_r \wti{\uU}^{(2)}_r \Big)   \lb{6.78}                 
\end{equation}
is also of measure zero. Applying $\underline{\alpha}_{P_0}^{-1}$ to the
complement of the set in \eqref{6.78} then yields a set $\calM_1$ of
full measure in  $\symgg$. In particular, $\calM_1$ is
necessarily dense in $\symgg$.
\end{proof}
%%%%%%%%%%%%%%%%%%%%%%%%%%%%%%%%%%%%%%%%%%

%%%%%%%%%%%%%%%%%%%%%%%%%%%%%%%%%%%%%%%%%% 
\begin{theorem}  \lb{t6.6}
Let $\calD_{\humu(n_0,t_{0,r})}\in\calM_1$ be an initial divisor as in 
Lemma \ref{l6.5}. Then the sequences $a, b$ constructed from
$\humu(n_0,t_{0,r})$ as described in Theorem \ref{t6.4} satisfy
Hypothesis \ref{h3.4.1}. In particular, the solution $a, b$ of the
algebro-geometric initial value problem \eqref{6.65}, \eqref{6.66} is
global in $(n,t_r)\in\bbZ\times\bbR$.
\end{theorem}
%%%%%%%%%%%%%%%%%%%%%%%%%%%%%%%%%%%%%%%%%%
\begin{proof}
Starting with $\calD_{\humu(n_0,t_{0,r})}\in\calM_1$, the procedure
outlined in this section and summarized in Theorem \ref{t6.4} leads to 
$\calD_{\humu(n,t_{r})}$ for all $(n,t_r)\in \bbZ \times 
(t_{0,r}-T_0,t_{0,r}+T_0)$ such that \eqref{3.6.55}holds. But if
$a, b$ should blow up, then $\calD_{\humu(n,t_{r})}$ must hit $P_{\infty_+}$
which is impossible by our choice of initial condition.
\end{proof}
%%%%%%%%%%%%%%%%%%%%%%%%%%%%%%%%%%%%%%%%%%

Note, however, that in general (i.e., unless one is, e.g., in the special
periodic or self-adjoint case), $\calD_{\humu(n,t_{r})}$ will get
arbitrarily close to $P_{\infty_+}$ since straight motions on the torus
are generically dense (see e.g. \cite[Sect.\ 51]{Ar89} or \cite[Sects.\ 1.4, 1.5]{KH95}) and hence no uniform bound on the sequences $a(n,t_r), b(n,t_r)$ exists as $(n,t_r)$ vary in $\bbZ\times\bbR$. In particular, these complex-valued algebro-geometric solutions of the Toda hierarchy initial value problem, in general, will not be quasi-periodic (cf.\ the usual definition of quasi-periodic functions, e.g., in \cite[p.\ 31]{PF92}) with respect to $n$ or $t_r$.

%%%%%%%%%%%%%%%%%%%%%%%%%%%%%%%%%%%%%%%%%%%%
%%%%%%%%%%% appendices %%%%%%%%%%%%%%%%%%%%%%%%%%
%%%%%%%%%%%%%%%%%%%%%%%%%%%%%%%%%%%%%%%%%%%%
\appendix
\section{Hyperelliptic Curves of the Toda-Type} \lb{sA}
\renewcommand{\theequation}{A.\arabic{equation}}
\renewcommand{\thetheorem}{A.\arabic{theorem}}
\setcounter{theorem}{0} 
\setcounter{equation}{0}
%%%%%%%%%%%%%%%%%%%%%%%%%%%%%%%%%%%%%%%%%%%%
%%%%%%%%%%%%%%%%%%%%%%%%%%%%%%%%%%%%%%%%%%%%

We provide a brief summary of some of the fundamental notations
needed {}from the theory of hyperelliptic Riemann surfaces. More
details can be found in some of the standard textbooks \cite{FK92}
and \cite{Mu84} as well as in monographs and surveys dedicated to 
integrable systems such as \cite[Ch.\ 2]{BBEIM94}, \cite{Du81}, 
\cite[App.\ A, B]{GH03}, \cite[App.\ A]{Te00}.

Fix $\gg \in \bbN$. We intend to describe the hyperelliptic
Riemann surface $\calK_\gg$ of genus $\gg$ of the Toda-type curve
\eqref{1.2.19}, associated with the polynomial
\begin{align}
\begin{split}
&\calF_\gg(z,y)=y^2-R_{2\gg+2}(z)=0, \lb{a1} \\
&R_{2\gg+2}(z)=\prod_{m=0}^{2\gg+1}(z-E_m), \quad
\{E_m\}_{m=0}^{2\gg+1}\subset\bbC.
\end{split}
\end{align}
To simplify the discussion we will assume that the affine part of
$\calK_\gg$ is nonsingular, that is, we assume that
\begin{equation}
E_m \neq E_{m'} \text{ for } m\neq m', \; m,m'=0,\dots,2\gg+1
\lb{a2}
\end{equation}
throughout this appendix. Next we introduce an appropriate set of
(nonintersecting) cuts $\calC_j$ joining $E_{m(j)}$ and
$E_{m^\prime(j)}$, $j=1,\dots,\gg+1$, and denote
\begin{equation}
\calC=\bigcup_{j=1}^{\gg+1} \calC_j, \quad
\calC_j\cap\calC_k=\emptyset, \quad j\neq k.\lb{a3}
\end{equation}
Define the cut plane
\begin{equation}
\Pi=\bbC\setminus\calC, \lb{a4}
\end{equation}
and introduce the holomorphic function
\begin{equation}
R_{2\gg+2}(\dott)^{1/2}\colon \Pi\to\bbC, \quad z\mapsto
\bigg(\prod_{m=0}^{2\gg+1}(z-E_m) \bigg)^{1/2}\lb{a5}
\end{equation}
on $\Pi$ with an appropriate choice of the square root branch in
\eqref{a5}. Next we define the set
\begin{equation}
\calM_{\gg}=\{(z,\sigma R_{2\gg+2}(z)^{1/2}) \mid z\in\bbC,\;
\sigma\in\{1,-1\} \}\cup \{P_{\infty_+},P_{\infty_-}\} \label{a6}
\end{equation}
by extending $R_{2\gg+2}(\dott)^{1/2}$ to $\calC$. The
hyperelliptic curve $\calK_\gg$ is then the set $\calM_{\gg}$ with
its natural complex structure obtained upon gluing the two sheets
of $\calM_{\gg}$ crosswise along the cuts. Moreover, we introduce
the set of branch points
\begin{equation}
\calB(\calK_\gg)=\{(E_m,0)\}_{m=0}^{2\gg+1}. \lb{a7}
\end{equation}
Points $P\in\calK_\gg\setminus\{P_{\infty_{\pm}}\}$ are denoted
by
\begin{equation}
P=(z,\sigma R_{2\gg+2}(z)^{1/2})=(z,y),  \lb{a8}
\end{equation}
where $y(\cdot)$ denotes the meromorphic function on $\calK_\gg$
satisfying $\calF_\gg(z,y)=y^2-R_{2\gg+2}(z)=0$ and
\begin{equation}
y(P)\underset{\zeta\to
0}{=}\mp\bigg(1-\f12\bigg(\sum_{m=0}^{2\gg+1}E_m\bigg)\zeta
+\Oh(\zeta^2)\bigg)\zeta^{-\gg-1} \text{  as $P\to
P_{\infty_\pm}$,} \; \zeta=1/z.     \lb{a55g}
\end{equation}

In addition, we introduce the holomorphic sheet exchange map
(involution)
\begin{equation}
*\colon\calK_\gg\to\calK_\gg, \quad P=(z,y)\mapsto P^*=(z,-y), \;
P_{\infty_\pm}\mapsto P^*_{\infty_\pm}=P_{\infty_\mp}  \lb{a9}
\end{equation}
and the two meromorphic projection maps
\begin{equation}
\tilde\pi\colon\calK_\gg\to\bbC\cup\{\infty\}, \quad
P=(z,y)\mapsto z, \; P_{\infty_\pm}\mapsto \infty \lb{a10}
\end{equation}
and
\begin{equation}
y\colon\calK_\gg\to\bbC\cup\{\infty\}, \quad P=(z,y)\mapsto y, \;
P_{\infty_\pm}\mapsto \infty.  \lb{a11}
\end{equation}
Thus the map $\tilde\pi$ has a pole of order 1 at $P_{\infty_\pm}$
and $y$ has a pole of order $\gg+1$ at $P_{\infty_\pm}$. Moreover,
\begin{equation}
\tilde\pi(P^*)=\tilde\pi(P), \quad y(P^*)=-y(P), \quad
P\in\calK_\gg. \lb{a12}
\end{equation}
As a result, $\calK_\gg$ is a two-sheeted branched covering of the
Riemann sphere $\bbC\bbP^1$ ($\cong\bbC\cup\{\infty\}$) branched
at the $2\gg+4$ points $\{(E_m,0)\}_{m=0}^{2\gg+1},
P_{\infty_\pm}$. $\calK_\gg$ is compact since $\tilde\pi$ is open
and $\bbC\bbP^1$ is compact. Therefore, the compact hyperelliptic
Riemann surface resulting in this manner has topological genus
$\gg$.

Next we introduce the upper and lower sheets $\Pi_{\pm}$ by
\begin{equation}
\Pi_{\pm}=\{(z,\pm  R_{2\gg+2}(z)^{1/2})\in \calM_\gg \mid
z\in\Pi\} \lb{a13}
\end{equation}
and the associated charts
\begin{equation}
\zeta_\pm \colon \Pi_\pm\to \Pi, \quad P\mapsto z.\lb{a14}
\end{equation}

Let $\{a_j,b_j\}_{j=1}^\gg$ be a homology basis for $\calK_\gg$
with intersection matrix of the cycles satisfying
\begin{equation}
a_j\circ b_k=\delta_{j,k}, \quad a_j\circ a_k=0, \quad b_j\circ
b_k=0, \quad j,k=1,\dots,\gg. \lb{a15}
\end{equation}
Associated with the homology basis $\{a_j,b_j\}_{j=1}^\gg$ we also
recall the canonical dissection of $\calK_\gg$ along its cycles
yielding the simply connected interior $\hatt \calK_ \gg$ of the
fundamental polygon $\partial {\hatt \calK}_\gg$ given by
\begin{equation}
\partial  {\hatt \calK}_\gg =a_1 b_1 a_1^{-1} b_1^{-1}
a_2 b_2 a_2^{-1} b_2^{-1} \cdots a_\gg^{-1} b_\gg^{-1}. \lb{a16}
\end{equation}

Let $\calM (\calK_n)$ and $\calM^1 (\calK_n)$ denote the set of meromorphic
functions (0-forms) and meromorphic differentials (1-forms) on $\calK_n$. 
The residue of a meromorphic differential
$\nu\in \calM^1 (\calK_n)$ at a
point $Q \in \calK_n$ is defined by
\begin{equation}
\text{res}_{Q}(\nu)
=\frac{1}{2\pi i} \int_{\gamma_{Q}} \nu,
\lb{a33}
\end{equation}
where $\gamma_{Q}$ is a counterclockwise oriented
smooth simple closed
contour encircling $Q$ but no other pole of
$\nu$.  Holomorphic
differentials are also called Abelian differentials
of the first kind. Abelian differentials of the
second kind $\omega^{(2)} \in \calM^1 (\calK_n)$ are characterized
by the property that all their residues vanish.  They will 
usually be normalized by demanding that all their $a$-periods
vanish, that is,
\begin{equation}
\int_{a_j} \omega^{(2)} =0, \quad  j=1,\dots,n.
\lb{a34}
\end{equation}
If $\omega_{P_1, n}^{(2)}$ is a differential of the second kind on $\calK_n$
whose only pole is $P_1 \in \hatt \calK_n$ with principal part
$\zeta^{-n-2}\,d\zeta$, $n\in\bbN_0$ near
$P_1$ and $\omega_j =
 \big(\sum_{m=0}^\infty d_{j,m} (P_1) \zeta^m\big)\, d\zeta$
near $P_1$, then
\begin{equation}
\frac{1}{2\pi i} \int_{b_j} \omega_{P_1, m}^{(2)} =
 \frac{d_{j,m} (P_1)}{m+1}, \quad m=0,1,\dots
\lb{a35}
\end{equation}

Using local charts one infers that $d z/y$ is a holomorphic
differential on $\calK_\gg$ with zeros of order $\gg-1$ at
$P_{\infty_\pm}$ and hence
\begin{equation}
\eta_j=\frac{z^{j-1}d z}{y}, \quad j=1,\dots,\gg, \lb{a22}
\end{equation}
form a basis for the space of holomorphic differentials on
$\calK_\gg$. Introducing the invertible matrix $C$ in $\bbC^\gg$
\begin{align}
C & =\big(C_{j,k}\big)_{j,k=1,\dots,\gg}, \quad C_{j,k}
= \int_{a_k} \eta_j, \lb{a23} \\
\underline{c} (k) & = (c_1(k), \dots, c_\gg(k)), \quad c_j (k)
=\big(C^{-1}\big)_{j,k}, \quad j,k=1,\dots,\gg, \lb{a24}
\end{align}
the normalized differentials $\ome_j$ for $j=1,\dots,\gg$,
\begin{equation}
\ome_j = \sum_{\ell=1}^\gg c_j (\ell) \eta_\ell, \quad \int_{a_k}
\ome_j = \delta_{j,k}, \quad j,k=1,\dots,\gg, \lb{a25}
\end{equation}
form a canonical basis for the space of holomorphic differentials
on $\calK_\gg$.

In the chart $(U_{P_{\infty_\pm}}, \zeta_{P_{\infty_\pm}})$
induced by $1/\tilde\pi$ near $P_{\infty_\pm}$ one infers,
\begin{align}
{\ul \omega} & = (\omega_1,\dots,\omega_\gg)=
     \mp \sum_{j=1}^\gg \f{\uc (j)
\zeta^{\gg-j}d\zeta}{\big(\prod_{m=0}^{2\gg+1}
(1-\zeta E_m) \big)^{1/2}} \lb{a26} \\
& = \pm \bigg( \uc (\gg) +\zeta\bigg( \frac12 \uc (\gg)
\sum_{m=0}^{2\gg+1} E_m +\uc (\gg-1) \bigg)  + \Oh(\zeta^2) \bigg)
d\zeta \text{ as $P\to P_{\infty_\pm}$,}
\no \\
& \hspace*{9.35cm} \zeta=1/z. \no
\end{align}

The matrix $\tau=\big(\tau_{j,\ell}\big)_{j,\ell=1}^\gg$  in
$\bbC^{\gg\times\gg}$ of $b$-periods defined by
\begin{equation}
\tau_{j,\ell}=\int_{b_j}\omega_\ell, \quad j,\ell=1, \dots,\gg,  
\label{a28}
\end{equation}
satisfies
\begin{equation}
\Im(\tau)>0 \, \text{ and } \, \tau_{j,\ell}=\tau_{\ell,j},
\;\, j,\ell =1,\dots,\gg.  \lb{a29}
\end{equation}

Associated with the matrix $\tau$ one introduces the period
lattice
\begin{equation}
L_\gg = \{ \ul z \in\bbC^\gg \mid \ul z = \ul m + \ul n\tau, \;
\ul m, \ul n \in\bbZ^\gg\}   \lb{a30}
\end{equation}
and the Riemann theta function associated with $\calK_\N$ and
the given homology basis $\{a_j,b_j\}_{j=1,\dots,\N}$,
\begin{equation}
\theta(\ul z)=\sum_{\ul n\in\bbZ^\N}\exp\big(2\pi
i(\ul n,\ul z)+\pi i(\ul n, \ul n\tau)\big),
\quad \ul z\in\bbC^\N, \label{b9}
\end{equation}
where $(\ul u, \ul v)=\ol {\ul u}\, \ul v^\top=\sum_{j=1}^\N \ol{u_j}\, v_j$
denotes the scalar product in $\bbC^\N$. It has the fundamental properties
\begin{align}
& \theta(z_1, \ldots, z_{j-1}, -z_j, z_{j+1},
\ldots, z_\N) =\theta
(\ul z), \lb{a27}\\
& \theta (\ul z +\ul m + \ul n\tau)
=\exp \big(-2 \pi i (\ul n,\ul z) -\pi i (\ul n, \ul n\tau) \big) \theta
(\ul z), \quad \ul m, \ul n \in\bbZ^\N.
\lb{aa51}
\end{align}

Next, fixing a base point $Q_0\in\calK_\gg\setminus\{P_{\infty_\pm}\}$,
one denotes by $J(\calK_\gg) = \bbC^\gg/L_\gg$ the Jacobi variety
of $\calK_\gg$, and defines the Abel map $\underline{A}_{Q_0}$ by
\begin{equation}
\underline{A}_{Q_0} \colon \calK_n \to J(\calK_\gg), \quad
\underline{A}_{Q_0}(P)= \bigg(\int_{Q_0}^P
\omega_1,\dots,\int_{Q_0}^P \omega_\gg \bigg) \pmod{L_\gg}, \quad
P\in\calK_\gg. \label{a42}
\end{equation}

Similarly, one introduces
\begin{equation}
\ul \alpha_{Q_0}  \colon \Div(\calK_\gg) \to J(\calK_\gg),\quad
\calD \mapsto \ul \alpha_{Q_0} (\calD) =\sum_{P \in \calK_\gg}
\calD (P) \ul A_{Q_0} (P), \label{a43}
\end{equation}
where $\Div(\calK_\gg)$ denotes the set of divisors on
$\calK_\gg$. Here a map $\calD \colon \calK_\gg \to \bbZ$ is
called a divisor on $\calK_\gg$ if $\calD(P)\neq0$ for only
finitely many $P\in\calK_\gg$. (In the main body of this paper we
will choose $Q_0$ to be one of the branch points, i.e.,
$Q_0\in\calB(\calK_\gg)$, and for simplicity we will always choose
the same path of integration from $Q_0$ to $P$ in all Abelian
integrals.) 

In connection with divisors on $\calK_\gg$ we will employ the
following (additive) notation,
\begin{align}
&\calD_{Q_0\ul Q}=\calD_{Q_0}+\calD_{\ul Q}, \quad \calD_{\ul
Q}=\calD_{Q_1}+\cdots +\calD_{Q_m}, \lb{a46} \\
& {\ul Q}=\{Q_1, \dots ,Q_m\} \in \sym^m \calK_\gg, \quad
Q_0\in\calK_\gg, \;\, m\in\bbN, \no
\end{align}
where for any $Q\in\calK_\gg$,
\begin{equation} \lb{a47}
\calD_Q \colon  \calK_\gg \to\bbN_0, \quad P \mapsto  \calD_Q (P)=
\begin{cases} 1 & \text{for $P=Q$},\\
0 & \text{for $P\in \calK_\gg\setminus \{Q\}$}, \end{cases}
\end{equation}
and $\sym^m \calK_\gg$ denotes the $m$th symmetric product of
$\calK_\gg$. In particular, $\sym^m \calK_\gg$ can be identified
with the set of nonnegative divisors $0 \leq \calD \in
\Div(\calK_\gg)$ of degree $m\in\bbN$. A divisor $\calD_{\ul
Q}=\calD_{Q_1}+\cdots +\calD_{Q_m}$ will be called finite if $Q_k\in 
\calK_p\setminus\{\Pinfp,\Pinfm\}$, $k=1,\dots,m$.

For $f\in \calM (\calK_\gg) \setminus \{0\}$, $\omega \in \calM^1
(\calK_\gg) \setminus \{0\}$ the divisors of $f$ and $\omega$ are
denoted by $(f)$ and $(\omega)$, respectively.  Two divisors
$\calD$, $\calE\in \Div(\calK_\gg)$ are called equivalent, denoted
by $\calD \sim \calE$, if and only if $\calD -\calE =(f)$ for some
$f\in\calM (\calK_\gg) \setminus \{0\}$.  The divisor class
$[\calD]$ of $\calD$ is then given by $[\calD] =\{\calE \in
\Div(\calK_{\gg})\mid\calE \sim \calD\}$.  We recall that
\begin{equation}
\deg ((f))=0, \quad \deg ((\omega)) =2(\gg-1), \quad f\in\calM (\calK_\gg)
\setminus \{0\}, \;  \omega\in \calM^1 (\calK_\gg) \setminus
\{0\}, \lb{a48}
\end{equation}
where the degree $\deg (\calD)$ of $\calD$ is given by $\deg
(\calD) =\sum_{P\in \calK_\gg} \calD (P)$.  It is customary to
call $(f)$ (respectively, $(\omega)$) a principal (respectively,
canonical) divisor.

Introducing the complex linear spaces
\begin{align}
\calL (\calD) & =\{f\in \calM (\calK_\gg)\mid f=0
      \text{ or } (f) \geq \calD\}, \quad 
r(\calD) =\dim_\bbC \calL (\calD),
\lb{a49}\\
\calL^1 (\calD) & =
      \{ \omega\in \calM^1 (\calK_\gg)\mid \omega=0
      \text{ or } (\omega) \geq
\calD\},\quad  i(\calD) =\dim_\bbC \calL^1 (\calD)  \lb{a50}
\end{align}
(with $i(\calD)$ the index of specialty of $\calD$), one infers
that $\deg (\calD)$, $r(\calD)$, and $i(\calD)$ only depend on the
divisor class $[\calD]$ of $\calD$.  Moreover, we recall the
following fundamental facts.

%%%%%%%%%%%%%%%%%%%%%%%%%%%%%%%%%%%%%%%%%%
\begin{theorem} \lb{thm1}
Let $\calD \in \Div(\calK_\gg)$, $\omega \in \calM^1 (\calK_\gg)
\setminus \{0\}$. Then,
\begin{equation}
      i(\calD) =r(\calD-(\omega)), \quad \gg\in\bbN_0.
\lb{a51}
\end{equation}
The Riemann-Roch theorem reads
\begin{equation}
r(-\calD) =\deg (\calD) + i (\calD) -\gg+1, \quad n\in\bbN_0.
\lb{a52}
\end{equation}
By Abel's theorem, $\calD\in \Div(\calK_\gg)$, $\gg\in\bbN$, is
principal if and only if
\begin{equation}
\deg (\calD) =0 \text{ and } \ul \alpha_{Q_0} (\calD) =\ul{0}.
\lb{a53}
\end{equation}
Finally, assume $\gg\in\bbN$. Then $\ul \alpha_{Q_0} :
\Div(\calK_\gg) \to J(\calK_\gg)$ is surjective $($Jacobi's
inversion theorem$)$.
\end{theorem}
%%%%%%%%%%%%%%%%%%%%%%%%%%%%%%%%%%%%%%%%%%

%%%%%%%%%%%%%%%%%%%%%%%%%%%%%%%%%%%%%%%%%%
\begin{theorem} \lb{thm2}
Let $\calD_{\ul Q} \in \sym^{\gg} \calK_\gg$, $\ul Q=\{Q_1,
\ldots, Q_\gg\}$.  Then,
\begin{equation}
1 \leq i (\calD_{\ul Q} ) =s \lb{a54}
\end{equation}
if and only if there are $s$ pairs of the type $\{P,
P^*\}\subseteq \{Q_1,\ldots, Q_\gg\}$ $($this includes, of course,
branch points for which $P=P^*$$)$. Obviously, one has $s\leq
\gg/2$.
\end{theorem}
%%%%%%%%%%%%%%%%%%%%%%%%%%%%%%%%%%%%%%%%%%
Next, we denote by $\ul \Xi_{Q_0}=(\Xi_{Q_{0,1}}, \dots,
\Xi_{Q_{0,\gg}})$ the vector of Riemann constants,
\begin{equation}
\Xi_{Q_{0,j}}=\frac12(1+\tau_{j,j})- \sum_{\substack{\ell=1 \\
\ell\neq j}}^\gg\int_{a_\ell} \omega_\ell(P)\int_{Q_0}^P\omega_j,
\quad j=1,\dots,\gg. \lb{a55}
\end{equation}

%%%%%%%%%%%%%%%%%%%%%%%%%%%%%%%%%%%%%%%%%%
\begin{theorem} \lb{thm3}
Let $\ul Q =\{Q_1,\dots,Q_\gg\}\in \sym^{\gg} \calK_\gg$ and
assume $\calD_{\ul Q}$ to be nonspecial, that is, $i(\calD_{\ul
Q})=0$. Then,
\begin{equation}
\theta\big(\ul {\Xi}_{Q_0} -\ul {A}_{Q_0}(P) + \alpha_{Q_0}
(\calD_{\ul Q})\big)=0 \text{ if and only if }
P\in\{Q_1,\dots,Q_\gg\}. \lb{a56}
\end{equation}
\end{theorem}
%%%%%%%%%%%%%%%%%%%%%%%%%%%%%%%%%%%%%%%%%%

%%%%%%%%%%%%%%%%%%%%%%%%%%%%%%%%%%%%%%%%%%
%%%%%%%%%%% appendix B %%%%%%%%%%%%%%%%%%%%%%%%%
\section{Some Interpolation Formulas} \lb{sB}
\renewcommand{\theequation}{B.\arabic{equation}}
\renewcommand{\thetheorem}{B.\arabic{theorem}}
\setcounter{theorem}{0}
\setcounter{equation}{0}
%%%%%%%%%%%%%%%%%%%%%%%%%%%%%%%%%%%%%%%%%%
%%%%%%%%%%%%%%%%%%%%%%%%%%%%%%%%%%%%%%%%%%

In this appendix we recall a useful interpolation formula which goes
beyond the standard Lagrange interpolation formula for polynomials in the
sense that the zeros of the interpolating polynomial need not be distinct.

%%%%%%%%%%%%%%%%%%%%%%%%%%%%%%%%%%%%%%%%%%
\begin{lemma} \lb{tB.1}
Let $p\in\bbN$ and $S_{p-1}$ be a polynomial of degree $p-1$. In
addition, let $F_p$ be a monic polynomial of degree $p$ of the form
\begin{equation}
F_p(z)=\prod_{k=1}^q (z-\mu_k)^{p_k}, \quad p_j \in\bbN, \; 
\mu_j\in\bbC, \; j=1,\dots,q, \quad \sum_{k=1}^q p_k =p.   \lb{B.1}
\end{equation}
Then,
\begin{align}
S_{p-1}(z)&=F_p(z)\sum_{k=1}^q \sum_{\ell=0}^{p_k -1} 
\f{S_{p-1}^{(\ell)}(\mu_k)}{\ell! (p_k-\ell-1)!}  \lb{B.2} \\
& \quad \times \bigg(\f{d^{p_k-\ell-1}}{d \zeta^{p_k-\ell-1}}
\Bigg((z-\zeta)^{-1}\prod_{k'=1, \, k'\neq k}^q 
(\zeta-\mu_{k'})^{-p_{k'}}\Bigg)
\Bigg)\Bigg|_{\zeta=\mu_k}, \quad z\in\bbC.   \no
\end{align}
In particular, $S_{p-1}$ is uniquely determined by prescribing the $p$
values
\begin{equation}
S_{p-1}(\mu_k), S_{p-1}'(\mu_k),\dots,S_{p-1}^{(p_{k}-1)}(\mu_k), \quad 
k=1,\dots,q,   \lb{B.3}
\end{equation}
at the given points $\mu_1.\dots,\mu_q$. \\
Conversely, prescribing the $p$ complex numbers
\begin{equation}
\alpha_k^{(0)}, \alpha_k^{(1)},\dots,\alpha_k^{(p_k-1)}, \quad
k=1,\dots,q, 
\end{equation}
there exists a unique polynomial $T_{p-1}$ of degree $p-1$,
\begin{align}
T_{p-1}(z)&=F_p(z)\sum_{k=1}^q \sum_{\ell=0}^{p_k -1} 
\f{\alpha_k^{(\ell)}}{\ell! (p_k-\ell-1)!}  \lb{B.3a} \\
& \quad \times \bigg(\f{d^{p_k-\ell-1}}{d \zeta^{p_k-\ell-1}}
\Bigg((z-\zeta)^{-1}\prod_{k'=1, \, k'\neq k}^q 
(\zeta-\mu_{k'})^{-p_{k'}}\Bigg)
\Bigg)\Bigg|_{\zeta=\mu_k}, \quad z\in\bbC,  \no
\end{align}
such that
\begin{equation}
T_{p-1}(\mu_k)=\alpha_k^{(0)}, T_{p-1}'(\mu_k)=\alpha_k^{(1)},
\dots,T_{p-1}^{(p_{k}-1)}(\mu_k)=\alpha_k^{(p_k-1)}, \quad 
k=1,\dots,q.   \lb{B.3b}
\end{equation}
\end{lemma}
%%%%%%%%%%%%%%%%%%%%%%%%%%%%%%%%%%%%%%%%
\begin{proof}
Our starting point for proving \eqref{B.2} is the following formula
derived, for instance, by Markushevich \cite[Part 2, Sect.\ 2.11, p.\
68]{Ma85},
\begin{equation}
S_{p-1}(z)=\f{1}{2\pi i} \oint_{\Gamma} 
\f{d\zeta \, S_{p-1}(\zeta)}{F_p(\zeta)} \f{F_p(\zeta)-F_p(z)}{\zeta -z},
\quad z\in\bbC,   \lb{B.4}
\end{equation}
where $\Gamma$ is a simple, smooth, counterclockwise oriented curve
encircling $\mu_1,\dots,\mu_q$ strictly in its interior. Since the
integrand in \eqref{B.4} is analytic at the point $\zeta=z$, we may,
without loss of generality, assume that $\Gamma$ does not encircle $z$.
With this assumption one obtains 
\begin{equation}
\f{1}{2\pi i} \oint_{\Gamma} \f{d\zeta \, S_{p-1}(\zeta)}{\zeta-z}=0 
\lb{B.5}
\end{equation}
and hence deforming $\Gamma$ into sufficiently small counterclockwise
oriented circles $\Gamma_k$ with center at $\mu_k$, $k=1,\dots,q$, such
that no $\mu_{k'}$, $k'\neq k$, is encircled by $\Gamma_k$, one obtains 
\begin{align}
S_{p-1}(z) & =-\f{F_p(z)}{2\pi i}\oint_{\Gamma} 
\f{d\zeta \, S_{p-1}(\zeta)}{F_p(\zeta)(\zeta-z)}  \no \\
& = -\f{F_p(z)}{2\pi i} \sum_{k=1}^q \oint_{\Gamma_k} 
\f{d\zeta \, S_{p-1}(\zeta)}{F_p(\zeta)(\zeta-z)}  \no \\
& = -\f{F_p(z)}{2\pi i} \sum_{k=1}^q \sum_{\ell=0}^{p-1} 
\f{S_{p-1}^{(\ell)}(\mu_k)}{\ell!} \oint_{\Gamma_k} 
\f{d\zeta \, (\zeta-\mu_k)^\ell}{F_p(\zeta)(\zeta-z)}  \no \\
& = -\f{F_p(z)}{2\pi i} \sum_{k=1}^q \sum_{\ell=0}^{p-1} 
\f{S_{p-1}^{(\ell)}(\mu_k)}{\ell!} \oint_{\Gamma_k} 
\f{d\zeta \, (\zeta-\mu_k)^\ell}{(\zeta-z)\prod_{k'=1}^q
(\zeta-\mu_{k'})^{p_{k'}}}  \no \\ 
& = -\f{F_p(z)}{2\pi i} \sum_{k=1}^q \sum_{\ell=0}^{p-1} 
\f{S_{p-1}^{(\ell)}(\mu_k)}{\ell!} \oint_{\Gamma_k} 
\f{d\zeta \,
(\zeta-\mu_k)^{\ell-p_k}}{(\zeta-z)\prod_{\substack{k'=1\\k'\neq k}}^q
(\zeta-\mu_{k'})^{p_{k'}}}   \no \\ 
& = -\f{F_p(z)}{2\pi i} \sum_{k=1}^q \sum_{\ell=0}^{p_k-1} 
\f{S_{p-1}^{(\ell)}(\mu_k)}{\ell!} \oint_{\Gamma_k} 
\f{d\zeta \,
(\zeta-\mu_k)^{\ell-p_k}}{(\zeta-z)\prod_{\substack{k'=1\\k'\neq k}}^q
(\zeta-\mu_{k'})^{p_{k'}}},   \lb{B.6}
\end{align}
where we used
\begin{equation}
\oint_{\Gamma_k} d\zeta \, (\zeta-\mu_k)^{\ell-p_k} f(\zeta)=0 \, 
\text{ if } \, \ell\geq p_k, \, \ell\in\bbN,   \lb{B.7}
\end{equation}
for any function $f$ analytic in a neighborhood of the disk $D_k$ with
boundary $\Gamma_k$, $k=1,\dots,q$, to arrive at the last line of
\eqref{B.6}. An application of Cauchy's formula for derivatives of
analytic functions to \eqref{B.6} then yields
\begin{align}
S_{p-1}(z) & = -F_p(z) \sum_{k=1}^q \sum_{\ell=0}^{p_k-1} 
\f{S_{p-1}^{(\ell)}(\mu_k)}{\ell!}   \no \\
& \quad \times \f{1}{2\pi i} \oint_{\Gamma_k} 
d\zeta \f{1}{(\zeta-\mu_k)^{(p_k-\ell-1)+1}}
\f{1}{(\zeta-z)\prod_{k'=1, \, k'\neq k}^q
(\zeta-\mu_{k'})^{p_{k'}}}  \no \\
& = F_p(z)\sum_{k=1}^q \sum_{\ell=0}^{p_k -1} 
\f{S_{p-1}^{(\ell)}(\mu_k)}{\ell! (p_k-\ell-1)!}  \no \\
& \quad \times \bigg(\f{d^{p_k-\ell-1}}{d \zeta^{p_k-\ell-1}}
\Bigg(\f{1}{(z-\zeta)\prod_{k'=1, \, k'\neq k}^q 
(\zeta-\mu_{k'})^{p_{k'}}}\Bigg)
\Bigg)\Bigg|_{\zeta=\mu_k}, \quad z\in\bbC,   \lb{B.8}
\end{align}
and hence \eqref{B.2}. Conversely, a linear algebraic argument shows that 
any polynomial $T_{p-1}$ of degree $p-1$ is uniquely determined by data
of the type
\begin{equation}
T_{p-1}(\mu_k), T_{p-1}'(\mu_k),\dots,T_{p-1}^{(p_{k}-1)}(\mu_k), \quad 
k=1,\dots,q.   \lb{B.9}
\end{equation}
Uniqueness of the representation \eqref{B.2} then proves \eqref{B.3a}.
\end{proof}
%%%%%%%%%%%%%%%%%%%%%%%%%%%%%%%%%%%%%%%%%%

We briefly mention two special cases of \eqref{B.2}. First, assume the
generic case where all zeros of $F_p$ are distinct, that is,
\begin{equation}
q=p, \quad p_k=1, \quad \mu_k \neq \mu_{k'} \, 
\text{ for } \, k\neq k', \; k,k'=1,\dots,p. \lb{B.10}
\end{equation}
In this case \eqref{B.2} reduces to the classical Lagrange interpolation
formula
\begin{equation}
S_{p-1}(z)=F_p(z)\sum_{k=1}^p 
\f{S_{p-1}(\mu_k)}{((d
F_p(\zeta)/d\zeta)|_{\zeta=\mu_k})(z-\mu_k)}, \quad z\in\bbC.  \lb{B.11}
\end{equation}
Second, we consider the other extreme case where all zeros of $F_p$
coincide, that is,
\begin{equation}
q=1, \quad p_1=p, \quad F_p(z)=(z-\mu_1)^p, \quad z\in\bbC.  \lb{B.12}
\end{equation}
In this case \eqref{B.2} reduces of course to the Taylor expansion of
$S_{p-1}$ around $z=\mu_1$
\begin{equation}
S_{p-1}(z)=\sum_{\ell=0}^{p-1} \f{S_{p-1}^{(\ell)}(\mu_1)}{\ell!}
(z-\mu_1)^\ell, \quad z\in\bbC.   \lb{B.13}
\end{equation}

%%%%%%%%%%%%%%%%%%%%%%%%%%%%%%%%%%%%%%%%%%%
%%%%%%%%%%% appendix C %%%%%%%%%%%%%%%%%%%%%%%%%
\section{Asymptotic Spectral Parameter Expansions and Nonlinear Recursion
Relations} \lb{sC}
\renewcommand{\theequation}{C.\arabic{equation}}
\renewcommand{\thetheorem}{C.\arabic{theorem}}
\setcounter{theorem}{0}
\setcounter{equation}{0}
%%%%%%%%%%%%%%%%%%%%%%%%%%%%%%%%%%%%%%%%%%%
%%%%%%%%%%%%%%%%%%%%%%%%%%%%%%%%%%%%%%%%%%%

In this appendix we discuss asymptotic spectral parameter expansions for
$F_p/y$ and $G_{p+1}/y$ as well as nonlinear recursion relations for the
corresponding homogeneous coefficients $\hat f_\ell$ and $\hat g_\ell$ 
and analogous quantities fundamental to the polynomial recursion formalism
for the Toda hierarchy.

We start by recalling the following elementary results (which are
consequences of the  binomial expansion). Let
\begin{align}
&\{E_m\}_{m=0,\dots,2p+1}\subset\bbC \; \text{ for some } \;
p\in\bbN_0 \lb{h.1} \\
& \text{and } \; \eta\in\bbC \; \text{ such that } \;
|\eta|<\min\{|E_0|^{-1},\dots, |E_{2p+1}|^{-1}\}. \no
\end{align}
Then
\begin{equation}
\bigg(\prod_{m=0}^{2p+1} \big(1-E_m\eta \big)
\bigg)^{-1/2}=\sum_{k=0}^{\infty}\hat c_k(\ul E)\eta^{k}, \lb{sbB2.26d}
\end{equation}
where
\begin{align}
\hat c_0(\ul E)&=1,\no \\
\hat c_k(\ul E)&=\!\!\!\!\!\!\!\sum_{\substack{j_0,\dots,j_{2p+1}=0\\
   j_0+\cdots+j_{2p+1}=k}}^{k}\!\!
\f{(2j_0)!\cdots(2j_{2p+1})!}
{2^{2k} (j_0!)^2\cdots (j_{2p+1}!)^2}E_0^{j_0}\cdots E_{2p+1}^{j_{2p+1}},
\quad  k\in\bbN. \label{T2.26e}
\end{align}
The first few coefficients explicitly read
\begin{align}
\hat c_0(\ul E)&=1, \quad
\hat c_1(\ul E)=\f12\sum_{m=0}^{2p+1} E_m,\no \\
\hat c_2(\ul E)&=\f14\sum_{\substack{m_1,m_2=0\\ m_1< m_2}}^{2p+1}
E_{m_1} E_{m_2}+\f38 \sum_{m=0}^{2p+1} E_m^2,
\quad \text{etc.} \lb{T2.26f}
\end{align}
Similarly,
\begin{equation}
\bigg(\prod_{m=0}^{2p+1} \big(1-E_m \eta \big)
\bigg)^{1/2}=\sum_{k=0}^{\infty}c_k(\ul E)\eta^{k}, \lb{sbB2.26g}
\end{equation}
where
\begin{align}
c_0(\ul E)&=1,\no \\
c_k(\ul E)&=\!\!\!\!\!\!\!\!\sum_{\substack{j_0,\dots,j_{2p+1}=0\\
   j_0+\cdots+j_{2p+1}=k}}^{k}\!\!
\f{(2j_0)!\cdots(2j_{2p+1})!\, E_0^{j_0}\cdots E_{2p+1}^{j_{2p+1}}}
{2^{2k} (j_0!)^2\cdots (j_{2p+1}!)^2 (2j_0-1)\cdots(2j_{2p+1}-1)}, \quad
k\in\bbN. \label{T2.26h}
\end{align}
The first few coefficients are given explicitly by
\begin{align}
c_0(\ul E)&=1, \quad
c_1(\ul E)=-\f12\sum_{m=0}^{2p+1} E_m, \no \\
c_2(\ul E)&=\f14\sum_{\substack{m_1,m_2=0\\ m_1< m_2}}^{2p+1}
E_{m_1} E_{m_2}-\f18 \sum_{m=0}^{2p+1} E_m^2,
\quad \text{etc.} \lb{T2.26i}
\end{align}

%%%%%%%%%%%%%%%%%%%%%%%%%%%%%%%%%%%%%%%%%%
\begin{theorem} \lb{tE.8}
Assume \eqref{1.2.1}, $\sTl_\p(a,b)=0$, and suppose
$P=(z,y)\in\calK_\p\setminus\{\Pinfplus,\Pinfmin\}$. Then
$F_\p/y$ and $G_{\p+1}/y$ have the following
convergent expansions as $P\to\Pinfpm$,
\begin{equation}
\f{F_\p(z)}{y} = \mp\sum_{\ell=0}^\infty\hat f_\ell \zeta^{\ell+1}, \quad
\f{G_{\p+1}(z)}{y} = \mp\sum_{\ell=-1}^\infty\hat g_\ell
\zeta^{\ell+1}, \lb{TlH.8}
\end{equation}
where $\zeta=1/z$ is the local coordinate near $\Pinfpm$ and $\hat
f_\ell$ and $\hat g_\ell$ are the homogeneous versions of the coefficients $f_\ell$ and
$g_\ell$ introduced in \eqref{1.2.6}. In particular, $\hat
f_\ell$ and $\hat g_\ell$ can be computed from the nonlinear recursion
relations
\begin{align}
&\hat f_0=1, \quad \hat f_1=-b, \quad \hat f_2=a^2+(a^-)^2+b^2, \no \\
&\hat f_{\ell+2}=-\frac12\sum_{k=1}^{\ell+1}\hat f_{\ell+2-k}\hat f_k
-2b\sum_{k=0}^{\ell+1}\hat f_{\ell+1-k}\hat f_k\no \\
& \hspace*{1.1cm}
+\sum_{k=0}^{\ell}\big(-3b^2\hat f_{\ell-k}\hat f_k
+a^2\hat f_{\ell-k}^+\hat f_k
+(a^-)^2\hat f_{\ell-k}\hat f_k^-\big)\no \\
&  \hspace*{1.1cm}
+\sum_{k=0}^{\ell-1}\big(-2b^3 \hat f_{\ell-1-k}\hat f_k
+2a^2b \hat f_{\ell-1-k}^+\hat f_k
+2(a^-)^2b \hat f_{\ell-1-k}\hat f_k^-\big)\no \\
&  \hspace*{1.1cm}
+\sum_{k=0}^{\ell-2}\big(a^2b^2 \hat f_{\ell-2-k}^+\hat f_k
+(a^-)^2b^2 \hat f_{\ell-2-k}\hat f_k^-
+a^2 (a^-)^2 \hat f_{\ell-2-k}^+\hat f_k^- \no \\
&  \hspace*{2.15cm}
-\frac12 a^4 \hat f_{\ell-2-k}^+\hat f_k^+
-\frac12 (a^-)^4 \hat f_{\ell-2-k}^-\hat f_k^-
\big), \quad \ell\in\bbN, \lb{E.80} \\
\intertext{and}
&\hat g_{-1}=-1, \quad \hat g_0=0, \quad  \hat g_{1}=-2a^2, \no \\
&\hat g_{\ell+1}=
\frac12\sum_{k=-1}^{\ell}(b+b^+)\hat g_{\ell-1-k}\hat g_k
+\frac12\sum_{k=0}^{\ell}\hat g_{\ell-k}\hat g_k  \lb{E.81} \\
& \hspace*{1.05cm}
+\frac12\sum_{k=-1}^{\ell-1}\big(b b^+\hat g_{\ell-2-k}\hat g_k
-a^2(\hat g_{\ell-2-k}^- +\hat g_{\ell-2-k})(\hat g_{k}+\hat g_{k}^+)\big),
\quad \ell\in\bbN.\no
\end{align}
Moreover, one infers for the $E_m$-dependent summation constants
$c_\ell$, $\ell=0,\dots,\p+1$, in $F_\p$ and $G_{p+1}$ that
\begin{equation}
c_\ell=c_\ell(\ul E), \quad \ell=0,\dots,\p+1 \lb{E.12}
\end{equation}
and\footnote{$m\mini n=\min\{m,n\}$.}
\begin{align}
f_\ell&=\sum_{k=0}^\ell c_{\ell-k}(\ul E) \hat f_k,  \quad
\ell=0,\dots,p, \lb{E.12a} \\
%\begin{cases}g_\ell& \text{for $\ell=0,\dots,p-1$},\\
%g_p+f_{p+1}& \text{for $\ell=p$},
%\end{cases}&=\sum_{k=0}^\ell c_{\ell-k}(\ul E) \hat g_k - c_{\ell+1}(\ul E),
%\quad \ell=0,\dots,p+1, \lb{E.34} \\
g_\ell+f_{p+1}\delta_{p,\ell}&=\sum_{k=0}^\ell c_{\ell-k}(\ul E) \hat g_k
- c_{\ell+1}(\ul E),
\quad \ell=0,\dots,p, \lb{E.34a} \\
\hat f_\ell&=\sum_{k=0}^{\ell\wedge p} \hat c_{\ell-k}(\ul E) f_k, \quad \ell\in\bbN_0, \lb{E.35} \\
\hat g_\ell&=\sum_{k=0}^{\ell\wedge p} \hat c_{\ell-k}(\ul E) 
(g_k+f_{p+1}\delta_{p,k}) -\hat c_{\ell+1}(\ul E),  
\quad \ell\in\bbN_0. \lb{E.35a}
\end{align}
\end{theorem}
%%%%%%%%%%%%%%%%%%%%%%%%%%%%%%%%%%%%%%%%%%
\begin{proof}  Dividing $F_\p$ and $G_{\p+1}$ by $R_{2\p+2}^{1/2}$
    (temporarily fixing the branch of $R_{2\p+2}^{1/2}$ as $z^{p+1}$
    near infinity) one obtains
\begin{align}
\frac{F_\p(z)}{R_{2\p+2}(z)^{1/2}}&\underset{|z|\to\infty}{=}
\bigg(\sum_{k=0}^\infty\hat c_k(\ul E)z^{-k}\bigg)
\bigg(\sum_{\ell=0}^{\p} f_\ell z^{-\ell-1}\bigg)
= \sum_{\ell=0}^\infty {\check f}_{\ell} z^{-\ell-1}, \lb{E.38} \\
\f{G_{\p+1}(z)}{R_{2\p+2}(z)^{1/2}}&\underset{|z|\to\infty}{=}
\bigg(\sum_{k=0}^\infty \hat c_k(\ul E)z^{-k}\bigg)
\bigg(\sum_{\ell=0}^{\p+1} \tilde g_\ell z^{-\ell}\bigg)
=z^{-1} \sum_{\ell=-1}^\infty {\check g}_\ell z^{-\ell} \lb{E.39}
\end{align}
for some coefficients $\check f_\ell$ and $\check g_\ell$ to be
determined next. Here we have temporarily introduced the notation
\begin{equation}
G_{\p+1}(z)= -z^{p+1}+ \sum_{\ell=0}^p g_{p-\ell} z^\ell + f_{p+1}
=\sum_{\ell=0}^{p+1} \tilde g_{p-\ell} z^\ell.
\end{equation}
Dividing \eqref{3.2.35a} and
\eqref{3.2.35c} by $R_{2\p+2}$ and inserting
expansions \eqref{E.38} and \eqref{E.39} into the resulting
equations then yield the nonlinear recursion relations \eqref{E.80} and
\eqref{E.81} (with  $\hat f_\ell$ and $\hat g_\ell$ replaced by $\check
f_\ell$ and $\check g_\ell$, respectively). More precisely, one first
obtains $\abs{\check f_0}=\abs{\check g_{-1}}=1$ and upon choosing the
signs of $\check f_0$ and $\check g_{-1}$ such that $\check f_0=\hat
f_0=1$ and $\check g_{-1}=-1$ one obtains \eqref{E.80} and \eqref{E.81}.
Next, dividing \eqref{1.2.15b} and \eqref{1.2.16a} by
$R_{2p+2}^{1/2}$, inserting the expansions \eqref{E.38} and \eqref{E.39},
and comparing powers of $z^{-\ell}$ as $z\to\infty$, one infers that
$\check f_\ell$ and $\check g_\ell$ satisfy the linear recursion relations
\eqref{1.2.4a}--\eqref{1.2.4c}. Hence one concludes that
\begin{equation}
\check f_\ell=f_\ell, \quad \check g_\ell=g_\ell, \quad \ell\in\bbN_0
\end{equation}
for certain values of the summation constants $c_\ell$.  To show that
$\check f_\ell=\hat f_\ell$, $\check g_\ell=\hat g_\ell$, and hence all
$c_\ell$, $\ell\in\bbN$, vanish, we recall the notion of degree as used
in the proof of Lemma \ref{l3.4.3}, which serves as an efficient tool to
distinguish between homogeneous and nonhomogeneous quantities. To this end
we employ the notation
\begin{align}
f^{(r)}=S^{(r)}f, \quad f=\{f(n)\}_{n\in\bbZ}\subset\bbC, \quad
   S^{(r)}&=\begin{cases}(S^+)^r, &\text{$r\ge 0$},\\
(S^-)^{-r}, &\text{$r< 0$},\end{cases} \quad
r\in \bbZ,  \lb{sb2.1AA}
\end{align}
and introduce
\begin{equation}
\deg (a^{(r)})=\deg (b^{(r)})=1, \quad r\in\bbZ.
\end{equation}
This results in
\begin{equation}
\deg(\hat f_\ell)=\ell, \quad \deg(\hat g_\ell)=\ell+1, \quad
\ell\in\bbN.
\end{equation}
using induction in the linear recursion relations 
\eqref{1.2.4a}--\eqref{1.2.4c}. Similarly, the nonlinear recursion
relations \eqref{E.80} and \eqref{E.81} yield inductively that
\begin{equation}
\deg(\check f_\ell)=\ell, \quad \deg(\check g_\ell)=\ell+1, \quad
\ell\in\bbN.
\end{equation}
Hence one concludes that
\begin{equation}
\check f_\ell=\hat f_\ell, \quad \check g_\ell=\hat g_\ell, \quad
\ell\in\bbN_0.
\end{equation}
A comparison of coefficients in \eqref{E.38} proves \eqref{E.35}. Similarly, we use \eqref{E.39} to establish  \eqref{E.35a}. Next, multiplying \eqref{sbB2.26d} and \eqref{sbB2.26g}, a
comparison of coefficients of $\eta^{k}$ yields
\begin{equation}
\sum_{\ell=0}^k \hat c_{k-\ell}(\ul E)c_\ell(\ul E) =\delta_{k,0},
\quad k\in\bbN_0. \lb{sbB2.37}
\end{equation}
Thus, one computes
\begin{align}
\sum_{m=0}^\ell c_{\ell-m}(\ul E)\hat f_m&=\sum_{m=0}^\ell \sum_{k=0}^m
c_{\ell-m}(\ul E)\hat c_{m-k}(\ul E) f_k
=\sum_{k=0}^\ell \sum_{p=k}^{\ell} c_{\ell-p}(\ul E)\hat c_{p-k}(\ul
E)f_k \no \\
&=\sum_{k=0}^\ell\bigg(\sum_{m=0}^{\ell-k} c_{\ell-k-m}(\ul E)
\hat c_{m}(\ul E)\bigg)f_k=f_\ell, \quad \ell=0,\dots,p, \lb{E.22}
\end{align}
applying \eqref{sbB2.37}. Hence one obtains \eqref{E.12a} and
thus \eqref{E.12} (cf.\ \eqref{1.2.7}). The corresponding proof of
\eqref{E.34a} is similar to that of $f_\ell$.
\end{proof}
%%%%%%%%%%%%%%%%%%%%%%%%%%%%%%%%%%%%%%%%%%

\medskip

%%%%%%%%%%%%%%%%%%%%%%%%%%%%%%%%%%%%%%%% 
%%%%%%%%%%%%%%%%%%%%%%%%%%%%%%%%%%%%%%%% 
\noindent {\bf Acknowledgments.} We are indebted to Michael Gekhtman for  
discussions on this subject. Fritz Gesztesy gratefully acknowledges the extraordinary
hospitality of Helge Holden and the Department of Mathematical Sciences of
the Norwegian University of Science and Technology, Trondheim, during a
two-month stay in the summer of 2005, where parts of this paper
were written. He also gratefully acknowledges a research leave for the
academic year 2005/06 granted by the Research Council and the Office of
Research of the University of Missouri-Columbia.Fritz Gesztesy and Helge Holden are grateful for the  
hospitality of the Mittag-Leffler Institute, Sweden, creating a great  
working environment for research, during the Fall of 2005. 
Gerald Teschl gratefully acknowledges the hospitality of
the Department of Mathematics of the University of Missouri--Columbia and
the Department of Mathematical Sciences of the Norwegian University of
Science and Technology, Trondheim, respectively, during two one-week
stays in 2005.
%%%%%%%%%%%%%%%%%%%%%%%%%%%%%%%%%%%%%%%%%%%
%%%%%%%%%%%%%%%%%%%%%%%%%%%%%%%%%%%%%%%% 

%%%%%%%%%%%%%%%%%%%%%%%%%%%%%%%%%%%%%%%%%%%-------------------
% when bibtexing please active the next two lines
%\bibliographystyle{plainMOD}
%\bibliography{solitonref}
%------- end bibtex
% after bibtex, please paste file CH.bbl into this file below here
% and comment away the (first) two lines above.
%%%%%%%%%%%%%%%%%%%%%%%%%%%%%%%%%%%%%%%%%%
%%%%%%%%%%%%%%%%%%%%%%%%%%%%%%%%%%%%%%%%%%

\end{document}